\pdfoutput=1
\documentclass[aps,superscriptaddress,floatfix,nofootinbib,preprintnumbers,eqsecnum,amsmath,amssymb,twocolumn]{revtex4-1}

\usepackage{slashed, color}
\usepackage{graphicx}
\usepackage{dcolumn}
\usepackage{bm}

\begin{document}

\preprint{CTPU-PTC-20-25}

\title{Exploring the Universe with Dark Light Scalars}

\author{Bugeon Jo}
 \email{whqnrjs@gmail.com}
\author{Hyeontae Kim}
 \email{htkim428@gmail.com}
 \author{Hyung Do Kim}
 \email{hdkim@phya.snu.ac.kr}
\affiliation{
 Department of Physics and Astronomy,\\
 Seoul National University,\\
 Seoul 08826, Korea
}

\author{Chang Sub Shin}
 \email{csshin@ibs.re.kr}
\affiliation{
Center for Theoretical Physics of the Universe,\\
Institute for Basic Science (IBS),\\
Daejeon 34051, Korea
}

\date{\today}

\begin{abstract}
We study the cosmology of the dark sector consisting of (ultra) light scalars. Since the scalar mass is radiatively unstable, a special explanation is required to make the mass much smaller than the UV scale. There are two well-known mechanisms for the origin of scalar mass. The scalar can be identified as a pseudo-Goldstone boson, whose shift symmetry is explicitly broken by non- perturbative corrections, like the axion. Alternatively, it can be identified as a composite particle like the glueball, whose mass is limited by the confinement scale of the theory because no scalar degree of freedom exists at high scales. In both cases, the scalar can be naturally light, but interaction behavior is quite different. The lighter the axion (glueball), the weaker (stronger) it interacts. As the simplest non-trivial example, we consider the dark axion whose shift symmetry is anomalously broken by the hidden non-abelian gauge symmetry. After the confinement of the gauge group, the dark axion and the dark glueball get masses and both form multicomponent dark matter.
We carefully consider the effects of energy flow from the dark gluons to the dark axions and derive the full  equations of motion for the background and the perturbed variables.  The effect of the dark axion-dark gluon coupling on the evolution of the entropy and the isocurvature perturbations is also clarified. Finally, we discuss the gravo-thermal collapse of the glueball subcomponent dark matter after the halos form, in order to explore the potential to contribute to the formation of seeds for the supermassive black holes  observed at high redshifts. 
With the simplified assumptions, the glueball subcomponent dark matter with the mass of $0.01- 0.1 {\rm MeV}$, and the axion main dark matter component with  the decay constant $f_a={\cal O}(10^{15}-10^{16}){\rm GeV}$, the mass of ${\cal O}(10^{-14}-10^{-18})\,{\rm eV}$, can provide the hint on the origin of the supermassive black holes at high redshifts. 
 \end{abstract}

\maketitle


\section{\label{sec:1}Introduction}

The discovery of the Higgs boson   completed the Standard Model which explains most of the phenomena in the Universe including nuclei, atoms and their interactions. 
However, various studies based on precise measurements of the cosmic microwave background (CMB), velocity distributions of stars and galaxies, and large-scale structures (LSS) show that only 4\% of the Universe is understood by our knowledge of the Standard Models and 96\% must be filled with dark matter and dark energy that we do not know their origin \cite{Akrami:2018odb}. 

Among dark matter candidates, particles classified as `weakly interacting massive particle' (WIMP) were considered as the best candidate. This is because their freeze-out relic abundance can naturally explain the present amount of dark matter, and they can be tested by different on-going searches. Their mass scale can also be related to new physics that explains why the electroweak scale is stable against various quantum corrections. 

However, there is no conclusive evidence of WIMP dark matter so far, and the alternative candidates got more attention in recent years. The masses of these candidates are not limited to a narrow range, and various ideas have been proposed particularly focusing on the detection possibility \cite{Battaglieri:2017aum}.   
For such a broad range of dark matter mass, there is no clear guiding principle to specify its natural range. Especially, it is extremely unnatural to consider a light scalar compared to the probing scales unless there is a special reason for it. 

It is quite interesting to notice that an ultra-light scalar dark matter is allowed by cosmological and astrophysical observations, because the scalar can act as an oscillating classical field whose averaged equation of state is the same as that of the cold dark matter (CDM). Further interesting phenomena can arise from its field nature at small scales.   The fuzzy dark matter \cite{Hu:2000ke} and the QCD axion \cite{Dine:1981rt,Zhitnitsky:1980tq,Kim:1979if,Shifman:1979if} are good examples. Both candidates, the axion or axion like particle, can be naturally light by their approximate shift symmetry,  which  is explicitly broken by the controllable non-perturbative corrections. 
The consequence of the global symmetry is that the lighter the axion becomes, the weaker its interaction is. 
Therefore, a large occupation number of the (ultra) light axion is allowed, and it can be described by the evolution of  scalar condensate.

On the one hand, there is another natural way to obtain a light scalar dark matter. When the asymptotically free gauge group has confinement at low energy, the gauge fields are confined into the scalar particles, the glueball, whose mass is limited by the confining scale. Unlike in the case of the axion, the lighter the glueballs, the stronger the scattering cross-section among glueballs. The number changing interactions are active, so the occupation number is always limited by its temperature during its cosmological evolution. In this case, the relic density is determined by the freeze-out mechanism. Because of this property, the light dark glueball becomes a good candidate for self-interacting dark matter (SIDM) \cite{Carlson:1992fn,Boddy:2014yra}, which is one of the ways to make the cored density profile around the center of galaxies \cite{Spergel:1999mh}. As a subcomponent hot dark matter, it can also play the role to suppress small scale perturbations \cite{Buen-Abad:2018mas}.

These two mechanisms to obtain a light scalar dark matter provide completely different microscopic nature of dark matter, and yield different predictions for small scale evolution. 

In this paper we study the cosmology of light scalar dark matter, focusing on the origin of its mass and the consequences associated with it. 
We cover two mechanisms discussed above at the same time in a minimal set-up: 
the dark sector consisting of the axion and the confining hidden gauge symmetry without light  fermions. 
In this set-up, 
the axion's shift symmetry is non-perturbatively broken by a Chern-Simons type coupling between the axion and the dark gauge field, i.e., gluon.  
After the confinement of the gauge group, the axion and the glueball get masses and become a part of multicomponent dark matter. 
The idea that the mass of the ultra-light scalar dark matter originates from a confining hidden gauge symmetry was studied in  \cite{Davoudiasl:2017jke,Halverson:2018olu}, but comprehensive study on its cosmological evolution is still lacking. 

We derive a complete set of the equations of motion for the background and the perturbed variables of the dark axion and the dark gluon/glueball densities. Since the axion-gluon coupling provides energy transfer from the dark gluon to the dark axion,   we clarify its effect on thermodynamics of the dark gluon fluid and  entropy evolution. We also quantify the transfer of the isocurvature perturbation through the same coupling.

The multicomponent dark matter which simultaneously contains feebly interacting and strongly interacting particles has an interesting cosmological consequence.  When the glueball dark matter becomes a subcomponent, 
the stronger self-interaction between the glueballs is allowed and 
opens the possibility to form a black hole in the early Universe \cite{Pollack:2014rja,Choquette:2018lvq}. 
This may provide a possible answer to the question about the origin of observed supermassive black holes at high redshifts \cite{Mortlock_2011,DeRosa:2013iia,Wu:2015,Banados:2017unc}. We discuss the parameter space to provide the solution and possible caveats. 

The paper is organized as follows. 
In Sec.~\ref{sec:2}, we establish basic formalism from the Lagrangian to the dynamical equations of the background and perturbation variables of the coupled axion-gluon fluid. In Sec.~\ref{sec:3}, we focus on the background evolution of the glueball and the axion dark matter. The parametric dependence of the relic abundance of the glueball and the axion is also presented. Sec.~\ref{sec:4} is devoted to the evolution of the perturbed variables. For the initial conditions, we have three modes: adiabatic perturbation, isocurvature perturbation induced by the initial misalingnment of the axion and the temperature fluctuation of the dark gluon fluid.   
In Sec.~\ref{sec:5}, we discuss the  implication of the glueball subcomponent dark matter for the early formation of the supermassive black holes.  
Sec.~\ref{sec:6} is  conclusions.

\section{\label{sec:2}  axion dark matter and confining dark sector}

\subsection{General description of the model}

Our starting Lagrangian  of the dark sector is composed of the ultra-light axion $\phi$
whose field range is  $2\pi f_a$, and the dark gauge symmetry with the confinement scale $\Lambda$ ($\Lambda \ll f_a$).   The coupling between the axion and the dark gauge bosons are given as 
\begin{eqnarray}\label{Model}
 \hskip -0.6cm -\frac{\mathcal{L}_h}{\sqrt{-g}}= \frac{1}{2}(\partial_\mu\phi)^2  +\frac{1}{4} (G_{\mu\nu}^a)^2  + \frac{g_h^2 \phi }{32\pi^2 f_a} G^a_{\mu\nu}\tilde G^{a\mu\nu},
    \label{Lagrangian}
\end{eqnarray}
where  $G^a_{\mu\nu}$ is the dark gluon field strength and $g_h$ denotes the dark (hidden) gauge coupling.  For illustration, $SU(N)$ is taken as our dark gauge group. 
Although we will not consider an extremely large value of $N$ and only take $N^2={\cal O}(10)$ in concrete examples, we keep $N$-dependence explicitly in our discussion in order to organize the results using the large $N$ expansion.

Below the confinement scale,  the dynamics of the gauge fields can be described by the composite bosons, the glueballs.  The most relevant glueball for dark matter physics is the lightest glueball, $\varphi_g$. 
Considering the large $N$ limit (dominated by planar diagrams), and  the $4\pi$ factor from the naive dimensional analysis (with the cut-off of the order of $\Lambda$), 
the effective Lagrangian of $\varphi_g$ can be expanded in $(4\pi/N)(\varphi_g/m_g)$ as \cite{tHooft:1973alw,Witten:1979kh,Manohar:1983md,Coleman:1985rnk,Cohen:1997rt} 
\begin{eqnarray}\label{Leff}
-\frac{{\cal L}_{h{\rm eff}}}{\sqrt{-g}} 
&=& \frac{1}{2} (\partial_\mu \phi)^2 +  V(\phi)  +\frac{1}{2}(\partial_\mu \varphi_g)^2  + \frac{1}{2} m_g^2 \varphi_g^2   \nonumber\\
 && 
+ \frac{a_3}{3!} \Big(\frac{4\pi }{N}\Big)  m_g\varphi_g^3 + \frac{a_4}{4!}\Big(\frac{4\pi}{N}\Big)^2 \varphi_g^4 \nonumber\\
&&+ \frac{a_5}{5!}\Big(\frac{4\pi}{N}\Big)^3 \frac{\varphi_g^5}{m_g}  +\cdots,
\end{eqnarray}
where the lightest glueball's mass is denoted by  $m_g = {\cal O}(\Lambda)$, and the 
coefficients $a_i$ are expected to be ${\cal O}(1)$.

The axion also gets a scalar potential from the gluodynamics, which
can be written as a power series in  $(\phi/N f_a)^2$ 
around its CP conserving minimum \cite{Witten:1980sp, Witten:1998uka}
\begin{eqnarray} \label{axion_pot}
V(\phi) &=&  N^2 \Lambda^4\Big( \frac{c_2}{2}  \frac{\phi^2}{N^2 f_a^2} 
 + \frac{c_4}{4!} \frac{\phi^4}{N^4 f_a^4}  +\cdots \Big) \nonumber\\
 &=&\frac{1}{2} m_a^2 \phi^2 + \frac{c_4}{4! c_2}\frac{m_a^2}{ N^2 f_a^2} \phi^4 +\cdots.
\end{eqnarray}
In this expansion, the axion mass is given by  
\begin{eqnarray}\label{eq:axion_mass}
 m_a^2  &=& \frac{1}{f_a^2} \int d^4 x_{\textrm{{\tiny E}}} \left\langle \frac{g_h^2}{32\pi^2} G_{\textrm{{\tiny E}}}\tilde G_{\textrm{{\tiny E}}}(x_{\textrm{{\tiny E}}})\, \frac{g_h^2}{32\pi^2} G_{\textrm{{\tiny E}}}\tilde G_{\textrm{{\tiny E}}}(0)\right\rangle_{\phi=0}  
 \nonumber  \\ &=&  \Big(
 10^{-12}\sqrt{c_2}\,{\rm eV}\Big)^2
\Big(\frac{\Lambda}{{\rm MeV}}\Big)^4\Big(\frac{10^{15}\,{\rm GeV}}{f_a}\Big)^2,
\end{eqnarray}
where the integral is evaluated for Euclidean continuation of Eq.~(\ref{Model}). The $1/N^2$ factor for the quartic term of the axion potential leads to the suppression of anharmonic effects as long as the initial misalignment of the axion field is $\phi_i\lesssim f_a$.   

The glueball is not the lightest particle in our dark sector, and the symmetry allows the decay of the glueball to two axions as $\varphi_g\to \phi\phi$. We can infer the glueball life-time from the lattice calculation. 
The leading axion-glueball interaction can be obtained from the axion dependent glueball mass term:
\begin{eqnarray}\label{axion-dependent-mass}
m_g = m_g(\phi=0) \left(1  + g_2\frac{\phi^2}{N^2 f_a^2}+ {\cal O}\Big(\frac{\phi^4}{N^4 f_a^4}\Big)\right),
\end{eqnarray}
where $g_2\simeq -0.5$ ($-0.6$) for $N=3$ ($4$) \cite{DelDebbio:2006yuf,Vicari:2008jw}. Because the coefficient $g_2$ is not suppressed in the large $N$ limit \cite{DelDebbio:2006yuf}, its value is expected to remain as ${\cal O}(1)$ for all $N\geq 3$.
Through the interaction term $\phi^2\varphi_g^2$ from Eq.~(\ref{axion-dependent-mass}) and the cubic term $\varphi_g^3$ in Eq.~(\ref{Leff}), the one-loop diagram of the glueball provides 
the following effective Lagrangian, which is relevant for the glueball decay,
\begin{eqnarray} \label{eq:glueball_decay}
\left.\frac{\Delta {\cal L}_{\rm eff}}{\sqrt{-g}}\right|_\textrm{one-loop} = c_* \frac{  g_2 a_3 m_g^3}{4\pi N^3 f_a^2}  \phi^2 \varphi_g.
\end{eqnarray}  Here $c_*$ is the ${\cal O}(1)$ coefficient whose explicit value is not available at this moment.
From Eq.~(\ref{eq:glueball_decay}),
the life-time of the glueball is estimated as 
\begin{equation} 
\tau_{\varphi_g} \sim 10^{18}\,{\rm Gyr}\, \Big(\frac{N}{3}\Big)^6 \Big(\frac{f_a}{10^{13}\,{\rm GeV}}\Big)^4\Big(\frac{{\rm GeV}}{m_g}\Big)^5.
\end{equation}  
In the parameter space we will focus on, the glueball is cosmologically stable, so that 
both axion and glueball are dark matter of the Universe.

For cosmology, 
we consider the case that the dark sector and the visible sector are thermally decoupled at the beginning. In such a case, dark gluons/glueballs are thermalized by their own interactions at a  temperature $T_g$ that could be  different from the SM photon temperature $T_\gamma$.
Starting from the gluon fluid ($T_g\gg \Lambda$), 
as the Universe expands, $T_g$ drops and crosses the dark critical temperature $T_{g,c}={\cal O}( \Lambda)$,   
and the confining phase transition occurs.
Below $T_{g,c}$, all gluons are confined into the glueballs, and 
the evolution is described by the massive glueball fluids. 

The dark gluon temperature also affects the evolution of the dark axions.  
The leading term of the axion potential induced by the gluo-thermodynamics is 
\begin{equation}  \label{eq:pot_axion}
    V(T_g, \phi) =  
    \frac{1}{2} m_a^2(T_g) \phi^2.
\end{equation}
The axion mass $m_a(T_g)$ is well described by the dilute instanton gas approximation in the deconfining phase,
\begin{eqnarray} \label{eq:mains}
    m_a(T_g)\simeq  m_a \left(\frac{T_{g,c}}{T_g}\right)^{\eta_a}
    \quad {\rm for}\quad T_g\gtrsim T_{g,c},     
\end{eqnarray}
with $\eta_a=11 N/6 -2$. 
For  $N=3 (4)$, $\eta_a=3.5 (5.3)$ \cite{RevModPhys.53.43}.  After the confinement  ($T_g\lesssim T_{g,c}$), the axion mass is  saturated to its zero temperature value, $m_a(T_g)\simeq m_a$ \cite{Borsanyi:2016ksw}. 

Actually, the temperature dependence of the axion potential implies the existence of  the energy flow from the gluon fluid to the axions  as the temperature decreases.  
Then, a natural question is whether or not the entropy of the dark gluon also evolves during the energy transfer. 
In order to make it clearer, let us discuss the gluo-thermodynamics in more detail. The free energy density of the gluon/glueball fluid can be evaluated from the gluon partition function for a given temperature $T_g$. 
If the topological $\theta$-term 
\begin{equation} \theta\equiv \frac{\phi}{f_a}
\end{equation} is vanishing,  the free energy density $f_g$ is only the function of the temperature as 
$f_g(T_g)= - p_g$, where $p_g$ is the pressure of the gluon/glueball fluid. The energy ($\rho_g)$ and entropy ($s_g$) densities  are obtained by the thermodynamic relations, $s_g=  - df_g/dT_g$ and $\rho_g = T_g s_g- p_g$. 
On the other hand, the situation is a little bit different for the non-vanishing $\theta$-term. 
Because the gluon partition function also depends on $\theta$, 
the free energy density (the negative of the  pressure for the gluon fluid with the $\theta$-term) is evaluated as \cite{DelDebbio:2006yuf}
\begin{eqnarray}
f_{g+\theta}(T_g) =  - p_g + V(T_g, \phi), 
\end{eqnarray}  
where $p_g\equiv - f_g(T_g)$. The second term of the RHS represents the contribution of the vacuum energy density generated by the non-perturbative gluo-thermodynamics. 
The energy density of the gluon fluid with the $\theta$-term 
also can be decomposed into the sum of 
the vacuum energy density  and the pure gluonic contribution as
$\rho_{g+\theta}(T_g) =\rho_g 
+ V(T_g, \phi )$. Then, the thermodynamics relations provide 
\begin{eqnarray} \label{eq:gluoentropy}
s_{g+\theta} &=& -\frac{ d f_{g+\theta}(T_g)}{d T_g} =  \frac{dp_g}{dT_g} 
 - \frac{\partial V(T_g, \phi)}{\partial T_g} \nonumber\\
 &=& \frac{\rho_{g+\theta}(T_g) - f_{g+\theta}(T_g)}{T_g} = \frac{\rho_g + p_g}{T_g} = s_g.
\end{eqnarray} 
The relation $s_{g+\theta} = s_g$ implies that the entropy of the dark sector is mostly given by the gluonic excitations, not by the axionic excitations. This is the natural consequence because the axion is homogeneously distributed in space, and its time evolution is negligible compared to the thermal process of the gluon plasma. 
On the other hand, the entropy and the energy density of the gluon fluid depend not only on its temperature but also on the axion field value as  Eq.~(\ref{eq:gluoentropy}) and 
\begin{eqnarray} \label{eq:gluon}
\rho_g&=&  
T_g\frac{dp_g}{d T_g} - p_g 
- T_g\frac{\partial V(T_g,\phi)}{\partial T_g}.
\end{eqnarray}
From the continuity equation of the dark sector, 
we can explicitly show that the entropy of the dark sector in a comoving volume $s_g a^3$  is conserved for whatever value of $\phi$ during adiabatic evolution. 
 
Based on this observation, before discussing the explicit evolution of each component,  we address general equations of motion for axion and gluon as the fluids including their homogenous and perturbation parts.

\subsection{Dynamics of the axion-gluon/glueball fluids}

The axion dark matter is described by the evolution of the classical field, $\phi(x)$. The dark gluon/glueball densities and their perturbations can be parameterized by its temperature evolution $T_g(x)$ and $\phi(x)$ as discussed in the previous section. In this context,  $\{\phi(x), T_g(x)\}$ are good variables to derive full equations of motion of dark sector including their perturbations.  
Considering the fluid description, the evolution of energy densities and pressures are deduced from the evolution of $\phi$ and $T_g$ with the help of the Einstein equations, gluo-thermodynamics and the lattice calculation. 

We introduce the conformal Newtonian gauge  for the inhomogeneous part of the metric tensor  
\begin{eqnarray}
ds^2 =  a(\tau)^2\Big(- (1+ 2\Psi) d\tau^2 + (1 + 2 \Phi) d\vec x^2\Big),
\end{eqnarray}
where the conformal time $\tau$ and the conformal Hubble rate $\cal H$ 
are related with the proper time $t$ and the Hubble expansion rate $H=\dot a/a$ as
\begin{eqnarray}
\tau = \int \frac{dt}{a},\quad 
{\cal H} \equiv \frac{1}{a} \frac{da}{d\tau} = \frac{a'}{a} = a H .
\end{eqnarray}  
Here, we use the notation
\begin{eqnarray}
' \equiv \frac{d}{d\tau},\quad
\dot~\equiv \frac{d}{dt} 
\end{eqnarray}    for the time derivatives. 
For most of discussion in Sec.~\ref{sec:2} and \ref{sec:4}, we take the conformal time $\tau$ as the argument of the time-dependent variables, whereas the proper time $t$ is mainly used in Sec.~\ref{sec:5} when discussing the evolution after dark matter halos form.

Expanding $\phi$ and $T_g$ near the homogeneous solutions,
\begin{subequations}
    \begin{eqnarray}
         \phi(\tau, \vec{x}) &=& \phi(\tau)+\delta\phi(\tau, \vec{x})~,
    \end{eqnarray}  \begin{eqnarray}
         T_g(\tau, \vec{x}) &=&  T_g(\tau)+\delta T_g(\tau, \vec{x})~,
    \end{eqnarray}
\end{subequations}
the equations of motion of the background axion field 
is  given by 
    \begin{equation}\label{eq:EOMphi0}
       \phi''+2\mathcal{H}\phi'+a^2\frac{\partial V(T_g, \phi)}{\partial\phi}=0. 
    \end{equation} 
The corresponding background energy density and pressure are 
  \begin{eqnarray}\label{eq:rhoaandpa}
 \hskip -0.2cm    \rho_a=\frac{\phi'^2}{2a^2}+V(T_g, \phi),\quad 
        p_a=\frac{\phi'^2}{2a^2}-V(T_g, \phi).
    \end{eqnarray}
From Eq.~\eqref{eq:EOMphi0} and \eqref{eq:rhoaandpa}, the continuity equation for the background axion is obtained \begin{equation}\label{eq:conta}
    \rho'_a + 3\mathcal{H}(1+w_a)\rho_a=
    \frac{\partial V}{\partial T_g} T_g'.  \end{equation}
 $w_a\equiv p_a/\rho_a$ is the equation of state for the axion. Since the axion-gluon fluid is isolated from the visible sector, the total energy and pressure of dark sector should  satisfy the continuity equation without source terms
\begin{equation}\label{eq:contsum}
    \rho'_{a+g}+3\mathcal{H}(\rho_{a+g}+p_{a+g}) = 0.
\end{equation}
This leads to the evolution of the gluon/glueball fluid as\begin{align}\label{eq:contg}
    \rho'_g + 3\mathcal{H}(1+w_g)\rho_g  
    =-   \frac{\partial V}{\partial T_g} T_g', 
\end{align}
where $w_g= p_g/\rho_g$. 
Together with Eq.~(\ref{eq:gluoentropy}), (\ref{eq:gluon}),  we can derive
the conservation of the dark entropy as we claimed,
\begin{eqnarray}\label{eq:entropy_cons}
s_g' + 3 {\cal H} s_g = 0.
\end{eqnarray}

The evolution of the Fourier transformed perturbed axion field variable $\delta\phi$ 
for a given wave number $k$ is described by
     \begin{eqnarray}\label{eq:EOMdelphi}
       &&  \delta\phi''+2\mathcal{H}\delta\phi'+\bigg(k^2+a^2\frac{\partial^2 V }{\partial\phi^2}\bigg)\delta\phi + \phi'(-\Psi'+3 \Phi') \nonumber\\
        &&\quad+\,2a^2\frac{\partial V }{\partial \phi}\Psi+a^2\frac{\partial^2 V}{\partial\phi\partial T_g}\delta T_g=0.
    \end{eqnarray}
From this, the evolution of the fluid perturbation variables $\delta\rho_a$, $\delta p_a$, $v_a$ and $\pi_a$
\begin{subequations}\label{eq:pertvar}
    \begin{eqnarray}
  &&      \delta\rho_a =  \frac{ \phi'\delta\phi'+\phi'^2\Phi}{a^2}+\frac{\partial V}{\partial\phi}\delta\phi+\frac{\partial V}{\partial T_g}\delta T_g,\\
   &&     \delta p_a = \delta\rho_a-2\frac{\partial V}{\partial\phi}\delta\phi-2\frac{\partial V}{\partial T_g}\delta T_g, \\
&&  (\rho_a+p_a)v_a = a^{-2}k \phi'\delta\phi,\quad        p_a\pi_a=0
    \end{eqnarray}
\end{subequations}
can also be calculated. 
We follow the definition and convention of the variables in Refs.~\cite{Bartolo:2003ad,Hu:1998kj,Ma:1995ey}. 

On the one hand, 
from Eq.~(\ref{eq:gluon}), the  gluon/glueball fluid perturbations $\delta p_g$ and $\delta \rho_g$ can be related with $\delta T_g$ and $\delta\phi$ as  
\begin{eqnarray}\label{eq:pertglue}
 \delta \rho_g&=&  \left(T_g^2\frac{d^2p_g}{dT_g^2} 
-\frac{\partial^2 V}{\partial (\ln T_g)^2}\right)\frac{\delta T_g}{T_g} 
- \left(\frac{\partial^2 V}{\partial\phi\partial\ln T_g}\right)\delta\phi,  \nonumber\\
\delta p_g &=& \frac{dp_g}{d T_g} \delta T_g.
\end{eqnarray} 
Similarly, from Eqs.~\eqref{eq:EOMdelphi} and \eqref{eq:pertvar}, we derive the equations of motion for the fluid perturbations:
\begin{subequations}\label{eq:boltz}
    \begin{eqnarray}
        \delta'_a=&&-ku_a-3(1+w_a)\Phi'\nonumber\\
        &&-\bigg(3\mathcal{H}+\frac{1}{2}\frac{\partial\ln V}{\partial\ln T_g}\frac{T_g'}{T_g}\bigg)\bigg(\frac{\delta p_a}{\delta \rho_a}-w_a\bigg)\delta_a\nonumber\\
        &&+\frac{1}{2}(1-w_a)\frac{d}{d\tau}\bigg(\frac{\partial\ln V}{\partial\ln T_g}\frac{\delta T_g}{T_g}\bigg),\\
        u'_a=&&-\mathcal{H}(1-3w_a)u_a+k(1+w_a)\Psi+k\frac{\delta p_a}{\delta\rho_a}\delta_a\nonumber\\
        &&-\frac{1}{2}(1-w_a)\frac{\partial\ln V}{\partial\ln T_g}\bigg(\frac{T'_g}{T_g}u_a-k\frac{\delta T_g}{T_g}\bigg)~,  \\
            \delta'_g=&&-ku_g-3(1+w_g)\Phi'-3\mathcal{H}\bigg(\frac{\delta p_g}{\delta\rho_g}-w_g\bigg)\delta_g\nonumber\\
        &&+\frac{1}{2}\frac{\rho_a}{\rho_g}\bigg[\frac{\partial\ln V}{\partial\ln T_g}\frac{T'_g}{T_g}\bigg\{\bigg(\frac{\delta p_a}{\delta\rho_a}-1\bigg)\delta_a+(1-w_a)\delta_g\bigg\}\nonumber\\
        &&~~~~~~~~~~-(1-w_a)\frac{d}{d\tau}\bigg(\frac{\partial\ln V}{\partial\ln T_g}\frac{\delta T_g}{T_g}\bigg)\bigg],\label{eq:boltzdelg}\\
        u'_g=&&-\mathcal{H}(1-3w_g)u_g+(1+w_g)k\Psi+k\frac{\delta p_g}{\delta\rho_g}\delta_g\nonumber\\
        &&+\frac{1}{2}(1-w_a)\frac{\partial\ln V}{\partial\ln T_g}\bigg(\frac{T_g'}{T_g}u_g-k\frac{\delta T_g}{T_g}\bigg)~,
    \end{eqnarray}
\end{subequations}
where $\delta_a= \delta\rho_a/\rho_a$, 
$u_a = (1+ w_a) v_a$, and same definitions for $\delta_g$, $u_g$. 

Although  Eq.~(\ref{eq:boltz}) is not a closed form, 
it is straightforward to express $\{\delta p_a, \delta p_g, \delta T_g, \delta \phi\}$  in terms of $\{\delta_a, \delta_g, u_a\}$.
One can also take the perturbed variables as $\{\delta\phi, \delta T_g, u_g\}$ 
using Eqs.~(\ref{eq:pertvar})-(\ref{eq:pertglue}) for solving Eq.~(\ref{eq:boltz}). 
The nontrivial ingredients in our differential equations are the terms proportional to $\partial V/\partial T_g$, which originate from the non-perturbative interactions between the gluon and the axion. The effect of these terms becomes larger as $T_g$ approaches to $T_{g,c}$, and suddenly disappears after the confinement.

As the initial conditions for the cosmological evolution, the amount of dark gluons can be parameterized by the ratio between the entropies of the dark sector and the visible sector, $s_g/s_{SM}$.  Even though dealing with the entropy ratio between two sectors would be easier to trace the evolution of the densities,  in order to get a more intuitive picture about how cold the gluons are compared to the visible sector, we will use the ratio parameter between the temperatures. 
Taking a period around the phase transition, 
the photon temperature when $T_g$ arrives at $T_{g,c}$ is denoted by $T_{\gamma, c}$.  
Then, we define the ratio parameter $r$ as 
\begin{eqnarray} \label{ratio}
r \equiv \left(\frac{g_{*S}(T_{\gamma, c})}{2(N^2-1)}\frac{s_{g}}{s_{SM}}\right)^{1/3} 
\simeq  \frac{T_{g, c}}{T_{\gamma, c}} ,
\end{eqnarray}
where $s_g$ ($s_SM$) is the entropy density of the gluon fluid (the SM sector), and $g_{*S}$ is the effective number of degrees of freedom in entropy for the SM sector.

So far, we have ignored the effect of dissipation for the axion's motion induced by the background dark gluon plasma.  It is not crucial in our discussion, but let us clarify how small its effect is. 
Including the friction term ($\gamma_{\rm fr}$) induced by gluon plasma, 
the equation of motion of the background axion is written as \cite{McLerran:1990de}
\begin{eqnarray}
\ddot\phi+ \left(3 H + \gamma_{\rm fr}\right)\dot\phi + m_a^2(T_g) \phi = 0. 
\end{eqnarray}
In the deconfining phase, 
$\gamma_{\rm fr}= \Gamma_{\rm sph}(T_g)/2f_a^2 T_g$, where
the sphaleron rate is estimated by \cite{Moore:2010jd}:
\begin{eqnarray}
\hskip -1cm 
 \Gamma_{\rm sph}(T_g) &=& 
\int d^4 x \left\langle \frac{g_h^2}{32\pi^2} G\tilde G(x)\, \frac{g_h^2}{32\pi^2} G\tilde G(0)\right\rangle_{T_g}  \nonumber\\
\hskip -1cm
&=& {\cal O}(0.1-1)
\left(\frac{g^2_hN}{4\pi}\right)^5
\left(\frac{N^2-1}{N}\right) T_g^4,
\end{eqnarray}
for $g_h^2/4\pi \lesssim 0.1$.  Note that this is not an Euclidean correlator like Eq.~(\ref{eq:axion_mass}),
 but evaluated in real spacetime.
Around the critical temperature $T_g\sim T_{g,c}$, 
the gauge coupling can be large as $g_h^2N/4\pi ={\cal O}(1)$. 
In this regime,  
the sphaleron rate is expected as $\Gamma_{\rm sph}(T_{g,c}) \sim  T_{g,c}^4$ 
from the argument of dimensional analysis and the calculation using the AdS/CFT correspondence \cite{Son:2002sd}.
After the confining phase transition, no reliable calculation has been done so far. 
A crude estimation based on the dimensional analysis is that the dissipation rate is at most proportional to the entropy density (or number density) of the glueballs as  $\gamma_{\rm fr}\sim s_g/f_a^2$.
Because the time dependence of the Hubble rate and the dissipation rate are given as $H\propto a^{-2} (a^{-3/2})$ in radiation dominated era (matter dominated era) and  $\gamma_{\rm fr}\propto a^{-3}$, the gluon induced friction term is important only when the temperature of the visible sector becomes greater than \begin{eqnarray}
 T_\gamma > {\cal O}(1)\left(\frac{10^9\,{\rm GeV}}{N r^3}\right) \left(\frac{10^{14}\,{\rm GeV}}{f_a}\right)^2\left(\frac{g_h^2 N}{4\pi}\right)^{-5}.
 \end{eqnarray}
Comparing this with the temperature when the axion starts to oscillate in our scenario ($T_\gamma < {\rm TeV}$), it is always irrelevant.  

\section{\label{sec:3}Evolution history}

\subsection{\label{sec:3:1}Evolution of the background gluon and glueballs}

In Eqs.~(\ref{eq:conta})-(\ref{eq:boltz}), we establish the continuity equations for the background and perturbative variables based on the equations of motion of the axion field, and gluo-thermodynamics.   
In this section, some features of the background evolution of gluon and glueball fluids are discussed in more detail. 

The evolution of $\rho_g$ and $p_g$ is particularly easy for $T_g \gg T_{g, c}$ and $T_g \ll T_{g,c}$.  In the limit of $T_g\gg T_{g,c}$,  gluons are relativistic fluid, and the contribution of the axion to the evolution of $\rho_g$, $p_g$ is negligible because  from  Eq. (\ref{eq:gluon}) and $T_{g,c}\sim \Lambda$, 
\begin{eqnarray}
\frac{T_g}{\rho_g}\frac{\partial V(T_g,\phi)}{\partial T_g}
\sim\frac{\phi^2}{N f_a^2} \left(\frac{\Lambda}{T_g}\right)^{11N/3} \ll 1.
\end{eqnarray}  Therefore, the usual scaling relations hold as
\begin{eqnarray}\label{eq:relativistic}
        p_g&=&\frac{\pi^2}{45}(N^2-1)\, T_g^4 ,  \nonumber\\
     \rho_g&\approx&\frac{\pi^2}{15}(N^2-1)\,T_g^4 = 3 p_g .
    \end{eqnarray}  
As the gluon temperature approaches $T_{g,c}$, Eq.~\eqref{eq:relativistic} does not hold any more because 
the strong interactions among gluons become significant \cite{Boyd:1996bx} and the axion contribution in Eq.~\eqref{eq:gluon} is also increasing.
The true evolution can only be figured out by the lattice calculation.  
We adopt the lattice data for $\theta=0$ to evaluate $p_g(T_g)$ \cite{DENG1989334, Datta:2010sq, Borsanyi:2012ve}, and deduce the densities of the gluons for the nonzero $\theta$ using Eqs.~(\ref{eq:gluoentropy})-(\ref{eq:gluon}). 

At $T_g=T_{g,c}$, the dark gluons are combined into dark glueballs, whose masses are multiples of the confining scale $\Lambda$ \cite{Teper:1997am,Teper:1998kw,Morningstar:1999rf,Athenodorou:2020ani}.  Here, 
$\Lambda$ should be defined with a certain regularization scheme.  Taking the $ \overline{MS}$ scheme, 
\cite{Lucini:2008vi} shows the relation $\Lambda/\sqrt{\sigma} = 0.5 + 0.34/N^2$ for $N\geq3$, where $\sigma$ is the string tension. 
The relation between the critical temperature and the string tension is evaluated as $T_{g,c}/\sqrt{\sigma} = 0.59+ 0.46/N^2$ based on the study for $2\leq N\leq 8$ \cite{Lucini:2012wq}. As a result, we get
\begin{eqnarray} \label{eq:Tgc}
\frac{T_{g,c}}{\Lambda} \simeq  1.2 +\frac{0.1}{N^2}.
\end{eqnarray} 

The phase transition is first-order if $N>2$. 
In order to understand how long the phase transition happens, we can compare the energy density in the deconfined phase with that in the confined phase at $T_{g,c}$. 
The former is the energy density of the gluon plasma of ${\cal O}(0.1 N^2 T_{g,c}^4)$, while the latter is the sum of the glueball tower of ${\cal O}(0.1 T_{g,c}^4)$. 
Thus, as $N$ increases, larger latent heat is released and the transition period becomes longer.  

Let us shortly discuss how we evaluate the energy density of the glueballs.  
Since the lightest glueball mass is calculated as
\begin{eqnarray}
\frac{m_g}{T_{g,c}}\simeq 5.7 -\frac{1.2}{N^2},
\end{eqnarray} 
where we use the relation $m_g/\sqrt{\sigma}= 3.64$ for $N=3$ \cite{Teper:1998kw}, 
and $m_g/\sqrt{\sigma}=3.37 +1.93/N^2$ based on the results for $2\leq N\leq 5$ \cite{Lucini:2001ej},
all glueballs are non-relativistic at $T_{g,c}$. Using the spectral density $\hat\rho(m)$, the energy density of glueballs at $T_g \leq T_{g,c}$ can be written as 
\begin{eqnarray}\label{eq:spectraldensity}
\hskip -0.6cm 
 \rho_{g}(T_{g})=\int_0^\infty dm\, \hat\rho(m) 
\, m\left(\frac{m T_{g}}{2\pi}\right)^{3/2} e^{-m/T_g}.
\end{eqnarray}
For $\theta=0$ (the effect of the axion induced $\theta$-term will be discussed later), the glueballs are well described by the eigenstates of the spin ($J$), the parity ($P$), and the charge conjugation ($C$): $J^{PC}$. The lightest glueball corresponds to $0^{++}$ \cite{West:1995ym}. 
$\hat\rho(m)$  contains information about the tower of the glueballs.
It turns out that the spectral density can be successfully approximated by  
the sum of the discrete low-lying resonances with a mass $m_{J^{PC}}$ ($< m_{\rm th}$), and the continuum spectrum of the Hagedorn tower \cite{Hagedorn:1980kb,Caselle:2011fy,Caselle:2015tza} 
\begin{eqnarray}
\hat\rho(m) &\simeq& \sum_{m < m_{\rm th}}
(2J+1)\, \delta\left(m - m_{J^{PC}}\right)
\nonumber\\
&& +\,\frac{n_N}{m} \Big(\frac{2\pi T_H}{3\,m}\Big)^3  e^{m/T_H} \Theta(m- m_{\rm th}),
\end{eqnarray}
where $n_2=1$, $n_{N\geq 3}=2$. 
In the large $N$ limit, 
the Hagedorn temperature $T_H$ is related with  $T_{g,c}$ as \cite{Meyer:2004gx,Caselle:2011fy,Caselle:2015tza}
\begin{eqnarray}
\frac{T_H}{T_{g,c}} \simeq 1.16- \frac{0.9}{N^2}.
\end{eqnarray} 
We used the closed Nambu-Goto string model for glueballs to provide the relation between $T_H$ and the string tension, $T_H=\sqrt{3\sigma/2\pi}$ \cite{Caselle:2011fy}. 
The threshold mass $m_{\rm th}$ is not a physical quantity.  
 It is shown that taking $m_{\rm th} = 2m_g \simeq 10 T_H$ is good enough to reproduce the lattice results \cite{Caselle:2015tza}. Actually, from the closed string description of the glueball spectrum, 
the continuum approximation works well  for  $T_H \ll m$. 

Note that even if $N$ increases, the number of glueball degrees of freedom does not increase. Therefore, the contribution from the Hagedorn glueballs to $\rho_g$ near the confinement is insensitive to $N$ and becomes ${\cal O}(0.01-0.1) T_{g,c}^4$. The 
low-lying glueball contributions with $m_{J^{PC}}< 2m_g$ are ${\cal O}(10-50)\%$ of it.  
For this estimation,  the glueball spectrum of \cite{Teper:1998kw,Morningstar:1999rf} is used in the case of $N=3$. For the larger value of $N$, we adopt the spectrum of $0^{++(*)}$, $2^{++}$ of \cite{Lucini:2004my,Bennett:2020hqd}.  In fact, the glueball spectrum calculated using the lattice
has a large uncertainty  except for a few low-lying modes  in the case of $N>3$. However, 
we can reasonably assume that there is no significant change in the spectrum even for higher values of $N$ \cite{Lucini:2010nv}. 

Now we turn our attention to the effect of the axion on the glueball spectrum. First of all, the nonzero axion field value can change the glueball mass   as $\delta m/m = {\cal O}(\phi^2/N^2f_a^2)$ from Eq.~(\ref{axion-dependent-mass}). Its contribution is smaller than $10\%$ for $N\geq 3$, so is ignorable. 
Secondly, for the nonzero axion value, the parity is no longer good quantum number. This  leads to the mixing between the glueballs with different $P$ eigenvalues. For instance,  \cite{Gabadadze:2004jq} shows that the mixing between $0^{++}$ and $0^{-+}$ is not suppressed even in the large $N$ limit as  $\varphi_g= 0^{++}\to \varphi_g= 0^{++} + {\cal O}(\theta) 0^{-+}$.  It is also noticed that this mixing only shifts the lightest   state, while the heavier state $\varphi_{-+}=0^{-+}$ remains intact. Through the cubic interaction of $0^{++}$ as in Eq.~(\ref{Leff}), 
the interaction between $\varphi_g$ and $\varphi_{-+}$, 
\begin{eqnarray}
 \Delta{\cal L} \sim  \frac{4\pi \phi}{N f_a} m_g \varphi_g^2 \varphi_{-+},
\end{eqnarray} 
is induced, and may allows the decay of $\varphi_{-+}$ to two $\varphi_g$s when the axion has the nonzero expectation value. However,  $m_{\varphi_{-+}}\simeq 1.5 m_g < 2 m_g$ for $N=3$ \cite{Teper:1998kw,Morningstar:1999rf}, and this inequality is expected to hold for $N>3$  from the argument of the large $N$ expansion. 
Therefore, the axion-induced decay to the lightest glueballs seems to be  kinematically forbidden for $N\geq 3$. On the one hand, the existence of the mixing between $0^{-+}$ and $0^{++}$ in the presence of $\theta$-term can also imply the direct decay of $\varphi_{-+}$ to the lightest glueball and the axion from the interaction like $ (m_g^2/f_a)\,\phi\,  0^{-+} 0^{++}$. 
If this term is not canceled in the mass eigenbasis, the corresponding life-time of $\varphi_{-+}$
is estimated as $\tau_{\varphi_{-+}\to \varphi_g \phi}\sim  20\, {\rm Myr}\, (f_a/10^{15}\,{\rm GeV})^2({\rm MeV}/m_g)^{3}$, which can be shorter than the age of the Universe.  

Although there are large uncertainties in estimating the effect of the axion on the dynamics of heavier glueballs, 
we expect that 1) its effect on the glueball masses is small, 2) the heavier glueballs become less stable, and 3) they interact more with other glueballs. This means that the approach using the spectral density Eq.~(\ref{eq:spectraldensity}) works well near the critical temperature, $T_g\sim T_{g,c}$, even 
if the axion degree of freedom is included.

The lower limit of the actual transition period is given by the period obtained assuming the quasi-equilibrium transition. This is the case that the pressures of deconfining/confining phases are equal and  the  latent heat is released adiabatically as the Universe expands.  
In this situation, the temperatures of the confining and deconfining phases are the same and maintain at around $T_{g,c}$ during the transition, and the entropy is conserved. The duration of the phase transition  is estimated by the conservation of the dark entropy:
\begin{eqnarray} 
a_{cf} = a_{ci}\left(\frac{s_{\rm gluon}(T_{g,c})}{s_{\rm glueball}(T_{g,c})}\right)^{1/3}
\simeq a_{ci} N^{2/3},
\end{eqnarray}
where $a_{cf}$ ($a_{ci}$) is the scale factor  when the phase transition ends (starts), $s_{\rm gluon}(T_{g,c})$ ($s_{\rm glueball}(T_{g,c})$) denotes the entropy density of the dark gluon (glueball) at $T_g=T_{g,c}$. The $N$-dependence of the duration is obtained from $s_{\rm gluon}(T_{g,c})\simeq 0.1 N^2 T_{g,c}^3$$\simeq N^2 s_{\rm glueball}(T_{g,c})$. 

The phase transition becomes stronger first-order as $N$ increases. Thus, the additional entropy is generated during the transition and makes the glueballs hotter than the previous estimation.
However, this effect is negligible unless $N\gtrsim 4\pi$, because the nucleation temperature is just around $T_{g,c}$ and the strong interactions of the gluon and glueball fluids  provide a large friction coefficient for the bubble wall propagation. For $N=3$, $a_{cf}\simeq 2 a_{ci}$ is obtained  numerically, which is well matched with our parametric estimation $N^{2/3}$. 
The assumption of dark entropy conservation will be kept in the following discussion. 

After the phase transition, the evolution of the glueball temperature is calculated using the conservation of the dark entropy. The Hagedorn spectrum can slow down the decreasing rate of $T_g$, but numerically its contribution becomes gradually suppressed and negligible in comparison with  that of the lightest glueball when $a\gtrsim 10\, a_{cf}$ ($T\lesssim 0.55 T_{g,c}$).  
The detailed freeze-out process of the low-lying stable glueballs other than the lightest one is provided in \cite{Forestell:2016qhc} for $N=3$ with $\theta=0$.  Their abundance is also quite suppressed after the glueball temperature becomes smaller than $0.5 T_{g,c}$. 
As we already discussed, the mass spectrum of the heavier glueballs for $N>3$ is not much affected by the evolution of the axion. Furthermore, 
the role of the axion is generically to make the heavier glueballs unstable, so that the corresponding abundance 
could be further suppressed compared to the case with $\theta=0$. 
Therefore from now on, we will just focus on the evolution of the lightest glueball.

The relevant scattering processes to maintain thermal equilibrium of $\varphi_g$ is
two-to-two and three-to-two scatterings whose rates are estimated as 
\begin{eqnarray} \label{eq:g_scatter}
\hskip -0.5cm \sigma_{2\to2}v \sim \frac{  v_f (4\pi/N)^4}{32\pi m_g^2},\quad
 \sigma_{3\rightarrow2}v^2 \sim \frac{(4\pi/ N)^6}{(4\pi)^3  m_g^5},
\end{eqnarray} 
where $v_f$ is the relative velocity of the final particles from the scattering. 
 

As the Universe expands, the $3\rightarrow2$ process freezes out when the most of the $2\rightarrow2$ processes still active. 
This is because the interaction rate of the $3\rightarrow2$ process is proportional to the square of the number density of the glueballs, while that of the $2\rightarrow2$ processes is linearly proportional to the number density of the glueballs. 
Before the freeze-out of $3\to2$ interactions, the glueball density and pressure are the function of its temperature as 
\begin{eqnarray}\label{eq:rhoandpgb}
        \rho_g(T_g) &=& m_g\bigg(\frac{m_g T_g}{2\pi}\bigg)^{3/2}e^{-\frac{m_g}{T_g}}
        \Big(1+\frac{27}{8}\frac{T_g}{m_g}+\mathcal{O}\bigg(\frac{T_g^2}{m_g^2}\bigg)\Big),\nonumber\\
        p_g(T_g)&= &  T_g\bigg(\frac{m_g T_g}{2\pi}\bigg)^{3/2}e^{-\frac{m_g}{T_g}}\Big(1+\frac{15}{8}\frac{T_g}{m_g}+\mathcal{O}\bigg(\frac{T_g^2}{m_g^2}\bigg)\Big).\nonumber\\
\end{eqnarray}

\begin{figure*}[]
	\centering
\hskip-0.7cm	\includegraphics[width=0.5\textwidth]{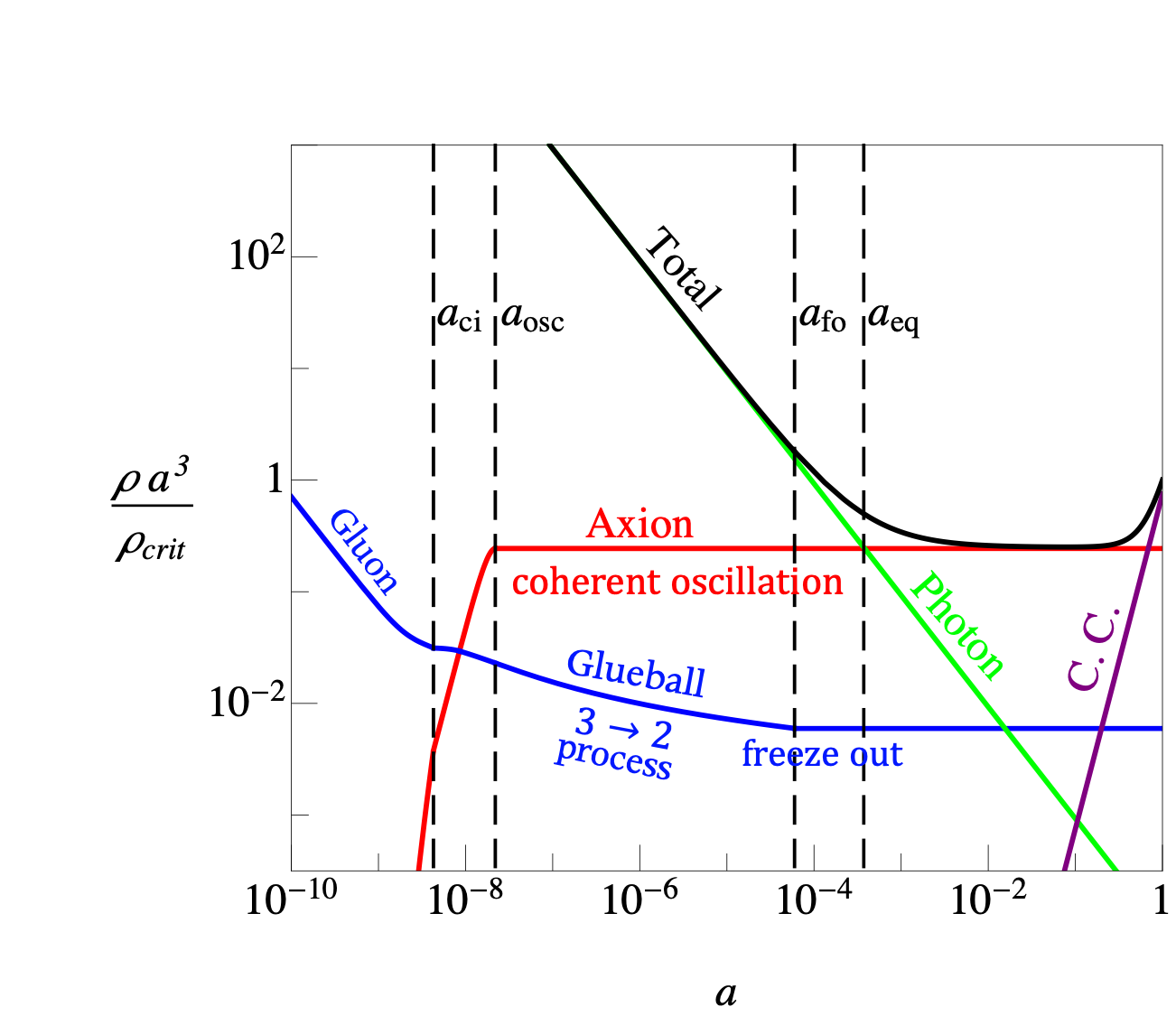}
	\hspace{-0.4cm}
	\includegraphics[width=0.5\textwidth]{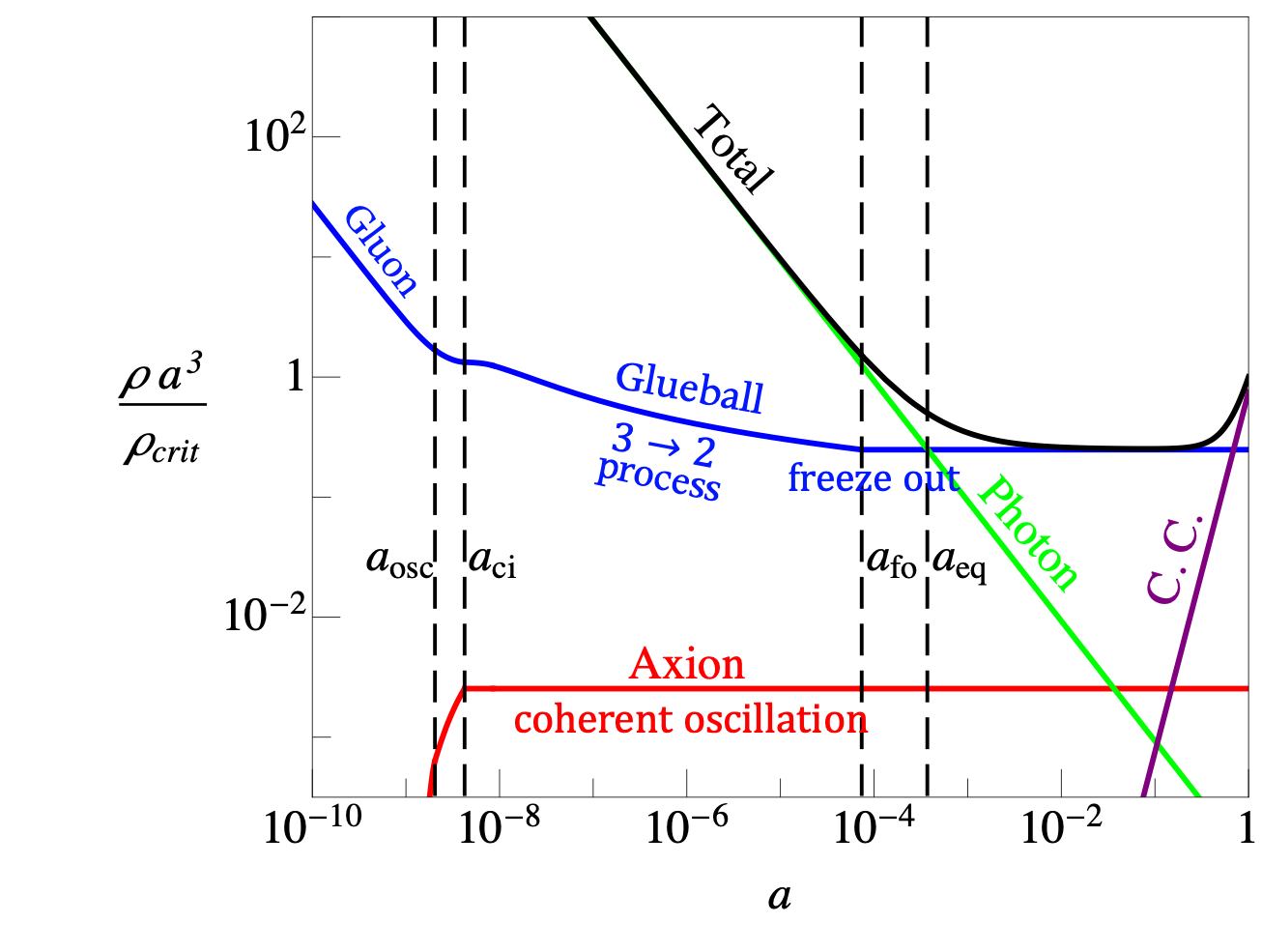}
	\hspace{1cm}
	\caption{\label{fig:density} The examples of dark matter density evolutions. Here we take $N=3$.
	The left panel corresponds to the axion dominated scenario, while the right panel is for the glueball dominated case. The parameters in the right panel are taken for illustration only because the corresponding value are ruled out by the self-interacting and warm dark matter constraints.}
\end{figure*}

\begin{figure}[tbp]
	\centering
	\includegraphics[width=0.5\textwidth]{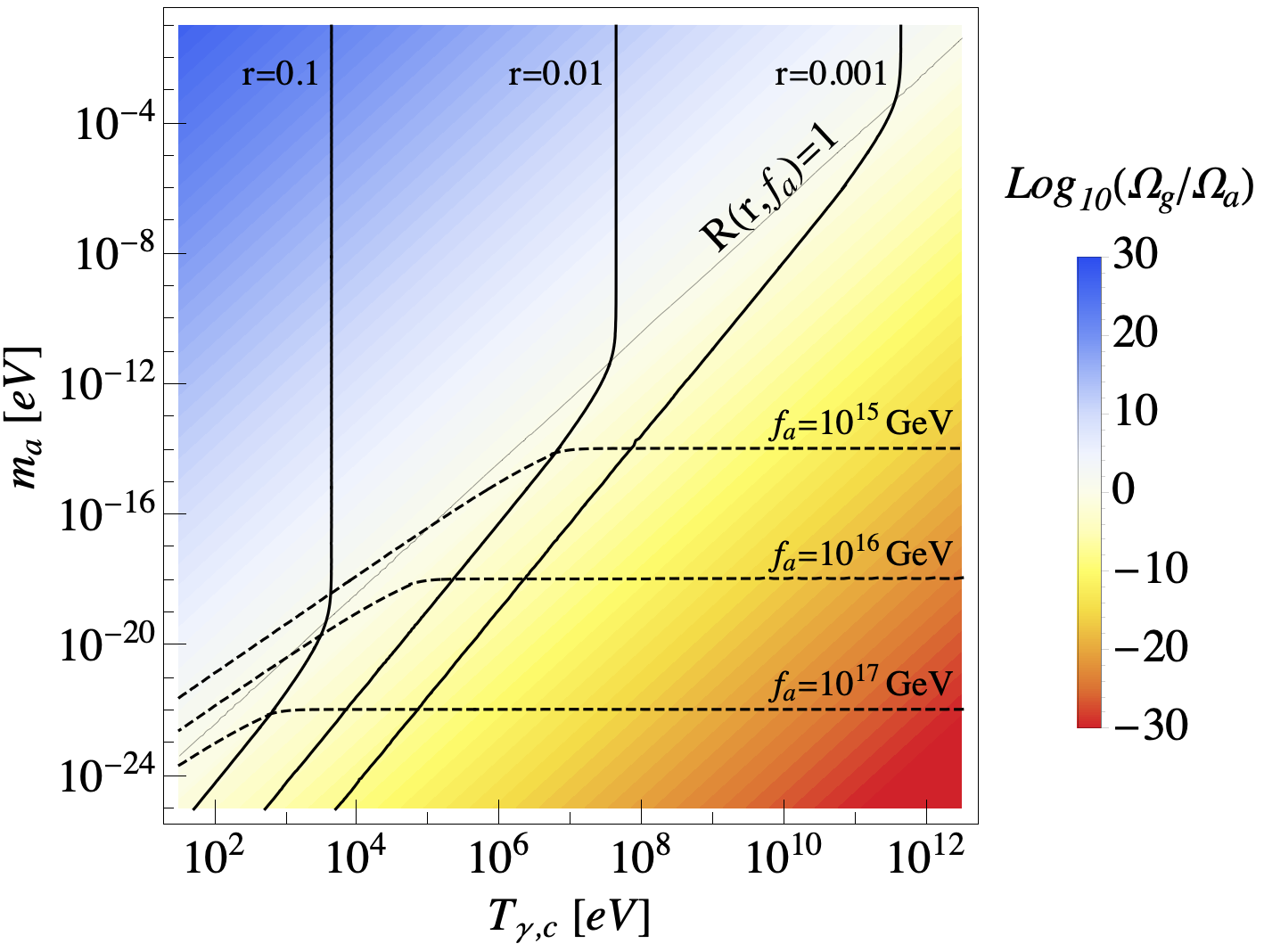}
	\caption{\label{fig:background} Parametric dependence of the relic abundance of the glueball and the axion for $\Omega_{\rm DM}h^2=0.11$. Here we take $N=3$. $m_{a}$ is the zero-temperature axion mass and $T_{\gamma,c}\simeq T_{g,c}/r$ is the photon temperature when the confining phase transition of the dark sector starts.  
	The parameters $\{$$r\,$(real lines), $f_a\,$(dashed lines)$\}$ are evaluated to give $\Omega_{\rm DM}h^2=0.11$ for each point in the plot. In the region above the line $R(r, f_a)=1$, the oscillation of the axion starts earlier than the confinement phase transition and the glueball dominates the dark matter abundance with the mass $m_g\simeq 6\, r T_{\gamma,c}$. Below $R(r,f_a)=1$, 	the axion starts to oscillate after the transition 
	and becomes the dominant component of dark matter. 
	Here, we did not impose the constraints from the current bound, which are discussed in text.}
\end{figure}

A distinguishing property of the chemical equilibrium maintained by the number changing self-interaction is that 
its temperature drops much slower than the photon temperature due to the entropy damping effect. 
From the entropy conservation $s_g\propto 1/a^3$,  the temperature scales as $T_g \sim  1/\ln a$ \cite{Carlson:1992fn}. 
As the consequence, the energy density drops faster than that of a cold dark matter 
\begin{eqnarray}
  \rho_g  \simeq T_g s_g \propto \frac{1}{a^3 \ln a},
\end{eqnarray} since the $3\to 2$ self-interaction converts the mass energy to the kinetic energy. This behavior ends when the process freezes out at $T_g= T_{g, fo}$ with 
\begin{equation}\label{eq:freezeoutcondition}
  \rho_{g}^2(T_{g, fo})
\simeq \frac{  (3  m_g T_{g, fo})(\left. H\right|_{T_g= T_{g, fo}})}{\langle\sigma_{3\to 2} v^2\rangle} .
\end{equation} 
After that, the glueballs still maintain kinetic equilibrium by the $2\to 2$ interactions, but they act as free-streaming particles for their background evolution. 
Using Eq.~(\ref{eq:freezeoutcondition}) and the dark entropy conservation, 
the freeze-out temperature is evaluated as 
\begin{eqnarray}
&& \frac{m_g}{T_{g,fo}} +\frac{5}{4}\ln \frac{m_g}{T_{g,fo}} + \frac{3}{4}\ln\frac{m_g} {{\rm MeV}} \nonumber\\
&& \simeq   28.2+\frac{3}{2}\ln\frac{r}{0.01}     -\frac{7}{2}\ln \frac{N}{3}, 
\end{eqnarray}
when it happens during radiation dominated era. 
As a specific example, 
for $N=3$, $r=0.01$, and $m_g=1\,{\rm MeV}$,  we get 
 \begin{eqnarray}
T_{g, fo} \simeq 0.04\,m_g \simeq 0.2\, T_{g, c}.
\end{eqnarray}
The relation $T_{g,fo} ={\cal O}(0.2) T_{g,c}$ is not much sensitive to the values of $r$ and $m_g$ that we are interested in.

Meanwhile, we can argue that the impact of the axion dynamics on the evolution of the lightest glueballs  is negligible. The main reason is that after the confining phase transition, the axion potential becomes independent of the glueball temperature, i.e. $\partial_{T_g} V(T_g, \phi)=0$. According to Eqs.~(\ref{eq:gluoentropy}) and (\ref{eq:gluon}), the gluo-thermodynamic quantities ($p_g, s_g, \rho_g$) are decoupled from the axion dynamics as long as the relaxation time for the number changing process $\Gamma_{3\to 2}^{-1} \sim m_g^{-1}(a/a_{cf})^6$ is shorter than the axion oscillation period $m_a^{-1}\sim m_g^{-1}(f_a/m_g)$.  If the axion starts oscillating before the glueball freeze-out, it is also plausible that the glueball energy density is modulated by the axion induced $\theta(t)$-term after $\Gamma_{3\to 2}^{-1}\gtrsim m_a^{-1}$. However, its maximum contribution, as in Eq.~(\ref{axion-dependent-mass}) for $\phi\sim f_a$,  is less than $10\%$ from the beginning, and the oscillating amplitude is redshifted such as $\delta m_g/m_g\sim \theta(t)^2/N^2 \propto \cos(m_a t)/a^{3}$. As its oscillation period becomes shorter than the relaxation time of the glueball's chemical process, its effect is also averaged out to be zero.
Hence, the impact of the axion is negligible over time and does not change our main results.

During the evolution of the glueballs, 
the photon temperature also evolves. When the dark glueballs freeze-out, the photon temperature becomes
\begin{eqnarray}\hskip -0.5cm
T_{\gamma,fo} \simeq  3\,{\rm keV} \left(\frac{N}{3}\right)^{1/2}\left(\frac{0.01}{r}\right)^{3/2} 
\left(\frac{m_g}{{\rm MeV}}\right)^{5/4}.
\end{eqnarray} 
One can also easily evaluate the case that the freeze-out of the dark glueball happens after the matter-radiation equality for $m_g < {\rm keV}$. 

So far, we have specified all history of the gluons and glueballs  in order to identify the time dependence of the glueball temperature $T_g(\tau)$, which is relevant to the evolution of the perturbative variables Eq.~(\ref{eq:boltz}). 
However, for the final relic density of the glueballs, it  can be evaluated in a much simpler way from the conservation of the entropy of dark sector as 
\begin{eqnarray}\label{eq:Omegag}
&&\Omega_g h^2 \simeq 0.014 \left(\frac{N^2-1}{10}\right) \left(\frac{r}{0.01}\right)^3\left(\frac{T_{g, fo}}{ 10\,{\rm keV}}\right)
\left(\frac{3.94}{g_{*S}(T_{\gamma, c})}\right) \nonumber\\
&&\ \simeq 0.014\left(\frac{N^2-1}{10}\right) \left(\frac{r}{0.01}\right)^4\left(\frac{T_{\gamma, c}}{ 5\,{\rm MeV}}\right)\left(\frac{3.94}{g_{*S}(T_{\gamma, c})}\right)\nonumber\\
&&\ \simeq 0.12\left(\frac{N^2-1}{10}\right) \left(\frac{r}{0.003}\right)^3\left(\frac{m_g}{100\,{\rm MeV}}\right)\left(\frac{3.94}{g_{*S}(T_{\gamma, c})}\right) .\nonumber \\
\end{eqnarray}

\subsection{\label{sec:3:2}Evolution of the background axion}

The dilute instanton gas approximation works well for the axion potential before the confining transition occurs. 
However, it is no longer valid to describe the axion potential in the confining phase. 
The lattice studies can provide a part of information for the axion potential, i.e. the coefficient of each term in perturbative expansion Eq.~(\ref{axion_pot}) as \cite{Bonati:2017pfq}. 
\begin{eqnarray}\label{eq:axion_coefficient}
c_2 &\simeq& 0.3+ \frac{1}{N^2}, \quad c_4 \simeq - 2.7 c_2.
\end{eqnarray}
For the evolution of the axion field, further information is necessary. An interesting feature of the axion potential induced by the confinement without light fermions is that 
it is not a single branch, but composed of  
multiple ($N$) branches, where for each branch the period of the scalar potential is $2\pi N f_a$ \cite{Gaiotto:2017yup}. The general expression of the scalar potential for a $k$th branch is 
\begin{eqnarray}
	V_k= N^2\Lambda^4 h \left(\frac{\phi}{N f_a} + \frac{2\pi (k-1)}{N}\right),
\end{eqnarray}
where $k=1,\cdots, N$ and  $h(\psi)$ is the $2\pi$-periodic function. 
The full shape of the axion potential is not available. However, the analytic form of $h(\psi)$ in a certain range of the axion was studied in the large $N$ limit using the holographic description of the pure $SU(N)$ gauge theory \cite{Dubovsky:2011tu,Bigazzi:2015bna}. 
The shape of the potential highly relies on the size of the  't~Hooft coupling $\lambda_h = g_h^2N$ at the KK scale. By comparing the axion potential in the dual gravity theory and that of the lattice calculation given by Eq.~(\ref{eq:axion_coefficient}), we find that $\lambda_h= 10-20$ gives a reasonable matching. 

At high temperatures of the gluons, the instanton approximation for the axion potential is valid, and there is a single branch. During the phase transition, branches will emerge, and the axion can be located in a different branch in a different patch of the Universe. 
If each branch provides a stable axion trajectory, we have to consider the effect of them seriously.

Following the approach of the holographic description \cite{Dubovsky:2011tu}, we can estimate the tunneling rate per volume between $k$th to $k-1$th branches as 
\begin{eqnarray}
\Gamma_{\rm tunneling} \sim \Lambda^4 e^{- S_{(k\to k-1)}},
\end{eqnarray}
where the Euclidean action is  
\begin{eqnarray}
 S_{(k\to k-1)}=   {\cal O}(10^{-11}) N\frac{ (N/k)^3}{(1+\frac{{\cal O}(1)k^2}{N^2})^2}.
\end{eqnarray}
This can be significantly large only when $N\gtrsim 10^3$.
Therefore, in our consideration with $N^2={\cal O}(10)$, all branches with higher energy densities are quite unstable, and the transition to the lowest energy state will occur almost immediately.  
As a result, the effective potential of the axion is well described by 
\begin{eqnarray} \label{effective_pot}
V(\phi) = \min_k V_k(\phi),
\end{eqnarray}
and one can think the evolution of the axion within the range $ 2\pi f_a$. 

Without worrying about the effect of other branches, 
Eq.~\eqref{eq:EOMphi0} gives
\begin{eqnarray}
  \phi''+2\mathcal{H} \phi'+a^2 m_a^2(T_g)\phi =0, 
\end{eqnarray}
for $\phi\lesssim f_a$.
If the second term of the LHS is much larger than the third term, the axion field is approximately constant because of the large Hubble friction. This is the slow-roll limit. In the opposite case, the axion field oscillates with the oscillation frequency $m_a(T_g)$.  
Such evolution can be well approximated by the simple transition at $a = a_{osc}$, where $a_{osc}$ is the scale factor to give $3{\cal H}=a m_a(T_g)$ ($3 H = m_a(T_g)$). For each epoch,
\begin{eqnarray}\label{eq:phisol}
    \phi(\tau) &\simeq &
    \phi_i \equiv f_a \theta_i \quad (a < a_{osc})    \\
    &\simeq&
    \mathcal{A}(\tau) \cos\Big(\int^\tau d\tilde\tau a(\tilde\tau) m_a(T_g(\tilde\tau)) \Big).
    \quad  (a > a_{osc}) \nonumber \end{eqnarray}
Here, $\theta_i$ is the initial misalignment angle of the axion field,  $\mathcal{A}(\tau)$ is slowly varying function with $\mathcal{A}'/\mathcal{A} \ll  a m_a(T_g)$.  
The axion acts like dark energy during $a< a_{osc}$, while for $a>a_{osc}$, the axion plays the role of cold dark matter because  $\langle w_a\rangle \simeq 0$ by averaging out the fast oscillation.  
The initial axion value $\phi_i$ is not deterministic. Since both $\theta_i \ll 1$, and 
$|\theta_i - \pi| \ll 1$ need some tuning or special model building, here we take 
\begin{eqnarray} 
\theta_i = {\cal O}(1).
\end{eqnarray}

Since the axion's mass depends on the history of the dark gluons  (Eq.~(\ref{eq:mains})), there are two characteristic scales which determine the evolution history of the axion: $a_{osc}$ (onset of the axion oscillation) and $a_{ci}$ (onset of the confining phase transition). As the scale factor approaches $a_{ci}$, the contribution of the gluons to the axion's potential becomes substantial, and the axion mass is saturated. 
The evolution of the axion mass is smooth compared to the gluon thermodynamic relaxation time scale, i.e. $\dot m_a/m_a \sim N H \ll  T_g$,  unless $N$ is very large.

Using Eq.~(\ref{eq:mains}) and the definition of $a_{osc}$, we find that the following quantity 
\begin{eqnarray} \label{R(r,f)}
 R(r, f_a)\equiv\Big(\frac{r}{0.01}\Big)^2\Big(\frac{6\times 10^{13}\,{\rm GeV}}{f_a}\Big)
\end{eqnarray}
determines 
whether or not the axion starts to oscillate before the confining transition. 
If $R(r, f_a)>1$, the axion starts to oscillate in the deconfining phase. 
The corresponding photon temperature is
\begin{eqnarray}
  T_{\gamma, osc} \simeq   R(r, f_a)^\frac{1}{2+\eta_a} T_{\gamma, c}.
\end{eqnarray}
Otherwise ($R(r, f_a) < 1$),  the axion oscillation occurs after the phase  transition. 
It happens when the photon temperature becomes
\begin{eqnarray}
 T_{\gamma, osc} \simeq   R(r, f_a)^{\frac{1}{2}} T_{\gamma, c}.
\end{eqnarray}

The initial energy density of the axion at $T_\gamma = T_{\gamma, osc}$ is approximated as 
\begin{eqnarray}
\rho_a \simeq \frac{1}{2}m_a^2(T_{g,osc}) f_a^2 \theta_i^2.
\end{eqnarray}
After that, the axion field oscillates with the time dependent frequency.  
We notice that for the combination
\begin{eqnarray} 
N_a = \frac{ a^3 \rho_a }{m_a(T_{g})}, 
\end{eqnarray}
Eq.~(\ref{eq:conta}) gives 
\begin{eqnarray}
N_a' + \left(3 {\cal H} + \frac{ m_a'(T_g(\tau))}{m_a(T_g(\tau))}\right) w_a N_a =0.
\end{eqnarray}
For $N{\cal H}\ll am_a(T_g)$, 
the mass of the axion changes much slowly compared to the oscillation time scale. 
Such a fast oscillation implies a vanishing averaged equation of state $\langle w_a\rangle =0$ during the cosmological evolution. 
Therefore, $N_a$ is nearly conserved and $\rho_a/m_a(T_g)\propto 1/a^3$. 

In summary, if $R(r, f_a)> 1$, the axion starts to oscillate before the confining phase transition ($T_{\gamma, osc} > T_{\gamma, c}$), and  the present relic density of the axion dark matter becomes 
\begin{eqnarray}\label{axion_DM1}
\Omega_a h^2 &\simeq & 0.8\times 10^{-3}\theta_i^2\left(\frac{r}{0.01}\right)^4
\left(\frac{T_{\gamma, c}}{5\,{\rm MeV}}\right) 
R(r, f_a)^{-\frac{3+\eta_a}{2+\eta_a}}. \nonumber\\ 
\end{eqnarray}  
If $R(r, f_a)< 1$, the axion oscillates after the confining phase transition ($T_{\gamma, osc} < T_{\gamma, c}$). The corresponding axion dark matter density is estimated as 
\begin{eqnarray}\label{axon_DM2}
\hskip -0.3cm \Omega_a h^2 &\simeq&   0.8\times 10^{-3}\theta_i^2\left(\frac{r}{0.01}\right)^4
\left(\frac{T_{\gamma, c}}{5\,{\rm MeV}}\right)R(r, f_a)^{-\frac{3}{2}}\nonumber\\
\hskip -0.3cm  &\simeq&0.05\theta_i^2 \left(\frac{r}{0.01}\right)\left(\frac{T_{\gamma, c}}{5\,{\rm MeV}}\right) \left(\frac{f_a}{10^{15}\,{\rm GeV}}\right)^{3/2} \nonumber\\
&\simeq& 0.15 \left(\frac{m_a}{10^{-22}\,{\rm eV}}\right)^{1/2} \left(\frac{f_a}{10^{17}\,{\rm GeV}}\right)^2.
\end{eqnarray} 

As shown in Eq.~(\ref{axion_DM1}) and Eq.~(\ref{eq:Omegag}), the glueballs dominate the dark matter density if the axion oscillates earlier than the confining phase transition.  The reason is simply that the initial axion energy density is bounded by the confining scale $\Lambda^4\sim T_{g,c}^4$.  On the one hand, when $r^2/f_a$ is small enough, so that the axion starts to oscillate after the phase transition, the axion becomes a dominant component of dark matter. 

The left and right panels of Fig.~\ref{fig:density} show the evolution of energy densities for axion and glueball dominated dark matter scenarios, respectively. 
 Fig.~\ref{fig:background} shows the parametric dependence of the dark matter which give the correct relic density. The axes are represented by $T_{\gamma,c}$, the photon temperature when the confining phase transition of the dark gauge sector starts, and $m_a$, the zero temperature axion mass, defined in Eq.~\eqref{eq:axion_mass} with Eq.~\eqref{eq:axion_coefficient}, respectively. As we discussed, the dark matter today is dominated by the axion in the region $T_{\gamma, osc}< T_{\gamma, c}$, and by the glueball in the opposite region $T_{\gamma, c} < T_{\gamma, osc}$.
 
The initial amount of the dark gluon plasma is limited by the constraint on the effective extra relativistic degrees of freedom $\Delta N_{\rm eff}$ as \cite{Aghanim:2018eyx}
 \begin{eqnarray}
\Delta N_{\rm eff} &=& \left(\frac{\rho_g}{\rho_{\nu_e}} \right)_{\rm BBN}=  \frac{2(N^2 -1)r^4}{(7/4)(4/11)^{4/3}} \nonumber\\
&=&0.07\left(\frac{N^2-1}{10}\right) \left(\frac{r}{0.2}\right)^4 \lesssim 0.3.
 \end{eqnarray}

There are various astrophysical observations to constrain the mass of the glueball and axion dark matter. 
We shortly summarize the relevant bounds.
When the glueball dominates dark matter,  
its self-interaction gives observable effects if the scattering rate is large enough to reach the isothermal profile around the center of the halo. 
The self-interaction can also be detectable from the merger of dark matter halos, because the glueballs will be slowed down during the collision if their scattering cross-section is large enough. This leads to the offset between the dark matter and the collisionless components like stars.
From these observations, 
the cross-section of the glueball like self-interacting dark matter  is bounded as  (\cite{Tulin:2017ara} and references therein)
\begin{eqnarray}
\hskip -0.5cm
\frac{\sigma_{2\to2}}{m_g} &\simeq& \Big(\frac{4}{N}\Big)^4\frac{1}{m_g^3}\simeq  
\Big(\frac{4}{N}\Big)^4\Big(\frac{60{\rm MeV}}{m_g}\Big)^{3} {\rm  cm}^2/\textrm{g} \nonumber \\ 
&\lesssim& {\cal O}(0.5-5)\,{\rm cm}^2/\textrm{g}. 
\end{eqnarray}
In terms of the glueball mass, it should be  greater than ${\cal O}(50){\rm MeV}$ 
if it is the dominant component of dark matter. 
The phenomenology of heavier glueball dark matter was studied in \cite{Acharya:2017szw}.

If the axion is the dominant component of dark matter, 
there is  the  lower bound on the axion mass due to its fuzziness. 
The de Broglie wavelength becomes astrophysical scale ($\sim\rm kpc$) if $m_a$ is around $10^{-22}\,{\rm eV}$, 
and suppresses the structure formation.
The ultra-light axion can act like wave dark matter that are bound to or interact with each other by gravity inside the halo, which leads to the formation of solitonic cores and macroscopic quasiparticles moving around the center. 
These structures can have a great influence on the motion of stars. 
All these considerations give the strong constraint on the axion mass in the range $m_a\lesssim  10^{-22} - 10^{-20}{\rm eV}$  (\cite{Grin:2019mub,Ferreira:2020fam} and references therein). 
There is another constraint on the mass of the axion from the observation of highly spinning black holes.  
That is because if the axion mass is close to the inverse of the size of the spinning black hole, a superradiance phenomenon occurs and parts of black hole's mass and spin are removed by the superradiant axion cloud. 
Current observations of the spinning supermassive black holes with masses of $10^6-10^7\, M_\odot$ provide interesting constraints for the axion mass range $10^{-20}- 10^{-16} \, {\rm eV}$ (\cite{Brito:2015oca} and reference therein). Since the efficiency of the black hole superradiance depends on the axion self-interaction and the surrounding environment,  the constraint is rather model dependent.

When the dark matter is mostly composed of the axions ($\Omega_a h^2\simeq 0.11$), the fraction of the dark glueball subcomponent dark matter becomes
\begin{eqnarray}\label{fraction_g}
f_g &\equiv&\frac{\Omega_g}{\Omega_{\rm DM}} 
\simeq 0.28\left(\frac{N^2-1}{10\,\theta_i^2}\right)\left(\frac{r}{0.01}\right)^3
\left(\frac{10^{15}\,{\rm GeV}}{f_a}\right)^{3/2} \nonumber\\
&\simeq& 0.02\left(\frac{N^2-1}{10}\right)
\left(\frac{r}{0.01}\right)^3\left(\frac{m_g}{0.05\,{\rm MeV}}\right).
\end{eqnarray}
One can think that there is no strong constraint on the self-interactions of the glueball dark matter if $f_g \lesssim 0.1$.   However,  as discussed in Sec.~\ref{sec:5},  the evolution of the glueball dark matter after structures form may alter the cosmological history of the Universe from $z=7-15$.

\section{\label{sec:4}Perturbations}

We now study the evolution of the cosmological perturbations for the axion and glueball dark matter.  
Generically, both have non-trivial features compared to the CDM. 
For example, the late time transition of the axion from dark energy to dark matter modifies the early ISW effect  \cite{Hlozek:2017zzf}. The perturbation at scales smaller than the effective de Broglie wavelength of the axion is suppressed by its wave nature \cite{Hu:2000ke,Hwang:2009js,Irsic:2017yje}.  
For glueballs, the number-changing self-interaction also disturbs the growth of the density perturbation at  scales which enter the horizon well before the freeze-out \cite{Soni:2016gzf,Buen-Abad:2018mas}.

On the one hand, in our set-up, the dark sector is decoupled from the visible sector and 
the origin of their abundance can be totally different from that of the SM particles. 
Let us provide a simple example. 

 As the origin of Eq.~(\ref{Model}),  
 the axion field can be the phase of a complex scalar field $X$.
 The corresponding matter Lagrangian at high scales is 
\begin{eqnarray}
\hskip -0.3cm -\frac{{\cal L}_M}{\sqrt{-g}} &=&|\partial_\mu X|^2  + \bar Q i \gamma^\mu D_\mu Q+   \frac{\lambda}{4} \Big(|X|^2 - \frac{f_a^2}{2}\Big)^2 \nonumber\\
\hskip -0.3cm  && +\,  y X\, \bar Q P_L Q + h.c..
\end{eqnarray}
Here $Q$ is the vector-like fermion charged under the dark gauge group. 
The anomalous global symmetry
\begin{equation} 
U(1)_{\rm PQ}:\  X\to  e^{-2i\alpha} X ,\   Q \to  e^{ -i\gamma_5 \alpha} Q
\end{equation}  
is spontaneously broken by the nonzero vacuum expectation value of $X$, $f_a/\sqrt{2}$. 
Around the potential minimum, $X$ can be decomposed as 
\begin{eqnarray}
X(x) = \frac{f_a + s(x) }{\sqrt{2}} e^{ - i \phi(x)/f_a}.
\end{eqnarray}
The axion $\phi$ is identified as the Goldstone boson, so massless at perturbative level. 
The radial scalar $s$ gets a mass as $m_s = \sqrt{\lambda/2} f_a$, 
and the dark fermion mass is given by $M_Q= y f_a/\sqrt{2}$. 
Assuming the hierarchy $m_s\ll M_Q$, 
integrating out the heavy fermion yields the following effective Lagrangian 
\begin{eqnarray}\label{Model2}
\hskip -0.4 cm -\frac{{\cal L}_{M{\rm eff}}}{\sqrt{-g}} &\simeq& 
\frac{1}{2}(\partial_\mu s)^2 +   \frac{1}{2} m_s^2 s^2 + \frac{1}{4} \left( 1+ \frac{g_h^2 s}{8\pi^2 f_a}\right) (G^a_{\mu\nu})^2 \nonumber\\
\hskip -0.4 cm  &&+\frac{1}{2}  \left(1+ \frac{2 s}{f_a}\right) (\partial_\mu \phi)^2 
  + \frac{g_h^2 \phi}{32\pi^2 f_a} G^a_{\mu\nu}\tilde G^{a\mu\nu}.
\end{eqnarray}
The interaction between $s$ ($\phi$) and dark gluons is coming from the one-loop diagram mediated by $Q$. 
Finally, we obtain Eq.~(\ref{Model}) at scales well below the mass of the radial scalar $m_s$ which is much larger than the confining scale of the dark gauge symmetry.

If the inflation Hubble rate $H_I$ is given as $m_s \lesssim H_I \ll M_Q$, 
$U(1)_{\rm PQ}$ is not restored during inflation. 
In the case that the dominant scalar density is coming from the misalignment mechanism, 
the oscillation of the radial field happens after inflation around the potential minimum when $H\sim m_s$.  These scalars will eventually decay to axions with the decay rate  $\Gamma_s \sim m_s^3/8\pi f_a^2$ and gluons with the branching fraction ${\rm Br}(s\to gg) \sim (Ng_h^2/8\pi^2)^2$.
The gluons are quickly thermalized and form thermal bath with a temperature $T_g$, while the produced axions are just redshifted. These relativistic axions form dark radiation, whose abundance is negligible when $r$ is small enough.

Although this is just one of the production mechanisms of dark sector, it gives a good motivation to study the isocurvature perturbation of dark matter from the initial fluctuation of dark gluon temperature $\delta T_{g,i}$.
In this example, in addition to the adiabatic perturbation, there are two sources of the isocurvature perturbation. 
One is the fluctuation of the axion field during inflation $\delta\phi_{i}$, and  the other is  the fluctuation of the gluon temperature $\delta T_{g,i}$ induced by initial perturbation of the decaying scalar $s$.  
Because of the condition $m_s\lesssim H_I$, 
the super-horizon modes of both radial scalar and the axion fields get independent fluctuations as $\delta s_i(k)\simeq H_I/2\pi$ and $\delta\phi_i(k)\simeq H_I/2\pi$ during inflation. 
Therefore, $\delta s_i/s_i \sim (\delta \rho_s/\rho_s)_{\rm iso}\sim (\delta T_{g,i}/T_{g,i})_{\rm iso}$. 

Because the dark axions and the dark gluons are coupled with each other by the term $\partial V/\partial T_g$ as in Eq.~(\ref{eq:boltz}), both perturbations could be important for the final isocurvature perturbation of dark matter.  
Based on the evolution of the background dark axion and dark gluon/glueball, 
we solve the equations for the density perturbations focusing on the effect of isocurvature perturbation transfer and obtain the approximated solutions for the super-horizon modes ($k=0$), in order to understand the parametric dependence more clearly.

\subsection{Adiabatic perturbation}
For the evolution of the multicomponent fields or fluids, the perturbations can be decomposed into the curvature (adiabatic) perturbation and the isocurvature (entropy) perturbations. 
The adiabatic perturbation is the modes perturbed along the direction of the background evolution, so that   
\begin{eqnarray}
S_{XY} = {\cal H}\left(\frac{\delta \rho_X}{\rho_X'} -  \frac{\delta \rho_Y}{\rho'_Y}\right) = 0.
\end{eqnarray}
 for any different species $X$ and $Y$ \cite{Liddle:1999hq,Wands:2000dp, Weinberg:2003sw}.
 $S_{XY}$ is the relative entropy perturbation, whose name can be easily understood from thermodynamics. For an isolated species which satisfies the continuity equation 
 $\rho_X' + 3{\cal H}(\rho_X+ p_X)=0$, the perturbation of the entropy density $s_X$ 
 is given as 
 $ \delta s_X/s_X = {\cal H} \delta\rho_X/\rho_X' $, hence  $S_{XY} =3\,  \delta \ln (s_X/s_Y)$. 
 
 The adiabatic mode 
 can be described by the evolution of the comoving curvature perturbation   \cite{Mukhanov:1990me},
\begin{equation}\label{eq:initialR}
    \mathcal{R}=\Phi -  \frac{{\cal H}(\Phi'-\mathcal{H}\Psi)}{{\cal H}'- {\cal H}^2 }~.
\end{equation}
The corresponding initial condition for the adiabatic mode is derived as
\begin{subequations}
\begin{eqnarray}
 &&   \Psi_i=-\Phi_i=-\frac{2}{3}\mathcal{R}_i,\quad 
    \delta_{\gamma,i}=\frac{4}{3}\mathcal{R}_i,\\ 
  &&  \delta_{a,i}=-\frac{2\eta_a}{3}\mathcal{R}_i,\quad
    \delta_{g,i}=\frac{4}{3}\mathcal{R}_i~,
\end{eqnarray}
\end{subequations}
where $\delta_\gamma$ is for the photon fluid,  the index $i$ indicates the time at which the initial perturbation is defined. Here we use the axion potential and mass in Eq. \eqref{eq:mains}.

From Eq. \eqref{eq:conta} and \eqref{eq:contg}, we derive the solutions for the super-horizon modes in radiation-dominated era.
\begin{subequations}\label{eq:adisol}
\begin{eqnarray}
 \hskip -0.4cm   \Psi&=&-\Phi=-\frac{2}{3}\mathcal{R}_i,\quad
    \delta_\gamma=\frac{4}{3}\mathcal{R}_i~,\\
 \hskip -0.4cm    \delta_a&=&\bigg[(1+w_a)-  (1-w_a)\frac{m_a'(T_g(\tau))}{ 3\mathcal{H} m_a(T_g(\tau))}\bigg]\mathcal{R}_i~,\\
 \hskip -0.4cm    \delta_g&=&\bigg[(1+w_g)+\frac{(1-w_a)\rho_a}{\rho_g} \frac{m_a'(T_g(\tau))}{ 3\mathcal{H} m_a(T_g(\tau))}\bigg]\mathcal{R}_i~,
\end{eqnarray}
\end{subequations}
where  $m_a'(T_g(\tau))\equiv dm_a(T_g(\tau))/d\tau$.
As the scale factor becomes larger than $a_{ci}$, 
the terms proportional to $m_a'$ are rapidly vanishing. 
One can show that from the continuity equation for the coupled gluon-axion fluid, 
\begin{equation}
    \Delta(\rho_a\delta_a+\rho_g\delta_g)|_{a=a_{ci}^{\pm}}=0.
\end{equation}
The detailed evolution of $m_a(T_g)$ around $a=a_{ci}$  does not lead to the different final result. 

Eq. \eqref{eq:adisol} states that the adiabatic perturbation shares same form as $\delta_X = (1+w_X) {\cal R}_i$ after the confining phase transition of the dark sector, and no history dependence happens, because  $S_{XY} =0$ holds under the time evolution for the super-horizon modes. This is the characteristic feature of the adiabatic perturbation.

\subsection{Isocurvature perturbation}

The  isocurvature perturbation is a mode perturbed along a direction orthogonal to the direction of background evolution. 
Taking  $S_X$ as 
\begin{eqnarray}
S_X \equiv {\cal H} \left(\frac{\delta \rho_X}{\rho_X'} - \frac{\delta \rho_{\rm tot}}{\rho_{\rm tot}'}\right), 
\end{eqnarray}
we can trace the evolution of the individual component of the isocurvature perturbation.  
At the linear perturbation level, 
the curvature perturbation cannot generate the isocurvature perturbations, and it is also conserved on the super-horizon scales \cite{Wands:2000dp}. 
Thus, for the evolution of  isocurvature perturbations, we can safely take ${\cal R}_i=0$, so that $\Phi_i=0$, $\Psi_i=0$, and  $\delta \rho_{{\rm tot}, i}=0$ as the initial condition at high temperatures, and solve the perturbation equations for $X$ with the initial nonzero $\delta_{X, i}$

The actual evolution of the density perturbation can be numerically calculated based on Eq.~(\ref{eq:boltz}) and compared   with the CMB and matter power spectrum. 
The form of the perturbation becomes particularly simple, if both the axion and glueballs becomes CDM-like well before the matter-radiation equality. In this case, the constraints on the isocurvature perturbation can be easily provided by comparing the analytic formula in super-horizon limit with the criteria of the Planck 2018 \cite{Akrami:2018odb}. Because of this reason, let us focus on such a case. 

The isocurvature perturbation of dark matter is expressed as
\begin{eqnarray}
    (\hat \delta_{\rm DM})_{\rm iso}=\frac{\Omega_a}{\Omega_{\rm DM}}(\hat \delta_a)_{\rm iso}+\frac{\Omega_g}{\Omega_{\rm DM}} (\hat \delta_g)_{\rm iso}, 
\end{eqnarray} 
where `hat' denotes the Gaussian random variables satisfying $\langle \hat \delta_{a, i} \hat \delta_{g, i} \rangle = 0$, and $(\hat \delta_{a,g})_{\rm iso}$ are related with $\hat \delta_{a,g, i}$ as 
\begin{eqnarray}
\left[\begin{array}{c} 
(\hat \delta_a)_{\rm iso} \\ (\hat \delta_g)_{\rm iso} \end{array}\right]
= \left[\begin{array}{cc} 
{\cal T}_{aa} & {\cal T}_{ag}\\ {\cal T}_{ga} & {\cal T}_{gg} \end{array}\right]
\left[\begin{array}{c} 
\hat \delta_{a,i} \\ \hat \delta_{g,i}  \end{array}\right].
\end{eqnarray} 
For the super-horizon modes, 
the diagonal components of the transfer matrix are ${\cal T}_{aa}\simeq {\cal T}_{gg}\simeq 1$.  
The off-diagonal elements  are calculated as follows.

\subsubsection{Induced by the initial displacement of the axion field}

The evolution of the isocurvature perturbation induced by an initial density perturbation of the axion $\delta_{a, i} =2\, \delta\phi_i/\phi_i$ can be described by 
the input value of $\delta_{a, i}$ with the condition $\mathcal{R}_i=0$ and the associated solutions from Eq.~(\ref{eq:boltzdelg}). For $T_g \gg T_{g,c}$,  $\partial V/\partial T_g\approx0$ and 
\begin{equation}
    \Psi_i=-\Phi_i=0, \ \    \delta_{\gamma,i}=0,\ \   \delta_{g,i}=0.
\end{equation}
In the axion dominated dark matter scenario ($\Omega_a \gg \Omega_g$), the dominant contribution is trivial: $(\hat \delta_{\rm DM})_{\rm iso} \simeq \hat \delta_{a,i}$.
In the opposite case ($\Omega_a \ll \Omega_g$), the relevant equation for 
$(\hat\delta_g)_{\rm iso}$ induced by the axion is
\begin{eqnarray}\label{eq:eqdelgAObfCApprox}
\hskip -0.3cm
\hskip -0.6cm&&      \delta'_g+3\mathcal{H}(c_g^2-w_g)\delta_g  
          \nonumber\\
\hskip -0.6cm &&\simeq \frac{(w_a-1) \rho_a}{\rho_g} \frac{m_a'(T_g(\tau))}{m_a(T_g(\tau)}  \left(
1+ 3 {\cal H} c_g^2\frac{T_g}{T_g'}  \right) \delta_{a, i},
\end{eqnarray}
where $c_g^2 \equiv (dp_g/d\ln T_g)/(\partial\rho_g/\partial\ln T_g)$.  
Note that the combination of $1+ 3{\cal H} c_g^2 T_g/T_g'$ is vanishing in the limit of $\rho_a\ll\rho_g$ because \begin{eqnarray}
\frac{T_g'}{T_g} &\simeq& \Big(\frac{\partial \rho_g}{\partial \ln T_g}\Big)^{-1} \frac{d\rho_g}{d\tau}  \simeq - 3 {\cal H} c_g^2.
\end{eqnarray}
This  implies that the transfer matrix element 
${\cal T}_{ga}$ is of ${\cal O}(\Omega_a^2/\Omega_g^2)$. It is the consequence of the  entropy conservation of the dark sector.  So the dominant contribution to the isocurvature perturbation of dark matter is just that from the axion dark matter. 

In summary, 
\begin{eqnarray}
(\hat\delta_{\rm DM})_{\rm iso} & \simeq &\hat\delta_{a, i} \hskip 1.15cm {\rm for} \quad \Omega_g \ll \Omega_a,\nonumber\\
 (\hat\delta_{\rm DM})_{\rm iso} &\simeq&  
\frac{\Omega_a}{\Omega_{\rm DM}} \hat\delta_{a, i}\quad
 {\rm for} \quad \Omega_{a}\ll \Omega_g. 
\end{eqnarray}

\subsubsection{Induced by the initial fluctuation of the gluon temperature}

For the initial fluctuation of the gluon temperature $\delta_{g,i}\approx 4\, \delta T_{g,i}/T_{g,i}$, the condition $\mathcal{R}_i=0$ and $\delta\phi_i=0$ for the perturbative variables give the following initial values 
\begin{subequations}
\begin{eqnarray}
    \Psi_i&=&-\Phi_i=0,\\
    \delta_{\gamma,i}&=&-(N^2-1)r^4\delta_{g,i}~,\\
    \delta_{a,i}&=&-\frac{\eta_a}{2}\delta_{g,i}~.
\end{eqnarray}
\end{subequations}
In the glueball-dominated dark matter case, the contribution of the axion is suppressed by 
its energy density, so $(\hat\delta_{\rm DM})_{\rm iso} \simeq \hat\delta_{g,i}$.
In the opposite case ($\Omega_g \ll \Omega_a$), the effect of the gluon temperature fluctuation to the axion perturbation can be captured by  Eq.~\eqref{eq:EOMdelphi}.  
There are three stages of the axion evolution: 
(I) slow-rolling period ($H \gg m_a(T_g)$), (II) confining phase transition to give the saturation of the axion mass $m_a(T_g)=m_a$, (III) axion oscillating period ($H \ll m_a$).  For (I), the axion mass term is negligible and 
\begin{eqnarray}
\hskip -0.3cm  \delta\phi''+2\mathcal{H}\delta\phi'+ a^2\phi 
\left( \frac{3c_g^2}{4} \frac{d m_a^2(T_g)}{d\ln T_g}\right)\delta_{g,i} \simeq 0.
\end{eqnarray}  
The solution becomes 
\begin{eqnarray}\label{axion_sol1}
    \frac{\delta\phi}{\phi}\simeq&&  \frac{\eta_a }{2(2\eta_a+4)(2\eta_a+5)}
    \left(\frac{m_a^2(T_g)}{H^2}\right) \delta_{g,i}.
\end{eqnarray}
After the confining phase transition, the perturbation of the axion in the periods (II), (III) obeys the equation of motion without $\delta m_a$ term,
\begin{equation}\label{eq:delphiEOMnodelm}
    \delta\phi''+2\mathcal{H}\delta\phi'+a^2 m_a^2\delta\phi=0.
\end{equation}
The general solution to Eq. \eqref{eq:delphiEOMnodelm} can be written as the sum of the Bessel functions 
\begin{eqnarray}\label{eq:besseldelphi}
\hskip-0.6cm    \delta\phi&=&  \bigg(\frac{4H}{m_a}\bigg)^{\frac{1}{4}}
  \sum_{\lambda=\pm} \delta\phi_\lambda J_{\lambda/4}\bigg(\frac{1}{2}\frac{m_a^2}{H^2}\bigg), 
\end{eqnarray} 
with the  constant coefficients $\delta \phi_\pm$. 
Matching Eq.~(\ref{axion_sol1}) with Eq.~(\ref{eq:besseldelphi}) at $a= a_{ci}$ determines $\delta\phi_\pm$ and  gives the solution during the periods (II) and (III). In the period (III), $\delta_a$ is given by
\begin{eqnarray}
    \delta_a&\simeq&  \frac{\eta_a}{2\eta_a+4}\Big(\frac{m_{a}}{H_{ci}}\Big)^2\delta_{g,i},
\end{eqnarray}
where $H_{ci}$ is the Hubble rate at $a=a_{ci}$.  Note that the transfer matrix element ${\cal T}_{ag}$  is suppressed by the factor of $m_{a}^2/H_{ci}^2$ whenever the axion dominates the dark matter, but no further suppression happens.  
Therefore, 
\begin{eqnarray}\label{eq:delmgAO}
 \hskip -0.2cm  
(\hat\delta_{\rm DM})_{\rm iso}& \simeq& \hat\delta_{g, i} \hskip 4.3cm {\rm for} \ \Omega_a \ll \Omega_g,\nonumber\\
 \hskip -0.2cm   (\hat\delta_{\rm DM})_{\rm iso}
  &\simeq& \bigg(\frac{\eta_a}{2\eta_a+4}\Big(\frac{m_{a}}{H_{ci}}\Big)^2+ 
  \frac{\Omega_g}{\Omega_{\rm DM}}\bigg)\hat\delta_{g,i} \ 
 {\rm for}\  \Omega_g\ll \Omega_a. \nonumber \\
 \end{eqnarray}

\subsection{Bound on the isocurvature perturbation}

	Since $\delta\phi_i$ and $\delta T_{g,i}$ are independent random fluctuations, the power spectrum can be decomposed as 
\begin{eqnarray}
P(k,z) =  P_{{\cal RR}}(k, z)  +  \sum_{X=a,g} P_{{\cal II},X}(k,z), 
\end{eqnarray} 
where $X=a,g$ stand for the isocurvature perturbations induced by $\delta_{a, i}$ and $\delta_{g,i}$, respectively.
For the decomposition of the power spectrum as $P(k, z) = (2\pi^2/k^3) {\cal P}(k) T(k, z)$,   we can match the primordial spectrum ${\cal P}(k)$ with the values we obtained for the super-horizon modes in the previous section.
From the observations,  the adiabatic mode is nearly scale-independent. For the isocurvature perturbations, we can also naturally assume they  are nearly scale-invariant if 
they originate from the fluctuations of the scalar fields during inflation as discussed in the beginning of Sec.~\ref{sec:4}. Then,
\begin{eqnarray}
\label{PrrPII}
\hskip -0.3cm    \mathcal{P}_{\mathcal{R}\mathcal{R}}=A_s\bigg(\frac{k}{k_*}\bigg)^{n_s-1}, \ \mathcal{P}_{\mathcal{I}\mathcal{I},X}=A_X\bigg(\frac{k}{k_*}\bigg)^{n_X-1}~,
\end{eqnarray}
where $k_*$ is the pivot scale of the wave number, and parameters $\{A_X,n_X\}$ are constants.

Observation of the CMB presents the upper bound on the isocurvature perturbation~\cite{Crotty:2003rz,Li:2010yb,PhysRevLett.91.131302,Akrami:2018odb}. The constraint is expressed by the bound on the isocurvature fraction $\beta_\text{iso}$, which is defined by
\begin{equation}\label{eq:iso}
    \beta_\text{iso}(k)\equiv\frac{\mathcal{P_{II}}(k)}{\mathcal{P_{RR}}(k)+\mathcal{P_{II}}(k)}~,
\end{equation}
where $\mathcal{P_{RR}}$ and $\mathcal{P_{II}}$ are the power spectra defined in Eq. \eqref{PrrPII}, and related with the density perturbations as $\mathcal{P_{RR}}=k^3\langle\mathcal{R}_i^2\rangle/2\pi^2$, $\mathcal{P_{II}}=\sum_X k^3\langle(\hat\delta_{\rm DM})^2_{{\rm iso},X}\rangle/2\pi^2$. We focus on the large scales in order to constrain the primordial perturbations from the CMB data. The constraint on $\beta_\text{iso}$ for the pivot scale is given by \cite{Akrami:2018odb}
\begin{equation}\label{eq:beta35}
    \beta_\text{iso}(k_*=0.002\, {\rm Mpc}^{-1})<0.035.
\end{equation}
This can be compared with the value calculated in our scenario. 
  
If the axion dominates dark matter, i.e. $\Omega_g\ll \Omega_a\simeq \Omega_{\rm DM}$ ($R(r, f_a) < 1$ for Eq.~(\ref{R(r,f)})), the fraction $\beta_{\rm iso}$ can be expressed as 
\begin{eqnarray} \label{eq:betaiso1}
\hskip -0.3cm\beta_{\rm iso} &\simeq &  \frac{\delta_{a, i}^2}{{\cal R}_i^2}  +  \bigg(\frac{\eta_a}{2\eta_a+4}\Big(\frac{m_{a}}{H_{ci}}\Big)^2+ 
  \frac{\Omega_g}{\Omega_{\rm DM}}\bigg)^2 \frac{\delta_{g, i}^2}{{\cal R}_i^2}, 
\end{eqnarray} 
where 
\begin{eqnarray}
\hskip -0.4cm  
\frac{m_a}{H_{ci}}  \simeq  3 R(r, f_a), \quad
\frac{\Omega_g}{\Omega_{\rm DM}} \simeq (N^2-1) R(r, f_a)^{\frac{3}{2}}.
\end{eqnarray} 
Because $R(r, f_a) < 1$,  the $\Omega_g/\Omega_{\rm DM}$ part is  always dominant in the second term of the RHS in Eq.~(\ref{eq:betaiso1}).  

In the opposite case, if the glueball is the main dark matter component, i.e. $\Omega_a\ll \Omega_g \simeq \Omega_{\rm DM}$  ($R(r, f_a) > 1$), we have
\begin{eqnarray} 
\beta_{\rm iso} &\simeq &  \frac{\delta_{g, i}^2}{{\cal R}_i^2}  + 
  \Big(\frac{\Omega_a}{\Omega_{\rm DM}}\Big)^2 \frac{\delta_{a, i}^2}{{\cal R}_i^2},
\end{eqnarray}  and
\begin{eqnarray}
\frac{\Omega_a}{\Omega_{\rm DM}} \simeq  \frac{1}{(N^2-1) R(r, f_a)^{\frac{3+\eta_a}{2+\eta_a}}} .
\end{eqnarray}

Although the coupling between the axion and the gluon is the key for the amount of axion dark matter, the contributions of the same coupling to the isocurvature perturbation is always subdominant. The perturbation is just close to the sum of the independent elements, which allows a large isocurvature perturbation of the subcomponent dark matter.

For the initial isocurvature perturbation induced by the axion misalignment, we can naturally take the form of $\delta_{a, i} \simeq H_I/2\pi f_a$, where $H_I$ is the inflation Hubble rate.  In the axion dominated case, then, Eq.~(\ref{eq:beta35}) provides the constraint as 
\begin{eqnarray}
&&\left(\frac{N^2-1}{8}\right)^2\left(\frac{r}{0.004}\right)^6\left(\frac{10^{15}\,{\rm GeV}}{f_a}\right)^3 \left(\frac{\delta_{g,i}}{10^{-3}}\right)^2 \\
 &&+ \left(\frac{H_I}{5\times 10^{10}\,{\rm GeV}}\right)^2\left(\frac{10^{15}\,{\rm GeV}}{f_a}\right)^2 \lesssim 1.  \nonumber
\end{eqnarray}
In the glueball dominated case (taking $N=3$),
\begin{eqnarray}
&&\left(\frac{0.001}{r}\right)^{4.73}\left(\frac{f_a}{10^{10}\,{\rm GeV}}\right)^{2.36}\left(\frac{ H_I/2\pi f_a}{0.008}\right)^2\nonumber\\
&&+ \left(\frac{\delta_{g,i}}{4\times 10^{-6}}\right)^2 \lesssim 1. \end{eqnarray} 
Considering both cases, we find the upper bound on the inflation Hubble rate  as $H_I\lesssim 10^{11}$ GeV within the assumption of $H_I < f_a$. In the case of $H_I>f_a$, the axion cosmology is more UV dependent. 
We have to consider the restoration of $U(1)_{\rm PQ}$ symmetry during inflation, thermalization of the axions with the dark gluons, and the formation of dark axion strings. If the axions are thermalized with the gluons, the isocurvature perturbation will be mostly given by $\delta_{g,i}$. At the same time, the axion cosmic strings can leave the large density fluctuation at small scales. We relegate the study on the related cosmology to future work

The effect of subcomponent isocurvature perturbation is not clear yet. Since the glueballs are strongly self-interacting particles, it may provide non-trivial effects when the glueball is the subcomponent dark matter with a large isocurvature perturbation.

In the following section, we study the somewhat different aspect of the subcomponent glueball dark matter in the late time Universe.

\section{\label{sec:5}Subcomponent glueball DM: Formation of supermassive black hole}

The subcomponent self-interacting dark matter can play a certain role  in the formation of the supermassive black holes (SMBH). If the self-interaction is strong enough,  the gravo-thermal collapse of the subcomponent dark matter can occur at the center of the dark matter halo,  leading to the black hole formation at high redshifts $z\gtrsim 7$  \cite{Pollack:2014rja}.  
From the quasar observations, we have the list of SMBHs $(z_{\rm obs}, M_{\rm BH}$)  as
J1342+0928 ($7.54, 7.8\times 10^8M_\odot$), 
J1120+0641 ($7.09, 2.0\times 10^9M_\odot$), 
J2348-3054 ($6.89, 2.1\times 10^9M_\odot$) and also 
J0100+2802 ($6.3, 1.2\times 10^{10} M_\odot$) \cite{Mortlock_2011,DeRosa:2013iia,Wu:2015,Banados:2017unc}. 
The idea is that formation of these SMBHs can be explained by the evolution of the subcomponent dark matter.  

In the standard mechanism on the formation and growth of black holes, SMBHs can exponentially increase their mass by the accretion of baryonic material.
However, because the radiation pressure slows down the absorption of baryons, the rate  is limited. The maximal growth rate is captured by the Salpeter time based on the Eddington limit \cite{Salpeter:1964kb, Volonteri:2010},
\begin{eqnarray}
t_{\rm Sal} =  \frac{\epsilon \sigma_T}{4\pi G m_p} = \left(\frac{\epsilon}{0.1}\right) 45\,{\rm Myr},
\end{eqnarray} where  $m_p$ is the proton mass, $\sigma_T$ is the Thomson scattering cross-section, $G$ is the gravitational constant, and $\epsilon$ is the efficiency factor which depends on the environment of the black hole. 
If the seed black hole is generated at $t_i$ with a mass $M_{\rm seed}$,  the black hole mass is bounded as 
\begin{eqnarray} \label{blackhole_growth}
M_{\rm BH}(t) \lesssim M_{\rm seed} e^{\frac{t - t_i}{t_{\rm Sal}}}.
\end{eqnarray}
If the seed black hole is formed at $z=15$, 
the maximal black hole mass becomes $(2-6)\times 10^4 M_{\rm seed}$ at $z=7$. If the seed is formed at $z=30$, 
its mass becomes $(6-10)\times 10^5 M_{\rm seed}$. 
Therefore in order to explain the SMBHs with masses of ${\cal O}(10^9M_\odot)$ at $z\sim 7$, a seed mass should be greater than $(10^4 - 10^5) M_\odot$. This is quite challenging in the standard theory of the black hole formation. 

\begin{figure}[tbp]
\centering
\includegraphics[width=0.45\textwidth]{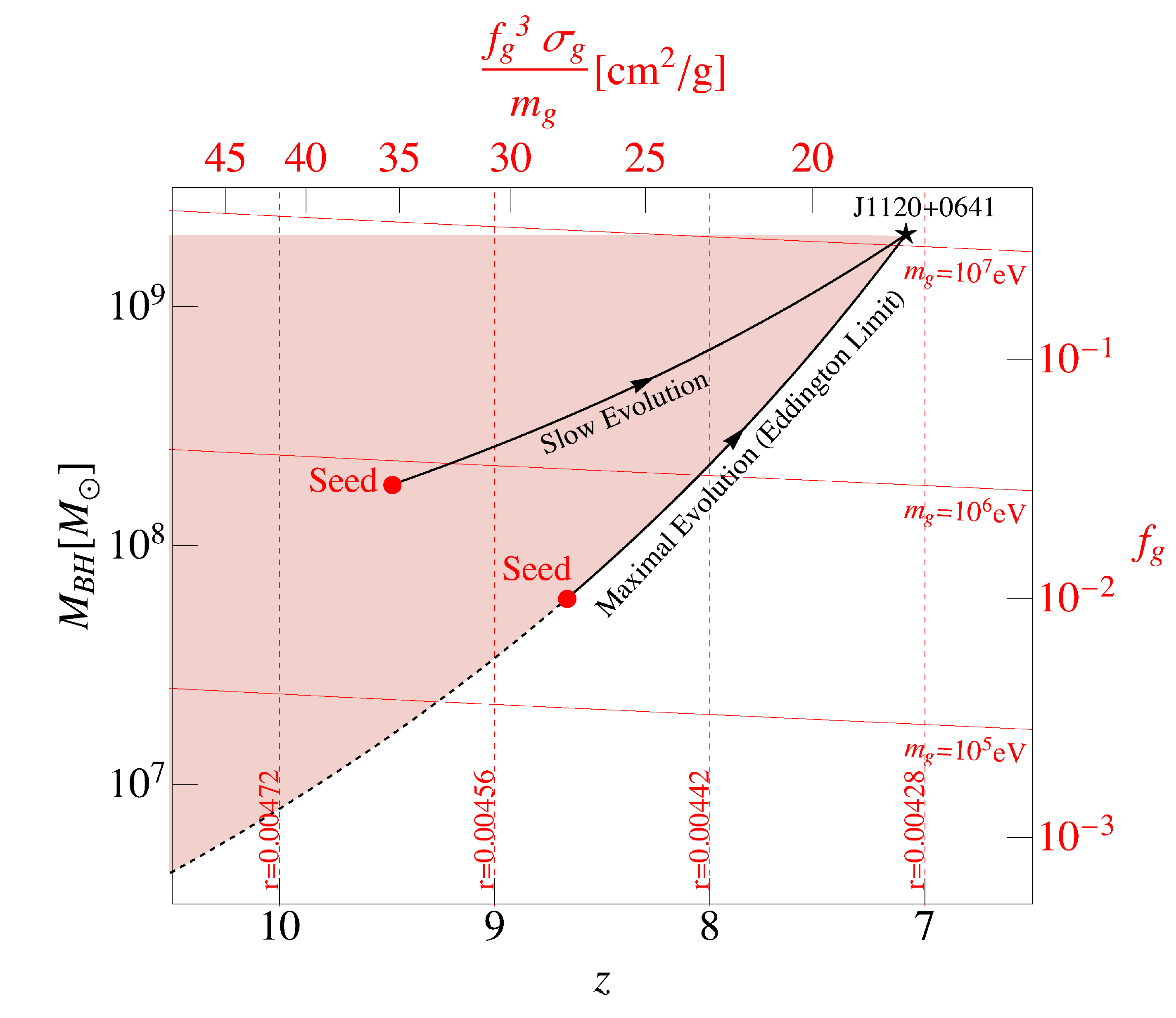}
\caption{\label{fig:SMBHEvolution} Illustration of the black hole growth history  for the observed high $z$ black hole $J1120+0641$ with the assumption of the isolated host halo ($M_h= 10^{12}M_\odot$) as \cite{Pollack:2014rja,Choquette:2018lvq}. All information in red illustrates parameter space for a seed black hole (red dot). The seed black hole can be on the Eddington curve or on the shaded area in which the observations are explained by slower growth of the seed black hole. The time of collapse ($z_{\rm col}$) and the mass of the seed black hole $M_\text{seed}$ are determined by model parameters \{$f_g$, $\sigma_g/m_g$\} or \{$m_g$, $r$\} for a given value of $N$. We take $N=3$.}
\end{figure}

On the other hand, by solving the gravo-thermal fluid equations \cite{Pollack:2014rja} and performing $N$-body simulation \cite{Choquette:2018lvq}  with the assumption that the host halo is isolated, it is shown that 
such a heavy seed black hole could be generated from the gravo-thermal collapse 
of the subcomponent dark matter. Given the NFW density profile 
\begin{eqnarray}
\rho(r) = \frac{\rho_s}{(r/r_s)(1 +r/r_s)^2},
\end{eqnarray} 
for the dominant DM component, the seed black hole is formed with the mass
\begin{eqnarray}
M_{\rm seed} \simeq  \beta_1 f_g M_h,
\end{eqnarray}   
when the age of the Universe becomes
\begin{eqnarray}
t(z_{\rm col}) = t(z_i) + \Delta t_{\rm col}.
\end{eqnarray}
Here, $M_h$ is the mass of the host halo,  $z$ corresponds to the redshift. 
$t(z_i)$ is the time when the virialized dark matter halo is isolated as we assume.  $\Delta t_{\rm col}$ is the duration of the gravo-thermal collapse of the subcomponent dark matter for given initial conditions.
$\beta_1$ and $\Delta t_{\rm col}$ are both calculated numerically.  The fraction factor  $\beta_1\simeq 0.025/(\ln(1+c) -c/(1+c))$  in \cite{Pollack:2014rja},
where $c$ is the concentration of the NFW profile ($M_h= 4\pi \rho_s r_s^3 (\ln(1+c) - c/(1+c))$), and 
$\beta_1\simeq 0.006$ in \cite{Choquette:2018lvq}. 
By comparing the dark matter halo density profiles of two papers, we 
find that both results are well matched.
The formation period $\Delta t_{\rm col}$ is estimated as the form 
\begin{eqnarray}\label{collapse}
\Delta t_{\rm col} \simeq \beta_2  f_g^{-p}  t_{\rm rel}
\end{eqnarray}  
where $\beta_2 \simeq 456\,(480)$, 
$p=0\, (2)$ in \cite{Pollack:2014rja} (\cite{Choquette:2018lvq}),
and the apparent relaxation time of the subcomponent dark matter at $t=t(z_i)$ is defined as
\begin{eqnarray}\label{relaxation}
&& t_{\rm rel} \equiv \frac{m_g}{f_g \sigma_g \rho_s v_s} \\
&&=
0.28\,\text{Myr}\bigg(\frac{10\,{\rm cm}^2/\text{g}}{f_g\sigma_{g}/m_g}\bigg)\bigg(\frac{10^{9}M_{\odot}/\text{kpc}^3}{\rho_s}\bigg)^{3/2}\bigg(\frac{3\,{\rm kpc}}{r_s}\bigg). \nonumber
\end{eqnarray} $\sigma_g$ is the elastic scattering cross-section between two subcomponent dark matters (dark glueballs in our case), $v_s$ is the virialized  velocity at $r=r_s$.
Then the seed black hole can form after the period
\begin{eqnarray}
 \Delta t_{\rm col} &\simeq& 130\,{\rm Myr}\bigg(\frac{10\,{\rm cm}^2/\text{g}}{f_g^{p+1}\sigma_{g}/m_g}\bigg)\nonumber\\
&&\times\bigg(\frac{10^{9}M_{\odot}/\text{kpc}^3}{\rho_s}\bigg)^{3/2}\bigg(\frac{3\,{\rm kpc}}{r_s}\bigg).
\end{eqnarray}
Note that  $\Delta t_{\rm col}$ can be shorter than the age of the Universe for a given $z$, $t(z)\simeq  550\,{\rm Myr}(\frac{10}{1+z})^{3/2}$.
Therefore,  for the isolated halo with a mass $M_h=10^{12} M_\odot$, 
$f_g^{p+1}\sigma_{g}/m_g \gtrsim (1-10)\,{\rm cm}^2/\textrm{g}$, 
and $f_g \lesssim 0.001-0.01$ can explain the SMBH around $z=7$. 
We illustrate the formation of the seed black hole and its growth history in Fig.~\ref{fig:SMBHEvolution} for the halo mass $M_h= 10^{12} M_\odot$. 

In our scenario, the dark glueball dark matter provides a such strongly interacting subcomponent dark matter. Since $\beta_2$ and $p$ are directly estimated in $N$-body simulation, we take the result of \cite{Choquette:2018lvq} ($\beta_2= 480,\, p=2$) as the benchmark value. 
Then, the relevant combination of the model parameters is $f_g^3 \sigma_g/m_g$, which is 
estimated as 
\begin{eqnarray}\label{crosssection}
\hskip -0.6cm \frac{f_g^3\sigma_{g}}{m_g} &=& \left(\frac{3}{N}\right)^4 \left(\frac{f_g}{m_g}\right)^3 \nonumber\\
&\simeq& 40\,{\rm cm^2}/\textrm{g}\left(\frac{3}{N}\right)^4\left(\frac{N^2-1}{10}\right)^3\left(\frac{r}{0.005}\right)^9. 
\end{eqnarray}
For the final expression,  Eq.~(\ref{fraction_g}) is used.
Because it is very sensitive to $r$, 
the ratio parameter is nearly predicted from the explanation of the SMBH  at high $z$. The corresponding  allowed range of the glueball mass is also provided as $m_g= {\cal O}(0.05){\rm MeV}$ for $f_g={\cal O}(0.01)$. 
As to the parameters of the dominant component of dark matter, the axion, 
its decay constant is  $f_a= {\cal O}(10^{15}-10^{16}\,{\rm GeV})$ and the axion mass becomes $m_a = {\cal O}(10^{-14}-10^{-18})\,{\rm eV}$. 
This is safe from the current  fuzzy dark matter constraints. 
Interestingly, this axion mass of $10^{-18}\,{\rm eV}$ is also related with the supermassive black hole with the mass of $M_{\rm BH}\sim 10^7 M_\odot$ through superradiance as we discussed before.
The axions can be efficiently generated from the spinning black hole by superradiant amplification. 
During the amplification, the axion also takes away the sizable amount of the black hole's angular momentum, which gives the contradiction to the observation \cite{Arvanitaki:2014wva}. 
However, if the self-interaction among the axions is sizable, they will collapse before the axion cloud is saturated \cite{Yoshino:2015nsa},  
and the loss of the angular momentum is limited. 
For $m_a\sim 10^{-18}\,{\rm eV}$, the GUT scale decay constant provides 
a sizable  axion self-interaction to trigger bosenovae. 
In our case, the situation is more subtle because of some nontrivial features of the axion potential. 
The perturbative quartic coupling of the axion could be suppressed by increasing $N$, 
but there is a kink structure of the potential at around $\phi\sim \pi f_a$. 
The more detailed study is necessary to provide the correct constraints on $f_a$ from superradiance.

Several simplications are used in the previous discussion. Let us discuss  possible caveats  and alternative history of the seed black hole formation. The host halo mass is taken as  $10^{12} M_\odot$. This is 
because  the halo mass is expected to be greater than ${\cal O}(10^3)$ times the mass of its SMBH \cite{Kormendy:1995er,Magorrian:1997hw}.  
In $N$-body simulations \cite{Heitmann:2006eu,Lukic:2007fc,Tacchella:2018qny}, the comoving number density of the cold dark matter halos with $M_h \geq10^{12}M_\odot$ is evaluated as  $(10^{-5}-10^{-6})({\rm Mpc})^{-3}$ at $z=7$. Thus, the halo is also heavy enough to coincide with the fact that observations of SMBH around $z =7$ are rare.

However, since we consider the formation of the seed black hole at higher redshifts ($z>7$), the existence of such (isolated) heavy halo is questionable.  If we extrapolate the halo mass function obtained by the $N$-body simulation \cite{Tacchella:2018qny},  the comoving number density of the halos with $M_h \geq10^{12}M_\odot$ becomes   
$(10^{-8}-10^{-9})({\rm Mpc})^{-3}$ at $z=10$, and  $10^{-15}({\rm Mpc})^{-3}$ at $z=15$.  In this context, the issue of formation of heavy seed black holes  is
just transferred to the problem of supermassive halo formation at high redshifts. 

On the one hand, based on $N$-body simulations, we can define $M_h(z)$ at a given $z$ in such a way that  the comoving number density of the halos  with  their masses greater than $M_{h}(z)$ is given by $10^{-6}({\rm Mpc})^{-3}$. Then, $M_{h}(z)$ is evaluated as $10^{12} M_\odot$ at $z=7$,  $10^{11} M_\odot$ at $z=10$, and $10^{10} M_\odot$ at $z=15$.  
It is more natural to think the possibility that  
when the seed black hole is formed at $z > 7$, the mass of the host halo is smaller than $10^{12}M_\odot$, 
although it is still one of the heaviest halos at $z_{i}$.  
These heaviest halos get bigger and bigger by mergers with nearby smaller halos or by accretion of the gases.  The actual merger history is quite complex, but the heaviest halo is likely to remain the heaviest. In this sense,  
we consider  $M_{h}(z)$ as the evolution of the host halo mass, and estimate the growth rate $\Gamma_h(z)$  as
\begin{eqnarray}
\Gamma_h(z) \equiv\frac{1}{M_h(z)} \frac{d M_{h}(z)}{dt} \simeq\frac{4}{t(z)}.
\end{eqnarray}  
The last equality holds numerically for $7\lesssim z\lesssim 15$.
The black hole growth rate by the accretion of baryons is much greater than the halo growth rate. However, the halo mass is still hierarchically larger than the black hole mass  during the evolution.

\begin{figure}[tbp]
\centering
\includegraphics[width=0.45\textwidth]{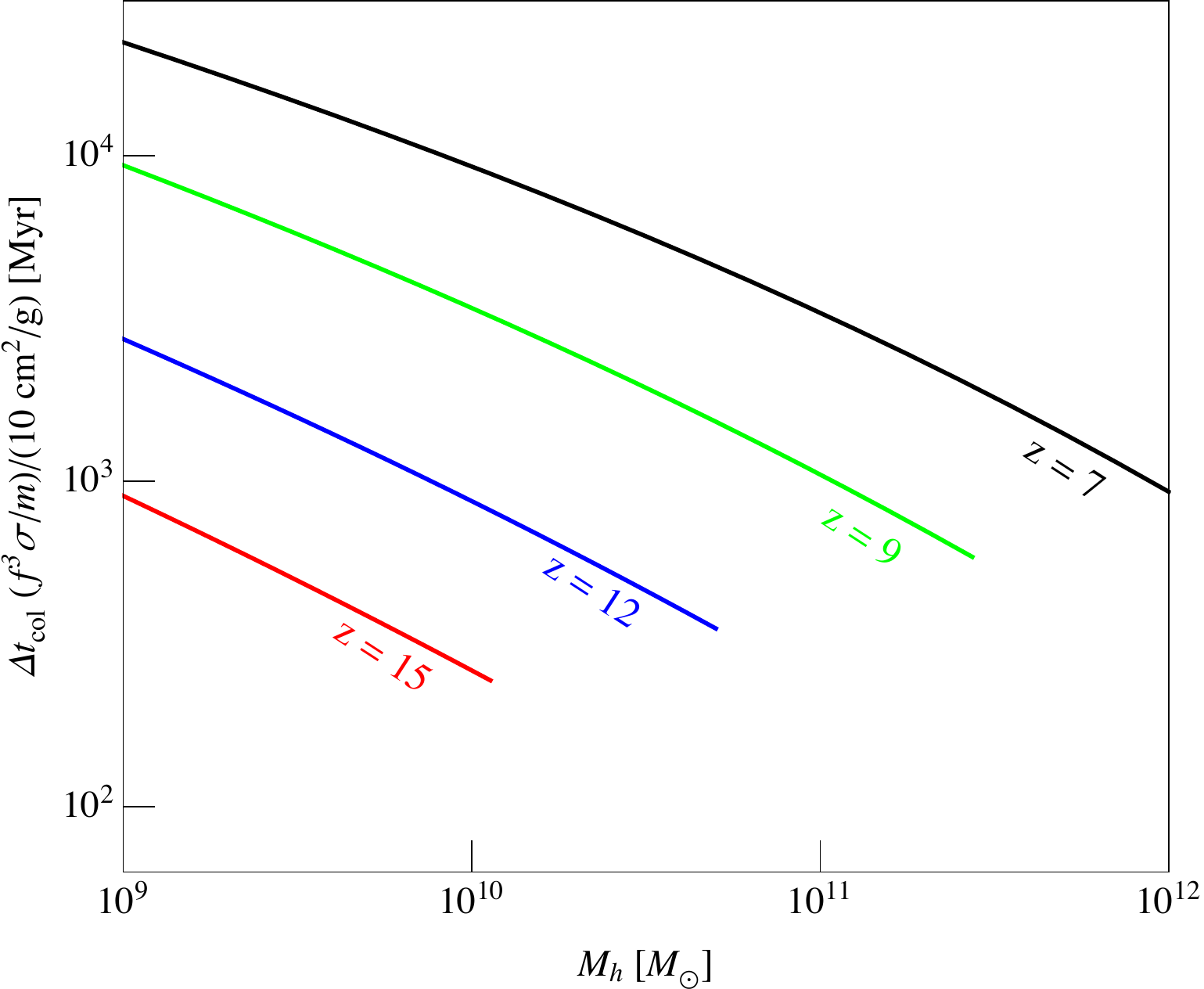}
\caption{\label{fig:collapse_time} 
The expected duration of the gravo-thermal collapse of the subcomponent dark matter $\Delta t_{\rm col}$ in the unit of $(f_g^3\sigma_g/m_g)^{-1}$, defined as Eqs.~(\ref{collapse}) and (\ref{relaxation}) 
with $\beta_2= 480$, $p=2$. It is plotted for the different halo masses and redshifts based on the NFW profile of the dominant dark matter component with the fitted concentration parameter $c(M_h, z)$ \cite{Ishiyama:2020vao}.
The end point of each line corresponds to the halo mass $M_h=M_{h}(z)$. 
The actual collapse time of the subcomponent dark matter will depend on the halo growth history.}
\end{figure}

Another important feature is that 
in terms of the halo mass, 
the relaxation time defined by Eq.~(\ref{relaxation})  depends on $z$, $c$ and $M_h$ as 
\begin{eqnarray}
t_{\rm rel} \propto \frac{(\ln(1+c)-\frac{c}{1+c})^{\frac{3}{2}}}{ (1+z)^{\frac{7}{2}} c^{\frac{7}{2}} M_h^{\frac{1}{3}}}.
\end{eqnarray}
The concentration parameter $c$ also depends on the halo mass and the redshift. 
The recent $N$-body simulation \cite{Ishiyama:2020vao} calculates the concentration parameter $c(M_h, z)$ as the function of $M_h$ and $z$  in a wide range of $M_h$ and $z$. With the reasonable extrapolation, we find
$c(10^{10} M_\odot, 13.8)\simeq c(10^{11} M_\odot, 9.5)\simeq c(10^{12} M_\odot, 7) = 4-5$. Thus,  
 $c(M_{h}(z), z)$  does not have significant $z$ dependence. 
Including all these considerations, Fig.~\ref{fig:collapse_time} shows the apparent gravo-thermal collapse period $\Delta t_{\rm col}$ as the function of $M_h$ and $z$ in the unit of $(f_g^3\sigma_g/m_g)^{-1}$. The formation of the seed black hole is more efficient for heavier halos at a given $z$. 
In order to see whether or not the early formation of the seed is preferred ($z$-dependence), we have to compare $\Delta t_{\rm col}$ with the Hubble time. Numerically, we find that the $z$ dependence of $\Delta t_{\rm col}$ for $M_h= M_{h} (z)$ approximately scales as $1/(1+z)^{1.5}$ in the range $z=7-15$ like the Hubble time. Therefore,  if the seed black hole can form, the formation happens at earlier time with a smaller mass. 
 
Even if $\Delta t_{\rm col}$ is shorter than the age of the Universe at $z_i$, the isolated halo assumption may not be valid if the period of the gravo-thermal collapse is longer than the halo growth time scale $1/\Gamma_h$. 
The general expectation is that the merger process will hinder the gravo-thermal collapse. We consider the conservative criterion for the formation of the seed black hole as 
\begin{eqnarray} \label{conservative}
 \Gamma_h(z) \Delta t_{\rm col}(z)  \lesssim   4\Delta t_{\rm col}(z)/t(z) \lesssim 1.
\end{eqnarray}
This condition means that the seed black hole can only form when the collapse process is faster than the growth rate of the halo mass.  We take $z_i=15$  as the initial redshift for the virialized heaviest host halo.  Then, Eq. \eqref{conservative} is satisfied if
\begin{equation} \label{40cmg}
    \frac{f_g^3\sigma_g}{m_g}\gtrsim 40\,{\rm cm^2/g}.
\end{equation}
After the seed black hole is formed around $z=15$, 
its mass is exponentially growing, and it becomes $M_{\rm BH}=10^{9}M_\odot$ at $z=7$ if the fraction of the glueball dark matter is given as $f_g=2\times10^{-4}$. This is the case of the fastest growth, so the lower bound of $f_g$ to explain current observations of the SMBHs is given by
\begin{equation}
    f_g\gtrsim2 \times10^{-4}~.
\end{equation}

So far, we have ignored the effect of the number-changing interactions of the dark glueballs during the gravo-thermal collapse. If the number-changing process becomes efficient as the density increases, the sizable pressure of the glueballs may disturb the collapse. 
To simplify our discussion, 
in terms of  the temperature of the glueball dark matter ($T_g$) inside the dark matter halo ($r< r_s$), there are two totally different sources to increase $T_g$.  One is the gravo-thermal collapse. Because the gravitationally bound system   has a negative specific heat, as heat flows outward, the glueballs become more and more concentrated in a smaller volume with a larger virial velocity. This results in temperature increasing, and leads to the collapse as the heat outflow accelerates.  
On the other hand, the $3\to 2$ scatterings directly produce the large kinetic energies of the daughter glueballs as $E_{\rm kin}\simeq m_g/2$, respectively.  
These energies will be redistributed among glueballs within the relaxation time, so that the overall glueball temperature will increase compared to the virial temperature, and  inhibit to collapse.  

In order to figure out the condition that the gravo-thermal collapse can start, we require a criterion that the rate of glueball temperature increase is small enough to satisfy  
\begin{eqnarray}
 \frac{\Delta t_{\rm col}}{T_g} \Big(\frac{d T_g}{dt}\Big)_{3\to 2} \ll  1.
\end{eqnarray}
The temperature increase rate by the three-to-two scatterings is estimated for the given glueball density $\rho_g$ and the velocity $v_g$
\begin{eqnarray}
\hskip -1cm \frac{1}{T_g}\Big( \frac{dT_g}{dt}\Big)_{3\to 2}&=& \xi_{\rm eff}  \langle \sigma_{3\to 2} v_g^2\rangle n_g^2 \frac{m_g}{T_g} \simeq  \frac{\langle \sigma_{3\to 2} v_g^2\rangle \rho_g^2}{m_g^2\langle v_g^2\rangle},
\end{eqnarray}
where $T_g = m_g \langle v_g^2\rangle$, 
$\xi_{\rm eff}$ is the efficiency factor of the energy redistribution.  $\xi_{\rm eff}$ could be suppressed if the mean-free path of the glueball  is much larger than the size of the core.  
In our case, most of the glueballs are trapped by the elastic scattering, so $\xi_{\rm eff}\simeq 1$.
Before the gravo-thermal collapse accelerates, the glueball density and the velocity are not much changed. For $\rho_g = f_g \rho_s$, $v_g = v_s$,
\begin{eqnarray}
\hskip -0.4cm \frac{\Delta t_{\rm col}}{T_g} \left(\frac{dT_g}{dt}\right)_{3\to 2} 
&\simeq & 0.06\Big(\frac{10^{-3}}{f_g}\Big)\Big(\frac{3}{N}\Big)^2
\Big(\frac{10^{-3}}{v_s}\Big)^2 \nonumber\\
\hskip -0.4cm  &&\times \Big(\frac{{\rm keV}}{m_g}\Big)^4
\Big(\frac{\rho_s}{10^{12} M_\odot/{\rm kpc}^3}\Big).
\end{eqnarray}
Therefore, 
we expect that the gravo-thermal collapse for the SMBH would not be triggered if $m_g \lesssim {\rm keV}$. 

If $m_g$ is much larger than ${\cal O}({\rm keV})$,  the three-to-two interaction is not effective before the gravo-thermal collapse happens. The gravo-thermal collapse begins to accelerate after $\Delta t_{\rm col}$. During the collapse, the diffusion of the dark matter mass is inefficient, and  the glueballs concentrate their mass of ${\cal O}(M_{\rm seed})$ around the center by increasing the core density and its temperature  \cite{Balberg:2002ue,Balberg:2001qg}.  
Then, the number changing interaction becomes gradually important. It is not clear how it affects the last stage of the gravo-thermal collapse (the formation of the seed black hole).  This is because the temperature increase rate caused by gravo-thermal collapse  is not known yet for such a high mass density of the core. 
We leave it for future work.

There is also the lower bound on the glueball mass from the cosmological  evolution. 
If the glueball is light enough,  it becomes a warm or hot dark matter, so that its speed around $z=7-15$ is greater than the escape velocity of the halo. This means that  the subcomponent dark matter is not clustered, and cannot provide a good initial condition. 
After the dark glueball freeze-out, its velocity scales as $1/a$. The corresponding redshifted glueball velocity at a given $z$ is  
\begin{eqnarray}
v_g(z) \simeq 10^{-3} \left(\frac{1+z}{16}\right)\left(\frac{r}{0.001}\right)^\frac{3}{2} 
\left(\frac{100\,{\rm eV}}{m_g}\right)^\frac{5}{4},
\end{eqnarray}
if the freeze-out happens before the epoch of matter-radiation equality, and 
\begin{eqnarray}
v_g(z) \simeq   10^{-3} \left(\frac{1+z}{16}\right)\left(\frac{r}{0.001}\right)^\frac{4}{3} 
\left(\frac{100\,{\rm eV}}{m_g}\right)^\frac{10}{9},
\end{eqnarray}
if the freeze-out happens in the dark matter dominated era. 
In order to explain the SMBH formation, this value should be hierarchically smaller than the virial velocity $v_s\sim 10^{-3}$ during the period $z=7-15$. 
In our scenario, $r$ is nearly fixed as $0.005$ see Eq.~(\ref{crosssection}). This implies the lower bound on $m_g$ as $100\,{\rm eV}$.

 \section{Conclusions} \label{sec:6}

 We have studied the cosmological evolution of dark light scalars, whose masses and interactions originate from the approximate global symmetry and the non-perturbative dynamics of the hidden gauge symmetry. 
 One is the feebly interacting dark axion, and   
 the other is the strongly interacting dark glueball.  Both can be dark matter if they are light enough. 
 The equations of motion are derived and evaluated to identify the dark matter abundance
 and the perturbation evolution induced by  the coupling between the axion and the dark gluon.  We also explore the possibility that the subcomponent glueball dark matter contributes to the formation of the supermassive black hole  at redshift $z\sim 7$. 
  
Although we have dealt with the problems as closely as possible, there are still many questions that have not been covered by this paper.  What would be the observable consequences of the first-order confining phase transition?  In our discussion, we ignore gravitational wave productions during the confining phase transition, because it is just weakly first-order unless $N$ is very large.  However, if  the phase transition happens around the recombination era, it may leave a footprint on the CMB.   What is the exact form of the axion scalar potential and the effect of self-interactions? 
The scalar potential of the axion is not a simple cosine form, and a multi-branch structure may provide the nontrivial effects if the axion is  produced around the spinning supermassive black hole by superradiant amplification. 
 What is the correct period of the gravo-thermal collapse when the fraction of the subcomponent dark matter is small enough? So far, there is no intensive study on the gravo-thermal collapse of the subcomponent dark matter for such a small fraction. The empirical form of the collapse time scale $\Delta  t_{\rm col}$ should be confirmed for $f_g\ll 10\%$ and higher scattering cross-sections. 
What is the effect of the number changing interactions of the glueball dark matter for the final stage of the black hole formation? During the gravo-thermal collapse, one may think of the possibility that the defining phase transition occurs, because of the large density of the glueball dark matter inside the core. It would be very interesting to study the implication of such a microscopic nature of the dark matter for the final formation of the black hole.    
\\ 
 \\
 \noindent {\bf Acknowledgement}
We would like to thank  Ayuki Kamada, Hee Jung Kim, Ji-hoon Kim, Jong-Wang Lee, Jiajun Zhang for useful discussions. 
BJ, HK, and HDK are supported by NRF 0426-20200008.
CSS is supported by IBS under the project code, IBS-R018-D1.

\bibliography{reference}

\begin{thebibliography}{90}%
\makeatletter
\providecommand \@ifxundefined [1]{%
 \@ifx{#1\undefined}
}%
\providecommand \@ifnum [1]{%
 \ifnum #1\expandafter \@firstoftwo
 \else \expandafter \@secondoftwo
 \fi
}%
\providecommand \@ifx [1]{%
 \ifx #1\expandafter \@firstoftwo
 \else \expandafter \@secondoftwo
 \fi
}%
\providecommand \natexlab [1]{#1}%
\providecommand \enquote  [1]{``#1''}%
\providecommand \bibnamefont  [1]{#1}%
\providecommand \bibfnamefont [1]{#1}%
\providecommand \citenamefont [1]{#1}%
\providecommand \href@noop [0]{\@secondoftwo}%
\providecommand \href [0]{\begingroup \@sanitize@url \@href}%
\providecommand \@href[1]{\@@startlink{#1}\@@href}%
\providecommand \@@href[1]{\endgroup#1\@@endlink}%
\providecommand \@sanitize@url [0]{\catcode `\\12\catcode `\$12\catcode
  `\&12\catcode `\#12\catcode `\^12\catcode `\_12\catcode `\%12\relax}%
\providecommand \@@startlink[1]{}%
\providecommand \@@endlink[0]{}%
\providecommand \url  [0]{\begingroup\@sanitize@url \@url }%
\providecommand \@url [1]{\endgroup\@href {#1}{\urlprefix }}%
\providecommand \urlprefix  [0]{URL }%
\providecommand \Eprint [0]{\href }%
\providecommand \doibase [0]{http://dx.doi.org/}%
\providecommand \selectlanguage [0]{\@gobble}%
\providecommand \bibinfo  [0]{\@secondoftwo}%
\providecommand \bibfield  [0]{\@secondoftwo}%
\providecommand \translation [1]{[#1]}%
\providecommand \BibitemOpen [0]{}%
\providecommand \bibitemStop [0]{}%
\providecommand \bibitemNoStop [0]{.\EOS\space}%
\providecommand \EOS [0]{\spacefactor3000\relax}%
\providecommand \BibitemShut  [1]{\csname bibitem#1\endcsname}%
\let\auto@bib@innerbib\@empty
\bibitem [{\citenamefont {Akrami}\ \emph {et~al.}(2018)\citenamefont {Akrami}
  \emph {et~al.}}]{Akrami:2018odb}%
  \BibitemOpen
  \bibfield  {author} {\bibinfo {author} {\bibfnamefont {Y.}~\bibnamefont
  {Akrami}} \emph {et~al.} (\bibinfo {collaboration} {Planck}),\ }\href@noop {}
  {\  (\bibinfo {year} {2018})},\ \Eprint {http://arxiv.org/abs/1807.06211}
  {arXiv:1807.06211 [astro-ph.CO]} \BibitemShut {NoStop}%
\bibitem [{\citenamefont {Battaglieri}\ \emph {et~al.}(2017)\citenamefont
  {Battaglieri} \emph {et~al.}}]{Battaglieri:2017aum}%
  \BibitemOpen
  \bibfield  {author} {\bibinfo {author} {\bibfnamefont {M.}~\bibnamefont
  {Battaglieri}} \emph {et~al.},\ }in\ \href@noop {} {\emph {\bibinfo
  {booktitle} {{U.S. Cosmic Visions: New Ideas in Dark Matter}}}}\ (\bibinfo
  {year} {2017})\ \Eprint {http://arxiv.org/abs/1707.04591} {arXiv:1707.04591
  [hep-ph]} \BibitemShut {NoStop}%
\bibitem [{\citenamefont {Hu}\ \emph {et~al.}(2000)\citenamefont {Hu},
  \citenamefont {Barkana},\ and\ \citenamefont {Gruzinov}}]{Hu:2000ke}%
  \BibitemOpen
  \bibfield  {author} {\bibinfo {author} {\bibfnamefont {W.}~\bibnamefont
  {Hu}}, \bibinfo {author} {\bibfnamefont {R.}~\bibnamefont {Barkana}}, \ and\
  \bibinfo {author} {\bibfnamefont {A.}~\bibnamefont {Gruzinov}},\ }\href
  {\doibase 10.1103/PhysRevLett.85.1158} {\bibfield  {journal} {\bibinfo
  {journal} {Phys. Rev. Lett.}\ }\textbf {\bibinfo {volume} {85}},\ \bibinfo
  {pages} {1158} (\bibinfo {year} {2000})},\ \Eprint
  {http://arxiv.org/abs/astro-ph/0003365} {arXiv:astro-ph/0003365} \BibitemShut
  {NoStop}%
\bibitem [{\citenamefont {Dine}\ \emph {et~al.}(1981)\citenamefont {Dine},
  \citenamefont {Fischler},\ and\ \citenamefont {Srednicki}}]{Dine:1981rt}%
  \BibitemOpen
  \bibfield  {author} {\bibinfo {author} {\bibfnamefont {M.}~\bibnamefont
  {Dine}}, \bibinfo {author} {\bibfnamefont {W.}~\bibnamefont {Fischler}}, \
  and\ \bibinfo {author} {\bibfnamefont {M.}~\bibnamefont {Srednicki}},\ }\href
  {\doibase 10.1016/0370-2693(81)90590-6} {\bibfield  {journal} {\bibinfo
  {journal} {Phys. Lett. B}\ }\textbf {\bibinfo {volume} {104}},\ \bibinfo
  {pages} {199} (\bibinfo {year} {1981})}\BibitemShut {NoStop}%
\bibitem [{\citenamefont {Zhitnitsky}(1980)}]{Zhitnitsky:1980tq}%
  \BibitemOpen
  \bibfield  {author} {\bibinfo {author} {\bibfnamefont {A.}~\bibnamefont
  {Zhitnitsky}},\ }\href@noop {} {\bibfield  {journal} {\bibinfo  {journal}
  {Sov. J. Nucl. Phys.}\ }\textbf {\bibinfo {volume} {31}},\ \bibinfo {pages}
  {260} (\bibinfo {year} {1980})}\BibitemShut {NoStop}%
\bibitem [{\citenamefont {Kim}(1979)}]{Kim:1979if}%
  \BibitemOpen
  \bibfield  {author} {\bibinfo {author} {\bibfnamefont {J.~E.}\ \bibnamefont
  {Kim}},\ }\href {\doibase 10.1103/PhysRevLett.43.103} {\bibfield  {journal}
  {\bibinfo  {journal} {Phys. Rev. Lett.}\ }\textbf {\bibinfo {volume} {43}},\
  \bibinfo {pages} {103} (\bibinfo {year} {1979})}\BibitemShut {NoStop}%
\bibitem [{\citenamefont {Shifman}\ \emph {et~al.}(1980)\citenamefont
  {Shifman}, \citenamefont {Vainshtein},\ and\ \citenamefont
  {Zakharov}}]{Shifman:1979if}%
  \BibitemOpen
  \bibfield  {author} {\bibinfo {author} {\bibfnamefont {M.~A.}\ \bibnamefont
  {Shifman}}, \bibinfo {author} {\bibfnamefont {A.}~\bibnamefont {Vainshtein}},
  \ and\ \bibinfo {author} {\bibfnamefont {V.~I.}\ \bibnamefont {Zakharov}},\
  }\href {\doibase 10.1016/0550-3213(80)90209-6} {\bibfield  {journal}
  {\bibinfo  {journal} {Nucl. Phys. B}\ }\textbf {\bibinfo {volume} {166}},\
  \bibinfo {pages} {493} (\bibinfo {year} {1980})}\BibitemShut {NoStop}%
\bibitem [{\citenamefont {Carlson}\ \emph {et~al.}(1992)\citenamefont
  {Carlson}, \citenamefont {Machacek},\ and\ \citenamefont
  {Hall}}]{Carlson:1992fn}%
  \BibitemOpen
  \bibfield  {author} {\bibinfo {author} {\bibfnamefont {E.~D.}\ \bibnamefont
  {Carlson}}, \bibinfo {author} {\bibfnamefont {M.~E.}\ \bibnamefont
  {Machacek}}, \ and\ \bibinfo {author} {\bibfnamefont {L.~J.}\ \bibnamefont
  {Hall}},\ }\href {\doibase 10.1086/171833} {\bibfield  {journal} {\bibinfo
  {journal} {Astrophys. J.}\ }\textbf {\bibinfo {volume} {398}},\ \bibinfo
  {pages} {43} (\bibinfo {year} {1992})}\BibitemShut {NoStop}%
\bibitem [{\citenamefont {Boddy}\ \emph {et~al.}(2014)\citenamefont {Boddy},
  \citenamefont {Feng}, \citenamefont {Kaplinghat},\ and\ \citenamefont
  {Tait}}]{Boddy:2014yra}%
  \BibitemOpen
  \bibfield  {author} {\bibinfo {author} {\bibfnamefont {K.~K.}\ \bibnamefont
  {Boddy}}, \bibinfo {author} {\bibfnamefont {J.~L.}\ \bibnamefont {Feng}},
  \bibinfo {author} {\bibfnamefont {M.}~\bibnamefont {Kaplinghat}}, \ and\
  \bibinfo {author} {\bibfnamefont {T.~M.~P.}\ \bibnamefont {Tait}},\ }\href
  {\doibase 10.1103/PhysRevD.89.115017} {\bibfield  {journal} {\bibinfo
  {journal} {Phys. Rev. D}\ }\textbf {\bibinfo {volume} {89}},\ \bibinfo
  {pages} {115017} (\bibinfo {year} {2014})},\ \Eprint
  {http://arxiv.org/abs/1402.3629} {arXiv:1402.3629 [hep-ph]} \BibitemShut
  {NoStop}%
\bibitem [{\citenamefont {Spergel}\ and\ \citenamefont
  {Steinhardt}(2000)}]{Spergel:1999mh}%
  \BibitemOpen
  \bibfield  {author} {\bibinfo {author} {\bibfnamefont {D.~N.}\ \bibnamefont
  {Spergel}}\ and\ \bibinfo {author} {\bibfnamefont {P.~J.}\ \bibnamefont
  {Steinhardt}},\ }\href {\doibase 10.1103/PhysRevLett.84.3760} {\bibfield
  {journal} {\bibinfo  {journal} {Phys. Rev. Lett.}\ }\textbf {\bibinfo
  {volume} {84}},\ \bibinfo {pages} {3760} (\bibinfo {year} {2000})},\ \Eprint
  {http://arxiv.org/abs/astro-ph/9909386} {arXiv:astro-ph/9909386} \BibitemShut
  {NoStop}%
\bibitem [{\citenamefont {Buen-Abad}\ \emph {et~al.}(2018)\citenamefont
  {Buen-Abad}, \citenamefont {Emami},\ and\ \citenamefont
  {Schmaltz}}]{Buen-Abad:2018mas}%
  \BibitemOpen
  \bibfield  {author} {\bibinfo {author} {\bibfnamefont {M.~A.}\ \bibnamefont
  {Buen-Abad}}, \bibinfo {author} {\bibfnamefont {R.}~\bibnamefont {Emami}}, \
  and\ \bibinfo {author} {\bibfnamefont {M.}~\bibnamefont {Schmaltz}},\ }\href
  {\doibase 10.1103/PhysRevD.98.083517} {\bibfield  {journal} {\bibinfo
  {journal} {Phys. Rev. D}\ }\textbf {\bibinfo {volume} {98}},\ \bibinfo
  {pages} {083517} (\bibinfo {year} {2018})},\ \Eprint
  {http://arxiv.org/abs/1803.08062} {arXiv:1803.08062 [hep-ph]} \BibitemShut
  {NoStop}%
\bibitem [{\citenamefont {Davoudiasl}\ and\ \citenamefont
  {Murphy}(2017)}]{Davoudiasl:2017jke}%
  \BibitemOpen
  \bibfield  {author} {\bibinfo {author} {\bibfnamefont {H.}~\bibnamefont
  {Davoudiasl}}\ and\ \bibinfo {author} {\bibfnamefont {C.~W.}\ \bibnamefont
  {Murphy}},\ }\href {\doibase 10.1103/PhysRevLett.118.141801} {\bibfield
  {journal} {\bibinfo  {journal} {Phys. Rev. Lett.}\ }\textbf {\bibinfo
  {volume} {118}},\ \bibinfo {pages} {141801} (\bibinfo {year} {2017})},\
  \Eprint {http://arxiv.org/abs/1701.01136} {arXiv:1701.01136 [hep-ph]}
  \BibitemShut {NoStop}%
\bibitem [{\citenamefont {Halverson}\ \emph {et~al.}(2018)\citenamefont
  {Halverson}, \citenamefont {Nelson}, \citenamefont {Ruehle},\ and\
  \citenamefont {Salinas}}]{Halverson:2018olu}%
  \BibitemOpen
  \bibfield  {author} {\bibinfo {author} {\bibfnamefont {J.}~\bibnamefont
  {Halverson}}, \bibinfo {author} {\bibfnamefont {B.~D.}\ \bibnamefont
  {Nelson}}, \bibinfo {author} {\bibfnamefont {F.}~\bibnamefont {Ruehle}}, \
  and\ \bibinfo {author} {\bibfnamefont {G.}~\bibnamefont {Salinas}},\ }\href
  {\doibase 10.1103/PhysRevD.98.043502} {\bibfield  {journal} {\bibinfo
  {journal} {Phys. Rev. D}\ }\textbf {\bibinfo {volume} {98}},\ \bibinfo
  {pages} {043502} (\bibinfo {year} {2018})},\ \Eprint
  {http://arxiv.org/abs/1805.06011} {arXiv:1805.06011 [hep-ph]} \BibitemShut
  {NoStop}%
\bibitem [{\citenamefont {Pollack}\ \emph {et~al.}(2015)\citenamefont
  {Pollack}, \citenamefont {Spergel},\ and\ \citenamefont
  {Steinhardt}}]{Pollack:2014rja}%
  \BibitemOpen
  \bibfield  {author} {\bibinfo {author} {\bibfnamefont {J.}~\bibnamefont
  {Pollack}}, \bibinfo {author} {\bibfnamefont {D.~N.}\ \bibnamefont
  {Spergel}}, \ and\ \bibinfo {author} {\bibfnamefont {P.~J.}\ \bibnamefont
  {Steinhardt}},\ }\href {\doibase 10.1088/0004-637X/804/2/131} {\bibfield
  {journal} {\bibinfo  {journal} {Astrophys. J.}\ }\textbf {\bibinfo {volume}
  {804}},\ \bibinfo {pages} {131} (\bibinfo {year} {2015})},\ \Eprint
  {http://arxiv.org/abs/1501.00017} {arXiv:1501.00017 [astro-ph.CO]}
  \BibitemShut {NoStop}%
\bibitem [{\citenamefont {Choquette}\ \emph {et~al.}(2019)\citenamefont
  {Choquette}, \citenamefont {Cline},\ and\ \citenamefont
  {Cornell}}]{Choquette:2018lvq}%
  \BibitemOpen
  \bibfield  {author} {\bibinfo {author} {\bibfnamefont {J.}~\bibnamefont
  {Choquette}}, \bibinfo {author} {\bibfnamefont {J.~M.}\ \bibnamefont
  {Cline}}, \ and\ \bibinfo {author} {\bibfnamefont {J.~M.}\ \bibnamefont
  {Cornell}},\ }\href {\doibase 10.1088/1475-7516/2019/07/036} {\bibfield
  {journal} {\bibinfo  {journal} {JCAP}\ }\textbf {\bibinfo {volume} {07}},\
  \bibinfo {pages} {036} (\bibinfo {year} {2019})},\ \Eprint
  {http://arxiv.org/abs/1812.05088} {arXiv:1812.05088 [astro-ph.CO]}
  \BibitemShut {NoStop}%
\bibitem [{\citenamefont {Mortlock}\ \emph {et~al.}(2011)\citenamefont
  {Mortlock}, \citenamefont {Warren}, \citenamefont {Venemans}, \citenamefont
  {Patel}, \citenamefont {Hewett}, \citenamefont {McMahon}, \citenamefont
  {Simpson}, \citenamefont {Theuns}, \citenamefont {Gonzáles-Solares},
  \citenamefont {Adamson},\ and\ \citenamefont {et~al.}}]{Mortlock_2011}%
  \BibitemOpen
  \bibfield  {author} {\bibinfo {author} {\bibfnamefont {D.~J.}\ \bibnamefont
  {Mortlock}}, \bibinfo {author} {\bibfnamefont {S.~J.}\ \bibnamefont
  {Warren}}, \bibinfo {author} {\bibfnamefont {B.~P.}\ \bibnamefont
  {Venemans}}, \bibinfo {author} {\bibfnamefont {M.}~\bibnamefont {Patel}},
  \bibinfo {author} {\bibfnamefont {P.~C.}\ \bibnamefont {Hewett}}, \bibinfo
  {author} {\bibfnamefont {R.~G.}\ \bibnamefont {McMahon}}, \bibinfo {author}
  {\bibfnamefont {C.}~\bibnamefont {Simpson}}, \bibinfo {author} {\bibfnamefont
  {T.}~\bibnamefont {Theuns}}, \bibinfo {author} {\bibfnamefont {E.~A.}\
  \bibnamefont {Gonzáles-Solares}}, \bibinfo {author} {\bibfnamefont
  {A.}~\bibnamefont {Adamson}}, \ and\ \bibinfo {author} {\bibnamefont
  {et~al.}},\ }\href {\doibase 10.1038/nature10159} {\bibfield  {journal}
  {\bibinfo  {journal} {Nature}\ }\textbf {\bibinfo {volume} {474}},\ \bibinfo
  {pages} {616–619} (\bibinfo {year} {2011})}\BibitemShut {NoStop}%
\bibitem [{\citenamefont {De~Rosa}\ \emph {et~al.}(2014)\citenamefont {De~Rosa}
  \emph {et~al.}}]{DeRosa:2013iia}%
  \BibitemOpen
  \bibfield  {author} {\bibinfo {author} {\bibfnamefont {G.}~\bibnamefont
  {De~Rosa}} \emph {et~al.},\ }\href {\doibase 10.1088/0004-637X/790/2/145}
  {\bibfield  {journal} {\bibinfo  {journal} {Astrophys. J.}\ }\textbf
  {\bibinfo {volume} {790}},\ \bibinfo {pages} {145} (\bibinfo {year}
  {2014})},\ \Eprint {http://arxiv.org/abs/1311.3260} {arXiv:1311.3260
  [astro-ph.CO]} \BibitemShut {NoStop}%
\bibitem [{\citenamefont {Wu}\ \emph {et~al.}(2015)\citenamefont {Wu},
  \citenamefont {Wang}, \citenamefont {Fan}, \citenamefont {Yi}, \citenamefont
  {Zuo}, \citenamefont {Bian}, \citenamefont {Jiang}, \citenamefont {McGreer},
  \citenamefont {Wang}, \citenamefont {Yang}, \citenamefont {Yang},
  \citenamefont {Thompson},\ and\ \citenamefont {Beletsky}}]{Wu:2015}%
  \BibitemOpen
  \bibfield  {author} {\bibinfo {author} {\bibfnamefont {X.-B.}\ \bibnamefont
  {Wu}}, \bibinfo {author} {\bibfnamefont {F.}~\bibnamefont {Wang}}, \bibinfo
  {author} {\bibfnamefont {X.}~\bibnamefont {Fan}}, \bibinfo {author}
  {\bibfnamefont {W.}~\bibnamefont {Yi}}, \bibinfo {author} {\bibfnamefont
  {W.}~\bibnamefont {Zuo}}, \bibinfo {author} {\bibfnamefont {F.}~\bibnamefont
  {Bian}}, \bibinfo {author} {\bibfnamefont {L.}~\bibnamefont {Jiang}},
  \bibinfo {author} {\bibfnamefont {I.}~\bibnamefont {McGreer}}, \bibinfo
  {author} {\bibfnamefont {R.}~\bibnamefont {Wang}}, \bibinfo {author}
  {\bibfnamefont {J.}~\bibnamefont {Yang}}, \bibinfo {author} {\bibfnamefont
  {Q.}~\bibnamefont {Yang}}, \bibinfo {author} {\bibfnamefont {D.}~\bibnamefont
  {Thompson}}, \ and\ \bibinfo {author} {\bibfnamefont {Y.}~\bibnamefont
  {Beletsky}},\ }\href {\doibase 10.1038/nature14241} {\bibfield  {journal}
  {\bibinfo  {journal} {Nature}\ }\textbf {\bibinfo {volume} {518}},\ \bibinfo
  {pages} {512} (\bibinfo {year} {2015})}\BibitemShut {NoStop}%
\bibitem [{\citenamefont {Banados}\ \emph {et~al.}(2018)\citenamefont {Banados}
  \emph {et~al.}}]{Banados:2017unc}%
  \BibitemOpen
  \bibfield  {author} {\bibinfo {author} {\bibfnamefont {E.}~\bibnamefont
  {Banados}} \emph {et~al.},\ }\href {\doibase 10.1038/nature25180} {\bibfield
  {journal} {\bibinfo  {journal} {Nature}\ }\textbf {\bibinfo {volume} {553}},\
  \bibinfo {pages} {473} (\bibinfo {year} {2018})},\ \Eprint
  {http://arxiv.org/abs/1712.01860} {arXiv:1712.01860 [astro-ph.GA]}
  \BibitemShut {NoStop}%
\bibitem [{\citenamefont {'t~Hooft}(1974)}]{tHooft:1973alw}%
  \BibitemOpen
  \bibfield  {author} {\bibinfo {author} {\bibfnamefont {G.}~\bibnamefont
  {'t~Hooft}},\ }\href {\doibase 10.1016/0550-3213(74)90154-0} {\bibfield
  {journal} {\bibinfo  {journal} {Nucl. Phys. B}\ }\textbf {\bibinfo {volume}
  {72}},\ \bibinfo {pages} {461} (\bibinfo {year} {1974})}\BibitemShut
  {NoStop}%
\bibitem [{\citenamefont {Witten}(1979)}]{Witten:1979kh}%
  \BibitemOpen
  \bibfield  {author} {\bibinfo {author} {\bibfnamefont {E.}~\bibnamefont
  {Witten}},\ }\href {\doibase 10.1016/0550-3213(79)90232-3} {\bibfield
  {journal} {\bibinfo  {journal} {Nucl. Phys. B}\ }\textbf {\bibinfo {volume}
  {160}},\ \bibinfo {pages} {57} (\bibinfo {year} {1979})}\BibitemShut
  {NoStop}%
\bibitem [{\citenamefont {Manohar}\ and\ \citenamefont
  {Georgi}(1984)}]{Manohar:1983md}%
  \BibitemOpen
  \bibfield  {author} {\bibinfo {author} {\bibfnamefont {A.}~\bibnamefont
  {Manohar}}\ and\ \bibinfo {author} {\bibfnamefont {H.}~\bibnamefont
  {Georgi}},\ }\href {\doibase 10.1016/0550-3213(84)90231-1} {\bibfield
  {journal} {\bibinfo  {journal} {Nucl. Phys. B}\ }\textbf {\bibinfo {volume}
  {234}},\ \bibinfo {pages} {189} (\bibinfo {year} {1984})}\BibitemShut
  {NoStop}%
\bibitem [{\citenamefont {Coleman}(1985)}]{Coleman:1985rnk}%
  \BibitemOpen
  \bibfield  {author} {\bibinfo {author} {\bibfnamefont {S.}~\bibnamefont
  {Coleman}},\ }\href {\doibase 10.1017/CBO9780511565045} {\emph {\bibinfo
  {title} {{Aspects of Symmetry}: {Selected Erice Lectures}}}}\ (\bibinfo
  {publisher} {Cambridge University Press},\ \bibinfo {address} {Cambridge,
  U.K.},\ \bibinfo {year} {1985})\BibitemShut {NoStop}%
\bibitem [{\citenamefont {Cohen}\ \emph {et~al.}(1997)\citenamefont {Cohen},
  \citenamefont {Kaplan},\ and\ \citenamefont {Nelson}}]{Cohen:1997rt}%
  \BibitemOpen
  \bibfield  {author} {\bibinfo {author} {\bibfnamefont {A.~G.}\ \bibnamefont
  {Cohen}}, \bibinfo {author} {\bibfnamefont {D.~B.}\ \bibnamefont {Kaplan}}, \
  and\ \bibinfo {author} {\bibfnamefont {A.~E.}\ \bibnamefont {Nelson}},\
  }\href {\doibase 10.1016/S0370-2693(97)00995-7} {\bibfield  {journal}
  {\bibinfo  {journal} {Phys. Lett. B}\ }\textbf {\bibinfo {volume} {412}},\
  \bibinfo {pages} {301} (\bibinfo {year} {1997})},\ \Eprint
  {http://arxiv.org/abs/hep-ph/9706275} {arXiv:hep-ph/9706275} \BibitemShut
  {NoStop}%
\bibitem [{\citenamefont {Witten}(1980)}]{Witten:1980sp}%
  \BibitemOpen
  \bibfield  {author} {\bibinfo {author} {\bibfnamefont {E.}~\bibnamefont
  {Witten}},\ }\href {\doibase 10.1016/0003-4916(80)90325-5} {\bibfield
  {journal} {\bibinfo  {journal} {Annals Phys.}\ }\textbf {\bibinfo {volume}
  {128}},\ \bibinfo {pages} {363} (\bibinfo {year} {1980})}\BibitemShut
  {NoStop}%
\bibitem [{\citenamefont {Witten}(1998)}]{Witten:1998uka}%
  \BibitemOpen
  \bibfield  {author} {\bibinfo {author} {\bibfnamefont {E.}~\bibnamefont
  {Witten}},\ }\href {\doibase 10.1103/PhysRevLett.81.2862} {\bibfield
  {journal} {\bibinfo  {journal} {Phys. Rev. Lett.}\ }\textbf {\bibinfo
  {volume} {81}},\ \bibinfo {pages} {2862} (\bibinfo {year} {1998})},\ \Eprint
  {http://arxiv.org/abs/hep-th/9807109} {arXiv:hep-th/9807109} \BibitemShut
  {NoStop}%
\bibitem [{\citenamefont {Del~Debbio}\ \emph {et~al.}(2006)\citenamefont
  {Del~Debbio}, \citenamefont {Manca}, \citenamefont {Panagopoulos},
  \citenamefont {Skouroupathis},\ and\ \citenamefont
  {Vicari}}]{DelDebbio:2006yuf}%
  \BibitemOpen
  \bibfield  {author} {\bibinfo {author} {\bibfnamefont {L.}~\bibnamefont
  {Del~Debbio}}, \bibinfo {author} {\bibfnamefont {G.~M.}\ \bibnamefont
  {Manca}}, \bibinfo {author} {\bibfnamefont {H.}~\bibnamefont {Panagopoulos}},
  \bibinfo {author} {\bibfnamefont {A.}~\bibnamefont {Skouroupathis}}, \ and\
  \bibinfo {author} {\bibfnamefont {E.}~\bibnamefont {Vicari}},\ }\href
  {\doibase 10.1088/1126-6708/2006/06/005} {\bibfield  {journal} {\bibinfo
  {journal} {JHEP}\ }\textbf {\bibinfo {volume} {06}},\ \bibinfo {pages} {005}
  (\bibinfo {year} {2006})},\ \Eprint {http://arxiv.org/abs/hep-th/0603041}
  {arXiv:hep-th/0603041} \BibitemShut {NoStop}%
\bibitem [{\citenamefont {Vicari}\ and\ \citenamefont
  {Panagopoulos}(2009)}]{Vicari:2008jw}%
  \BibitemOpen
  \bibfield  {author} {\bibinfo {author} {\bibfnamefont {E.}~\bibnamefont
  {Vicari}}\ and\ \bibinfo {author} {\bibfnamefont {H.}~\bibnamefont
  {Panagopoulos}},\ }\href {\doibase 10.1016/j.physrep.2008.10.001} {\bibfield
  {journal} {\bibinfo  {journal} {Phys. Rept.}\ }\textbf {\bibinfo {volume}
  {470}},\ \bibinfo {pages} {93} (\bibinfo {year} {2009})},\ \Eprint
  {http://arxiv.org/abs/0803.1593} {arXiv:0803.1593 [hep-th]} \BibitemShut
  {NoStop}%
\bibitem [{\citenamefont {Gross}\ \emph {et~al.}(1981)\citenamefont {Gross},
  \citenamefont {Pisarski},\ and\ \citenamefont {Yaffe}}]{RevModPhys.53.43}%
  \BibitemOpen
  \bibfield  {author} {\bibinfo {author} {\bibfnamefont {D.~J.}\ \bibnamefont
  {Gross}}, \bibinfo {author} {\bibfnamefont {R.~D.}\ \bibnamefont {Pisarski}},
  \ and\ \bibinfo {author} {\bibfnamefont {L.~G.}\ \bibnamefont {Yaffe}},\
  }\href {\doibase 10.1103/RevModPhys.53.43} {\bibfield  {journal} {\bibinfo
  {journal} {Rev. Mod. Phys.}\ }\textbf {\bibinfo {volume} {53}},\ \bibinfo
  {pages} {43} (\bibinfo {year} {1981})}\BibitemShut {NoStop}%
\bibitem [{\citenamefont {Borsanyi}\ \emph {et~al.}(2016)\citenamefont
  {Borsanyi} \emph {et~al.}}]{Borsanyi:2016ksw}%
  \BibitemOpen
  \bibfield  {author} {\bibinfo {author} {\bibfnamefont {S.}~\bibnamefont
  {Borsanyi}} \emph {et~al.},\ }\href {\doibase 10.1038/nature20115} {\bibfield
   {journal} {\bibinfo  {journal} {Nature}\ }\textbf {\bibinfo {volume}
  {539}},\ \bibinfo {pages} {69} (\bibinfo {year} {2016})},\ \Eprint
  {http://arxiv.org/abs/1606.07494} {arXiv:1606.07494 [hep-lat]} \BibitemShut
  {NoStop}%
\bibitem [{\citenamefont {Bartolo}\ \emph {et~al.}(2004)\citenamefont
  {Bartolo}, \citenamefont {Corasaniti}, \citenamefont {Liddle},\ and\
  \citenamefont {Malquarti}}]{Bartolo:2003ad}%
  \BibitemOpen
  \bibfield  {author} {\bibinfo {author} {\bibfnamefont {N.}~\bibnamefont
  {Bartolo}}, \bibinfo {author} {\bibfnamefont {P.~S.}\ \bibnamefont
  {Corasaniti}}, \bibinfo {author} {\bibfnamefont {A.~R.}\ \bibnamefont
  {Liddle}}, \ and\ \bibinfo {author} {\bibfnamefont {M.}~\bibnamefont
  {Malquarti}},\ }\href {\doibase 10.1103/PhysRevD.70.043532} {\bibfield
  {journal} {\bibinfo  {journal} {Phys. Rev. D}\ }\textbf {\bibinfo {volume}
  {70}},\ \bibinfo {pages} {043532} (\bibinfo {year} {2004})},\ \Eprint
  {http://arxiv.org/abs/astro-ph/0311503} {arXiv:astro-ph/0311503} \BibitemShut
  {NoStop}%
\bibitem [{\citenamefont {Hu}(1998)}]{Hu:1998kj}%
  \BibitemOpen
  \bibfield  {author} {\bibinfo {author} {\bibfnamefont {W.}~\bibnamefont
  {Hu}},\ }\href {\doibase 10.1086/306274} {\bibfield  {journal} {\bibinfo
  {journal} {Astrophys. J.}\ }\textbf {\bibinfo {volume} {506}},\ \bibinfo
  {pages} {485} (\bibinfo {year} {1998})},\ \Eprint
  {http://arxiv.org/abs/astro-ph/9801234} {arXiv:astro-ph/9801234} \BibitemShut
  {NoStop}%
\bibitem [{\citenamefont {Ma}\ and\ \citenamefont
  {Bertschinger}(1995)}]{Ma:1995ey}%
  \BibitemOpen
  \bibfield  {author} {\bibinfo {author} {\bibfnamefont {C.-P.}\ \bibnamefont
  {Ma}}\ and\ \bibinfo {author} {\bibfnamefont {E.}~\bibnamefont
  {Bertschinger}},\ }\href {\doibase 10.1086/176550} {\bibfield  {journal}
  {\bibinfo  {journal} {Astrophys. J.}\ }\textbf {\bibinfo {volume} {455}},\
  \bibinfo {pages} {7} (\bibinfo {year} {1995})},\ \Eprint
  {http://arxiv.org/abs/astro-ph/9506072} {arXiv:astro-ph/9506072} \BibitemShut
  {NoStop}%
\bibitem [{\citenamefont {McLerran}\ \emph {et~al.}(1991)\citenamefont
  {McLerran}, \citenamefont {Mottola},\ and\ \citenamefont
  {Shaposhnikov}}]{McLerran:1990de}%
  \BibitemOpen
  \bibfield  {author} {\bibinfo {author} {\bibfnamefont {L.~D.}\ \bibnamefont
  {McLerran}}, \bibinfo {author} {\bibfnamefont {E.}~\bibnamefont {Mottola}}, \
  and\ \bibinfo {author} {\bibfnamefont {M.~E.}\ \bibnamefont {Shaposhnikov}},\
  }\href {\doibase 10.1103/PhysRevD.43.2027} {\bibfield  {journal} {\bibinfo
  {journal} {Phys. Rev. D}\ }\textbf {\bibinfo {volume} {43}},\ \bibinfo
  {pages} {2027} (\bibinfo {year} {1991})}\BibitemShut {NoStop}%
\bibitem [{\citenamefont {Moore}\ and\ \citenamefont
  {Tassler}(2011)}]{Moore:2010jd}%
  \BibitemOpen
  \bibfield  {author} {\bibinfo {author} {\bibfnamefont {G.~D.}\ \bibnamefont
  {Moore}}\ and\ \bibinfo {author} {\bibfnamefont {M.}~\bibnamefont
  {Tassler}},\ }\href {\doibase 10.1007/JHEP02(2011)105} {\bibfield  {journal}
  {\bibinfo  {journal} {JHEP}\ }\textbf {\bibinfo {volume} {02}},\ \bibinfo
  {pages} {105} (\bibinfo {year} {2011})},\ \Eprint
  {http://arxiv.org/abs/1011.1167} {arXiv:1011.1167 [hep-ph]} \BibitemShut
  {NoStop}%
\bibitem [{\citenamefont {Son}\ and\ \citenamefont
  {Starinets}(2002)}]{Son:2002sd}%
  \BibitemOpen
  \bibfield  {author} {\bibinfo {author} {\bibfnamefont {D.~T.}\ \bibnamefont
  {Son}}\ and\ \bibinfo {author} {\bibfnamefont {A.~O.}\ \bibnamefont
  {Starinets}},\ }\href {\doibase 10.1088/1126-6708/2002/09/042} {\bibfield
  {journal} {\bibinfo  {journal} {JHEP}\ }\textbf {\bibinfo {volume} {09}},\
  \bibinfo {pages} {042} (\bibinfo {year} {2002})},\ \Eprint
  {http://arxiv.org/abs/hep-th/0205051} {arXiv:hep-th/0205051} \BibitemShut
  {NoStop}%
\bibitem [{\citenamefont {Boyd}\ \emph {et~al.}(1996)\citenamefont {Boyd},
  \citenamefont {Engels}, \citenamefont {Karsch}, \citenamefont {Laermann},
  \citenamefont {Legeland}, \citenamefont {Lutgemeier},\ and\ \citenamefont
  {Petersson}}]{Boyd:1996bx}%
  \BibitemOpen
  \bibfield  {author} {\bibinfo {author} {\bibfnamefont {G.}~\bibnamefont
  {Boyd}}, \bibinfo {author} {\bibfnamefont {J.}~\bibnamefont {Engels}},
  \bibinfo {author} {\bibfnamefont {F.}~\bibnamefont {Karsch}}, \bibinfo
  {author} {\bibfnamefont {E.}~\bibnamefont {Laermann}}, \bibinfo {author}
  {\bibfnamefont {C.}~\bibnamefont {Legeland}}, \bibinfo {author}
  {\bibfnamefont {M.}~\bibnamefont {Lutgemeier}}, \ and\ \bibinfo {author}
  {\bibfnamefont {B.}~\bibnamefont {Petersson}},\ }\href {\doibase
  10.1016/0550-3213(96)00170-8} {\bibfield  {journal} {\bibinfo  {journal}
  {Nucl. Phys. B}\ }\textbf {\bibinfo {volume} {469}},\ \bibinfo {pages} {419}
  (\bibinfo {year} {1996})},\ \Eprint {http://arxiv.org/abs/hep-lat/9602007}
  {arXiv:hep-lat/9602007} \BibitemShut {NoStop}%
\bibitem [{\citenamefont {Deng}(1989)}]{DENG1989334}%
  \BibitemOpen
  \bibfield  {author} {\bibinfo {author} {\bibfnamefont {Y.}~\bibnamefont
  {Deng}},\ }\href {\doibase https://doi.org/10.1016/0920-5632(89)90121-7}
  {\bibfield  {journal} {\bibinfo  {journal} {Nuclear Physics B - Proceedings
  Supplements}\ }\textbf {\bibinfo {volume} {9}},\ \bibinfo {pages} {334 }
  (\bibinfo {year} {1989})}\BibitemShut {NoStop}%
\bibitem [{\citenamefont {Datta}\ and\ \citenamefont
  {Gupta}(2010)}]{Datta:2010sq}%
  \BibitemOpen
  \bibfield  {author} {\bibinfo {author} {\bibfnamefont {S.}~\bibnamefont
  {Datta}}\ and\ \bibinfo {author} {\bibfnamefont {S.}~\bibnamefont {Gupta}},\
  }\href {\doibase 10.1103/PhysRevD.82.114505} {\bibfield  {journal} {\bibinfo
  {journal} {Phys. Rev. D}\ }\textbf {\bibinfo {volume} {82}},\ \bibinfo
  {pages} {114505} (\bibinfo {year} {2010})},\ \Eprint
  {http://arxiv.org/abs/1006.0938} {arXiv:1006.0938 [hep-lat]} \BibitemShut
  {NoStop}%
\bibitem [{\citenamefont {Borsanyi}\ \emph {et~al.}(2012)\citenamefont
  {Borsanyi}, \citenamefont {Endrodi}, \citenamefont {Fodor}, \citenamefont
  {Katz},\ and\ \citenamefont {Szabo}}]{Borsanyi:2012ve}%
  \BibitemOpen
  \bibfield  {author} {\bibinfo {author} {\bibfnamefont {S.}~\bibnamefont
  {Borsanyi}}, \bibinfo {author} {\bibfnamefont {G.}~\bibnamefont {Endrodi}},
  \bibinfo {author} {\bibfnamefont {Z.}~\bibnamefont {Fodor}}, \bibinfo
  {author} {\bibfnamefont {S.}~\bibnamefont {Katz}}, \ and\ \bibinfo {author}
  {\bibfnamefont {K.}~\bibnamefont {Szabo}},\ }\href {\doibase
  10.1007/JHEP07(2012)056} {\bibfield  {journal} {\bibinfo  {journal} {JHEP}\
  }\textbf {\bibinfo {volume} {07}},\ \bibinfo {pages} {056} (\bibinfo {year}
  {2012})},\ \Eprint {http://arxiv.org/abs/1204.6184} {arXiv:1204.6184
  [hep-lat]} \BibitemShut {NoStop}%
\bibitem [{\citenamefont {Teper}(1997)}]{Teper:1997am}%
  \BibitemOpen
  \bibfield  {author} {\bibinfo {author} {\bibfnamefont {M.~J.}\ \bibnamefont
  {Teper}},\ }in\ \href@noop {} {\emph {\bibinfo {booktitle} {{NATO Advanced
  Study Institute on Confinement, Duality and Nonperturbative Aspects of
  QCD}}}}\ (\bibinfo {year} {1997})\ pp.\ \bibinfo {pages} {43--74},\ \Eprint
  {http://arxiv.org/abs/hep-lat/9711011} {arXiv:hep-lat/9711011} \BibitemShut
  {NoStop}%
\bibitem [{\citenamefont {Teper}(1998)}]{Teper:1998kw}%
  \BibitemOpen
  \bibfield  {author} {\bibinfo {author} {\bibfnamefont {M.~J.}\ \bibnamefont
  {Teper}},\ }\href@noop {} {\  (\bibinfo {year} {1998})},\ \Eprint
  {http://arxiv.org/abs/hep-th/9812187} {arXiv:hep-th/9812187} \BibitemShut
  {NoStop}%
\bibitem [{\citenamefont {Morningstar}\ and\ \citenamefont
  {Peardon}(1999)}]{Morningstar:1999rf}%
  \BibitemOpen
  \bibfield  {author} {\bibinfo {author} {\bibfnamefont {C.~J.}\ \bibnamefont
  {Morningstar}}\ and\ \bibinfo {author} {\bibfnamefont {M.~J.}\ \bibnamefont
  {Peardon}},\ }\href {\doibase 10.1103/PhysRevD.60.034509} {\bibfield
  {journal} {\bibinfo  {journal} {Phys. Rev. D}\ }\textbf {\bibinfo {volume}
  {60}},\ \bibinfo {pages} {034509} (\bibinfo {year} {1999})},\ \Eprint
  {http://arxiv.org/abs/hep-lat/9901004} {arXiv:hep-lat/9901004} \BibitemShut
  {NoStop}%
\bibitem [{\citenamefont {Athenodorou}\ and\ \citenamefont
  {Teper}(2020)}]{Athenodorou:2020ani}%
  \BibitemOpen
  \bibfield  {author} {\bibinfo {author} {\bibfnamefont {A.}~\bibnamefont
  {Athenodorou}}\ and\ \bibinfo {author} {\bibfnamefont {M.}~\bibnamefont
  {Teper}},\ }\href@noop {} {\  (\bibinfo {year} {2020})},\ \Eprint
  {http://arxiv.org/abs/2007.06422} {arXiv:2007.06422 [hep-lat]} \BibitemShut
  {NoStop}%
\bibitem [{\citenamefont {Lucini}\ and\ \citenamefont
  {Moraitis}(2008)}]{Lucini:2008vi}%
  \BibitemOpen
  \bibfield  {author} {\bibinfo {author} {\bibfnamefont {B.}~\bibnamefont
  {Lucini}}\ and\ \bibinfo {author} {\bibfnamefont {G.}~\bibnamefont
  {Moraitis}},\ }\href {\doibase 10.1016/j.physletb.2008.08.047} {\bibfield
  {journal} {\bibinfo  {journal} {Phys. Lett. B}\ }\textbf {\bibinfo {volume}
  {668}},\ \bibinfo {pages} {226} (\bibinfo {year} {2008})},\ \Eprint
  {http://arxiv.org/abs/0805.2913} {arXiv:0805.2913 [hep-lat]} \BibitemShut
  {NoStop}%
\bibitem [{\citenamefont {Lucini}\ \emph {et~al.}(2012)\citenamefont {Lucini},
  \citenamefont {Rago},\ and\ \citenamefont {Rinaldi}}]{Lucini:2012wq}%
  \BibitemOpen
  \bibfield  {author} {\bibinfo {author} {\bibfnamefont {B.}~\bibnamefont
  {Lucini}}, \bibinfo {author} {\bibfnamefont {A.}~\bibnamefont {Rago}}, \ and\
  \bibinfo {author} {\bibfnamefont {E.}~\bibnamefont {Rinaldi}},\ }\href
  {\doibase 10.1016/j.physletb.2012.04.070} {\bibfield  {journal} {\bibinfo
  {journal} {Phys. Lett. B}\ }\textbf {\bibinfo {volume} {712}},\ \bibinfo
  {pages} {279} (\bibinfo {year} {2012})},\ \Eprint
  {http://arxiv.org/abs/1202.6684} {arXiv:1202.6684 [hep-lat]} \BibitemShut
  {NoStop}%
\bibitem [{\citenamefont {Lucini}\ and\ \citenamefont
  {Teper}(2001)}]{Lucini:2001ej}%
  \BibitemOpen
  \bibfield  {author} {\bibinfo {author} {\bibfnamefont {B.}~\bibnamefont
  {Lucini}}\ and\ \bibinfo {author} {\bibfnamefont {M.}~\bibnamefont {Teper}},\
  }\href {\doibase 10.1088/1126-6708/2001/06/050} {\bibfield  {journal}
  {\bibinfo  {journal} {JHEP}\ }\textbf {\bibinfo {volume} {06}},\ \bibinfo
  {pages} {050} (\bibinfo {year} {2001})},\ \Eprint
  {http://arxiv.org/abs/hep-lat/0103027} {arXiv:hep-lat/0103027} \BibitemShut
  {NoStop}%
\bibitem [{\citenamefont {West}(1996)}]{West:1995ym}%
  \BibitemOpen
  \bibfield  {author} {\bibinfo {author} {\bibfnamefont {G.~B.}\ \bibnamefont
  {West}},\ }\href {\doibase 10.1103/PhysRevLett.77.2622} {\bibfield  {journal}
  {\bibinfo  {journal} {Phys. Rev. Lett.}\ }\textbf {\bibinfo {volume} {77}},\
  \bibinfo {pages} {2622} (\bibinfo {year} {1996})},\ \Eprint
  {http://arxiv.org/abs/hep-ph/9603316} {arXiv:hep-ph/9603316} \BibitemShut
  {NoStop}%
\bibitem [{\citenamefont {Hagedorn}\ and\ \citenamefont
  {Rafelski}(1980)}]{Hagedorn:1980kb}%
  \BibitemOpen
  \bibfield  {author} {\bibinfo {author} {\bibfnamefont {R.}~\bibnamefont
  {Hagedorn}}\ and\ \bibinfo {author} {\bibfnamefont {J.}~\bibnamefont
  {Rafelski}},\ }\href {\doibase 10.1016/0370-2693(80)90566-3} {\bibfield
  {journal} {\bibinfo  {journal} {Phys. Lett. B}\ }\textbf {\bibinfo {volume}
  {97}},\ \bibinfo {pages} {136} (\bibinfo {year} {1980})}\BibitemShut
  {NoStop}%
\bibitem [{\citenamefont {Caselle}\ \emph {et~al.}(2011)\citenamefont
  {Caselle}, \citenamefont {Castagnini}, \citenamefont {Feo}, \citenamefont
  {Gliozzi},\ and\ \citenamefont {Panero}}]{Caselle:2011fy}%
  \BibitemOpen
  \bibfield  {author} {\bibinfo {author} {\bibfnamefont {M.}~\bibnamefont
  {Caselle}}, \bibinfo {author} {\bibfnamefont {L.}~\bibnamefont {Castagnini}},
  \bibinfo {author} {\bibfnamefont {A.}~\bibnamefont {Feo}}, \bibinfo {author}
  {\bibfnamefont {F.}~\bibnamefont {Gliozzi}}, \ and\ \bibinfo {author}
  {\bibfnamefont {M.}~\bibnamefont {Panero}},\ }\href {\doibase
  10.1007/JHEP06(2011)142} {\bibfield  {journal} {\bibinfo  {journal} {JHEP}\
  }\textbf {\bibinfo {volume} {06}},\ \bibinfo {pages} {142} (\bibinfo {year}
  {2011})},\ \Eprint {http://arxiv.org/abs/1105.0359} {arXiv:1105.0359
  [hep-lat]} \BibitemShut {NoStop}%
\bibitem [{\citenamefont {Caselle}\ \emph {et~al.}(2015)\citenamefont
  {Caselle}, \citenamefont {Nada},\ and\ \citenamefont
  {Panero}}]{Caselle:2015tza}%
  \BibitemOpen
  \bibfield  {author} {\bibinfo {author} {\bibfnamefont {M.}~\bibnamefont
  {Caselle}}, \bibinfo {author} {\bibfnamefont {A.}~\bibnamefont {Nada}}, \
  and\ \bibinfo {author} {\bibfnamefont {M.}~\bibnamefont {Panero}},\ }\href
  {\doibase 10.1007/JHEP07(2015)143} {\bibfield  {journal} {\bibinfo  {journal}
  {JHEP}\ }\textbf {\bibinfo {volume} {07}},\ \bibinfo {pages} {143} (\bibinfo
  {year} {2015})},\ \bibinfo {note} {[Erratum: JHEP 11, 016 (2017)]},\ \Eprint
  {http://arxiv.org/abs/1505.01106} {arXiv:1505.01106 [hep-lat]} \BibitemShut
  {NoStop}%
\bibitem [{\citenamefont {Meyer}(2004)}]{Meyer:2004gx}%
  \BibitemOpen
  \bibfield  {author} {\bibinfo {author} {\bibfnamefont {H.~B.}\ \bibnamefont
  {Meyer}},\ }\emph {\bibinfo {title} {{Glueball regge trajectories}}},\
  \href@noop {} {\bibinfo {type} {Other thesis}} (\bibinfo {year} {2004}),\
  \Eprint {http://arxiv.org/abs/hep-lat/0508002} {arXiv:hep-lat/0508002}
  \BibitemShut {NoStop}%
\bibitem [{\citenamefont {Lucini}\ \emph {et~al.}(2004)\citenamefont {Lucini},
  \citenamefont {Teper},\ and\ \citenamefont {Wenger}}]{Lucini:2004my}%
  \BibitemOpen
  \bibfield  {author} {\bibinfo {author} {\bibfnamefont {B.}~\bibnamefont
  {Lucini}}, \bibinfo {author} {\bibfnamefont {M.}~\bibnamefont {Teper}}, \
  and\ \bibinfo {author} {\bibfnamefont {U.}~\bibnamefont {Wenger}},\ }\href
  {\doibase 10.1088/1126-6708/2004/06/012} {\bibfield  {journal} {\bibinfo
  {journal} {JHEP}\ }\textbf {\bibinfo {volume} {06}},\ \bibinfo {pages} {012}
  (\bibinfo {year} {2004})},\ \Eprint {http://arxiv.org/abs/hep-lat/0404008}
  {arXiv:hep-lat/0404008} \BibitemShut {NoStop}%
\bibitem [{\citenamefont {Bennett}\ \emph {et~al.}(2020)\citenamefont
  {Bennett}, \citenamefont {Holligan}, \citenamefont {Hong}, \citenamefont
  {Lee}, \citenamefont {Lin}, \citenamefont {Lucini}, \citenamefont {Piai},\
  and\ \citenamefont {Vadacchino}}]{Bennett:2020hqd}%
  \BibitemOpen
  \bibfield  {author} {\bibinfo {author} {\bibfnamefont {E.}~\bibnamefont
  {Bennett}}, \bibinfo {author} {\bibfnamefont {J.}~\bibnamefont {Holligan}},
  \bibinfo {author} {\bibfnamefont {D.~K.}\ \bibnamefont {Hong}}, \bibinfo
  {author} {\bibfnamefont {J.-W.}\ \bibnamefont {Lee}}, \bibinfo {author}
  {\bibfnamefont {C.~J.~D.}\ \bibnamefont {Lin}}, \bibinfo {author}
  {\bibfnamefont {B.}~\bibnamefont {Lucini}}, \bibinfo {author} {\bibfnamefont
  {M.}~\bibnamefont {Piai}}, \ and\ \bibinfo {author} {\bibfnamefont
  {D.}~\bibnamefont {Vadacchino}},\ }\href {\doibase
  10.1103/PhysRevD.102.011501} {\bibfield  {journal} {\bibinfo  {journal}
  {Phys. Rev. D}\ }\textbf {\bibinfo {volume} {102}},\ \bibinfo {pages}
  {011501} (\bibinfo {year} {2020})},\ \Eprint
  {http://arxiv.org/abs/2004.11063} {arXiv:2004.11063 [hep-lat]} \BibitemShut
  {NoStop}%
\bibitem [{\citenamefont {Lucini}\ \emph {et~al.}(2010)\citenamefont {Lucini},
  \citenamefont {Rago},\ and\ \citenamefont {Rinaldi}}]{Lucini:2010nv}%
  \BibitemOpen
  \bibfield  {author} {\bibinfo {author} {\bibfnamefont {B.}~\bibnamefont
  {Lucini}}, \bibinfo {author} {\bibfnamefont {A.}~\bibnamefont {Rago}}, \ and\
  \bibinfo {author} {\bibfnamefont {E.}~\bibnamefont {Rinaldi}},\ }\href
  {\doibase 10.1007/JHEP08(2010)119} {\bibfield  {journal} {\bibinfo  {journal}
  {JHEP}\ }\textbf {\bibinfo {volume} {08}},\ \bibinfo {pages} {119} (\bibinfo
  {year} {2010})},\ \Eprint {http://arxiv.org/abs/1007.3879} {arXiv:1007.3879
  [hep-lat]} \BibitemShut {NoStop}%
\bibitem [{\citenamefont {Gabadadze}\ and\ \citenamefont
  {Iglesias}(2005)}]{Gabadadze:2004jq}%
  \BibitemOpen
  \bibfield  {author} {\bibinfo {author} {\bibfnamefont {G.}~\bibnamefont
  {Gabadadze}}\ and\ \bibinfo {author} {\bibfnamefont {A.}~\bibnamefont
  {Iglesias}},\ }\href {\doibase 10.1016/j.physletb.2005.01.042} {\bibfield
  {journal} {\bibinfo  {journal} {Phys. Lett. B}\ }\textbf {\bibinfo {volume}
  {609}},\ \bibinfo {pages} {167} (\bibinfo {year} {2005})},\ \Eprint
  {http://arxiv.org/abs/hep-th/0411278} {arXiv:hep-th/0411278} \BibitemShut
  {NoStop}%
\bibitem [{\citenamefont {Forestell}\ \emph {et~al.}(2017)\citenamefont
  {Forestell}, \citenamefont {Morrissey},\ and\ \citenamefont
  {Sigurdson}}]{Forestell:2016qhc}%
  \BibitemOpen
  \bibfield  {author} {\bibinfo {author} {\bibfnamefont {L.}~\bibnamefont
  {Forestell}}, \bibinfo {author} {\bibfnamefont {D.~E.}\ \bibnamefont
  {Morrissey}}, \ and\ \bibinfo {author} {\bibfnamefont {K.}~\bibnamefont
  {Sigurdson}},\ }\href {\doibase 10.1103/PhysRevD.95.015032} {\bibfield
  {journal} {\bibinfo  {journal} {Phys. Rev. D}\ }\textbf {\bibinfo {volume}
  {95}},\ \bibinfo {pages} {015032} (\bibinfo {year} {2017})},\ \Eprint
  {http://arxiv.org/abs/1605.08048} {arXiv:1605.08048 [hep-ph]} \BibitemShut
  {NoStop}%
\bibitem [{\citenamefont {Bonati}\ \emph {et~al.}(2017)\citenamefont {Bonati},
  \citenamefont {D'Elia}, \citenamefont {Rossi},\ and\ \citenamefont
  {Vicari}}]{Bonati:2017pfq}%
  \BibitemOpen
  \bibfield  {author} {\bibinfo {author} {\bibfnamefont {C.}~\bibnamefont
  {Bonati}}, \bibinfo {author} {\bibfnamefont {M.}~\bibnamefont {D'Elia}},
  \bibinfo {author} {\bibfnamefont {P.}~\bibnamefont {Rossi}}, \ and\ \bibinfo
  {author} {\bibfnamefont {E.}~\bibnamefont {Vicari}},\ }\href {\doibase
  10.22323/1.256.0348} {\bibfield  {journal} {\bibinfo  {journal} {PoS}\
  }\textbf {\bibinfo {volume} {LATTICE2016}},\ \bibinfo {pages} {348} (\bibinfo
  {year} {2017})},\ \Eprint {http://arxiv.org/abs/1702.01049} {arXiv:1702.01049
  [hep-lat]} \BibitemShut {NoStop}%
\bibitem [{\citenamefont {Gaiotto}\ \emph {et~al.}(2017)\citenamefont
  {Gaiotto}, \citenamefont {Kapustin}, \citenamefont {Komargodski},\ and\
  \citenamefont {Seiberg}}]{Gaiotto:2017yup}%
  \BibitemOpen
  \bibfield  {author} {\bibinfo {author} {\bibfnamefont {D.}~\bibnamefont
  {Gaiotto}}, \bibinfo {author} {\bibfnamefont {A.}~\bibnamefont {Kapustin}},
  \bibinfo {author} {\bibfnamefont {Z.}~\bibnamefont {Komargodski}}, \ and\
  \bibinfo {author} {\bibfnamefont {N.}~\bibnamefont {Seiberg}},\ }\href
  {\doibase 10.1007/JHEP05(2017)091} {\bibfield  {journal} {\bibinfo  {journal}
  {JHEP}\ }\textbf {\bibinfo {volume} {05}},\ \bibinfo {pages} {091} (\bibinfo
  {year} {2017})},\ \Eprint {http://arxiv.org/abs/1703.00501} {arXiv:1703.00501
  [hep-th]} \BibitemShut {NoStop}%
\bibitem [{\citenamefont {Dubovsky}\ \emph {et~al.}(2012)\citenamefont
  {Dubovsky}, \citenamefont {Lawrence},\ and\ \citenamefont
  {Roberts}}]{Dubovsky:2011tu}%
  \BibitemOpen
  \bibfield  {author} {\bibinfo {author} {\bibfnamefont {S.}~\bibnamefont
  {Dubovsky}}, \bibinfo {author} {\bibfnamefont {A.}~\bibnamefont {Lawrence}},
  \ and\ \bibinfo {author} {\bibfnamefont {M.~M.}\ \bibnamefont {Roberts}},\
  }\href {\doibase 10.1007/JHEP02(2012)053} {\bibfield  {journal} {\bibinfo
  {journal} {JHEP}\ }\textbf {\bibinfo {volume} {02}},\ \bibinfo {pages} {053}
  (\bibinfo {year} {2012})},\ \Eprint {http://arxiv.org/abs/1105.3740}
  {arXiv:1105.3740 [hep-th]} \BibitemShut {NoStop}%
\bibitem [{\citenamefont {Bigazzi}\ \emph {et~al.}(2015)\citenamefont
  {Bigazzi}, \citenamefont {Cotrone},\ and\ \citenamefont
  {Sisca}}]{Bigazzi:2015bna}%
  \BibitemOpen
  \bibfield  {author} {\bibinfo {author} {\bibfnamefont {F.}~\bibnamefont
  {Bigazzi}}, \bibinfo {author} {\bibfnamefont {A.~L.}\ \bibnamefont
  {Cotrone}}, \ and\ \bibinfo {author} {\bibfnamefont {R.}~\bibnamefont
  {Sisca}},\ }\href {\doibase 10.1007/JHEP08(2015)090} {\bibfield  {journal}
  {\bibinfo  {journal} {JHEP}\ }\textbf {\bibinfo {volume} {08}},\ \bibinfo
  {pages} {090} (\bibinfo {year} {2015})},\ \Eprint
  {http://arxiv.org/abs/1506.03826} {arXiv:1506.03826 [hep-th]} \BibitemShut
  {NoStop}%
\bibitem [{\citenamefont {Aghanim}\ \emph {et~al.}(2020)\citenamefont {Aghanim}
  \emph {et~al.}}]{Aghanim:2018eyx}%
  \BibitemOpen
  \bibfield  {author} {\bibinfo {author} {\bibfnamefont {N.}~\bibnamefont
  {Aghanim}} \emph {et~al.} (\bibinfo {collaboration} {Planck}),\ }\href
  {\doibase 10.1051/0004-6361/201833910} {\bibfield  {journal} {\bibinfo
  {journal} {Astron. Astrophys.}\ }\textbf {\bibinfo {volume} {641}},\ \bibinfo
  {pages} {A6} (\bibinfo {year} {2020})},\ \Eprint
  {http://arxiv.org/abs/1807.06209} {arXiv:1807.06209 [astro-ph.CO]}
  \BibitemShut {NoStop}%
\bibitem [{\citenamefont {Tulin}\ and\ \citenamefont
  {Yu}(2018)}]{Tulin:2017ara}%
  \BibitemOpen
  \bibfield  {author} {\bibinfo {author} {\bibfnamefont {S.}~\bibnamefont
  {Tulin}}\ and\ \bibinfo {author} {\bibfnamefont {H.-B.}\ \bibnamefont {Yu}},\
  }\href {\doibase 10.1016/j.physrep.2017.11.004} {\bibfield  {journal}
  {\bibinfo  {journal} {Phys. Rept.}\ }\textbf {\bibinfo {volume} {730}},\
  \bibinfo {pages} {1} (\bibinfo {year} {2018})},\ \Eprint
  {http://arxiv.org/abs/1705.02358} {arXiv:1705.02358 [hep-ph]} \BibitemShut
  {NoStop}%
\bibitem [{\citenamefont {Acharya}\ \emph {et~al.}(2017)\citenamefont
  {Acharya}, \citenamefont {Fairbairn},\ and\ \citenamefont
  {Hardy}}]{Acharya:2017szw}%
  \BibitemOpen
  \bibfield  {author} {\bibinfo {author} {\bibfnamefont {B.~S.}\ \bibnamefont
  {Acharya}}, \bibinfo {author} {\bibfnamefont {M.}~\bibnamefont {Fairbairn}},
  \ and\ \bibinfo {author} {\bibfnamefont {E.}~\bibnamefont {Hardy}},\ }\href
  {\doibase 10.1007/JHEP07(2017)100} {\bibfield  {journal} {\bibinfo  {journal}
  {JHEP}\ }\textbf {\bibinfo {volume} {07}},\ \bibinfo {pages} {100} (\bibinfo
  {year} {2017})},\ \Eprint {http://arxiv.org/abs/1704.01804} {arXiv:1704.01804
  [hep-ph]} \BibitemShut {NoStop}%
\bibitem [{\citenamefont {Grin}\ \emph {et~al.}(2019)\citenamefont {Grin},
  \citenamefont {Amin}, \citenamefont {Gluscevic}, \citenamefont {Hlǒzek},
  \citenamefont {Marsh}, \citenamefont {Poulin}, \citenamefont
  {Prescod-Weinstein},\ and\ \citenamefont {Smith}}]{Grin:2019mub}%
  \BibitemOpen
  \bibfield  {author} {\bibinfo {author} {\bibfnamefont {D.}~\bibnamefont
  {Grin}}, \bibinfo {author} {\bibfnamefont {M.~A.}\ \bibnamefont {Amin}},
  \bibinfo {author} {\bibfnamefont {V.}~\bibnamefont {Gluscevic}}, \bibinfo
  {author} {\bibfnamefont {R.}~\bibnamefont {Hlǒzek}}, \bibinfo {author}
  {\bibfnamefont {D.~J.}\ \bibnamefont {Marsh}}, \bibinfo {author}
  {\bibfnamefont {V.}~\bibnamefont {Poulin}}, \bibinfo {author} {\bibfnamefont
  {C.}~\bibnamefont {Prescod-Weinstein}}, \ and\ \bibinfo {author}
  {\bibfnamefont {T.~L.}\ \bibnamefont {Smith}},\ }\href@noop {} {\  (\bibinfo
  {year} {2019})},\ \Eprint {http://arxiv.org/abs/1904.09003} {arXiv:1904.09003
  [astro-ph.CO]} \BibitemShut {NoStop}%
\bibitem [{\citenamefont {Ferreira}(2020)}]{Ferreira:2020fam}%
  \BibitemOpen
  \bibfield  {author} {\bibinfo {author} {\bibfnamefont {E.~G.}\ \bibnamefont
  {Ferreira}},\ }\href@noop {} {\  (\bibinfo {year} {2020})},\ \Eprint
  {http://arxiv.org/abs/2005.03254} {arXiv:2005.03254 [astro-ph.CO]}
  \BibitemShut {NoStop}%
\bibitem [{\citenamefont {Brito}\ \emph {et~al.}(2015)\citenamefont {Brito},
  \citenamefont {Cardoso},\ and\ \citenamefont {Pani}}]{Brito:2015oca}%
  \BibitemOpen
  \bibfield  {author} {\bibinfo {author} {\bibfnamefont {R.}~\bibnamefont
  {Brito}}, \bibinfo {author} {\bibfnamefont {V.}~\bibnamefont {Cardoso}}, \
  and\ \bibinfo {author} {\bibfnamefont {P.}~\bibnamefont {Pani}},\ }\href
  {\doibase 10.1007/978-3-319-19000-6} {\emph {\bibinfo {title}
  {{Superradiance}: {Energy Extraction, Black-Hole Bombs and Implications for
  Astrophysics and Particle Physics}}}},\ Vol.\ \bibinfo {volume} {906}\
  (\bibinfo  {publisher} {Springer},\ \bibinfo {year} {2015})\ \Eprint
  {http://arxiv.org/abs/1501.06570} {arXiv:1501.06570 [gr-qc]} \BibitemShut
  {NoStop}%
\bibitem [{\citenamefont {Hlozek}\ \emph {et~al.}(2018)\citenamefont {Hlozek},
  \citenamefont {Marsh},\ and\ \citenamefont {Grin}}]{Hlozek:2017zzf}%
  \BibitemOpen
  \bibfield  {author} {\bibinfo {author} {\bibfnamefont {R.}~\bibnamefont
  {Hlozek}}, \bibinfo {author} {\bibfnamefont {D.~J.~E.}\ \bibnamefont
  {Marsh}}, \ and\ \bibinfo {author} {\bibfnamefont {D.}~\bibnamefont {Grin}},\
  }\href {\doibase 10.1093/mnras/sty271} {\bibfield  {journal} {\bibinfo
  {journal} {Mon. Not. Roy. Astron. Soc.}\ }\textbf {\bibinfo {volume} {476}},\
  \bibinfo {pages} {3063} (\bibinfo {year} {2018})},\ \Eprint
  {http://arxiv.org/abs/1708.05681} {arXiv:1708.05681 [astro-ph.CO]}
  \BibitemShut {NoStop}%
\bibitem [{\citenamefont {Hwang}\ and\ \citenamefont
  {Noh}(2009)}]{Hwang:2009js}%
  \BibitemOpen
  \bibfield  {author} {\bibinfo {author} {\bibfnamefont {J.-c.}\ \bibnamefont
  {Hwang}}\ and\ \bibinfo {author} {\bibfnamefont {H.}~\bibnamefont {Noh}},\
  }\href {\doibase 10.1016/j.physletb.2009.08.031} {\bibfield  {journal}
  {\bibinfo  {journal} {Phys. Lett. B}\ }\textbf {\bibinfo {volume} {680}},\
  \bibinfo {pages} {1} (\bibinfo {year} {2009})},\ \Eprint
  {http://arxiv.org/abs/0902.4738} {arXiv:0902.4738 [astro-ph.CO]} \BibitemShut
  {NoStop}%
\bibitem [{\citenamefont {Ir\v{s}i\v{c}}\ \emph {et~al.}(2017)\citenamefont
  {Ir\v{s}i\v{c}}, \citenamefont {Viel}, \citenamefont {Haehnelt},
  \citenamefont {Bolton},\ and\ \citenamefont {Becker}}]{Irsic:2017yje}%
  \BibitemOpen
  \bibfield  {author} {\bibinfo {author} {\bibfnamefont {V.}~\bibnamefont
  {Ir\v{s}i\v{c}}}, \bibinfo {author} {\bibfnamefont {M.}~\bibnamefont {Viel}},
  \bibinfo {author} {\bibfnamefont {M.~G.}\ \bibnamefont {Haehnelt}}, \bibinfo
  {author} {\bibfnamefont {J.~S.}\ \bibnamefont {Bolton}}, \ and\ \bibinfo
  {author} {\bibfnamefont {G.~D.}\ \bibnamefont {Becker}},\ }\href {\doibase
  10.1103/PhysRevLett.119.031302} {\bibfield  {journal} {\bibinfo  {journal}
  {Phys. Rev. Lett.}\ }\textbf {\bibinfo {volume} {119}},\ \bibinfo {pages}
  {031302} (\bibinfo {year} {2017})},\ \Eprint
  {http://arxiv.org/abs/1703.04683} {arXiv:1703.04683 [astro-ph.CO]}
  \BibitemShut {NoStop}%
\bibitem [{\citenamefont {Soni}\ and\ \citenamefont
  {Zhang}(2016)}]{Soni:2016gzf}%
  \BibitemOpen
  \bibfield  {author} {\bibinfo {author} {\bibfnamefont {A.}~\bibnamefont
  {Soni}}\ and\ \bibinfo {author} {\bibfnamefont {Y.}~\bibnamefont {Zhang}},\
  }\href {\doibase 10.1103/PhysRevD.93.115025} {\bibfield  {journal} {\bibinfo
  {journal} {Phys. Rev. D}\ }\textbf {\bibinfo {volume} {93}},\ \bibinfo
  {pages} {115025} (\bibinfo {year} {2016})},\ \Eprint
  {http://arxiv.org/abs/1602.00714} {arXiv:1602.00714 [hep-ph]} \BibitemShut
  {NoStop}%
\bibitem [{\citenamefont {Liddle}\ \emph {et~al.}(2000)\citenamefont {Liddle},
  \citenamefont {Lyth}, \citenamefont {Malik},\ and\ \citenamefont
  {Wands}}]{Liddle:1999hq}%
  \BibitemOpen
  \bibfield  {author} {\bibinfo {author} {\bibfnamefont {A.~R.}\ \bibnamefont
  {Liddle}}, \bibinfo {author} {\bibfnamefont {D.~H.}\ \bibnamefont {Lyth}},
  \bibinfo {author} {\bibfnamefont {K.~A.}\ \bibnamefont {Malik}}, \ and\
  \bibinfo {author} {\bibfnamefont {D.}~\bibnamefont {Wands}},\ }\href
  {\doibase 10.1103/PhysRevD.61.103509} {\bibfield  {journal} {\bibinfo
  {journal} {Phys. Rev. D}\ }\textbf {\bibinfo {volume} {61}},\ \bibinfo
  {pages} {103509} (\bibinfo {year} {2000})},\ \Eprint
  {http://arxiv.org/abs/hep-ph/9912473} {arXiv:hep-ph/9912473} \BibitemShut
  {NoStop}%
\bibitem [{\citenamefont {Wands}\ \emph {et~al.}(2000)\citenamefont {Wands},
  \citenamefont {Malik}, \citenamefont {Lyth},\ and\ \citenamefont
  {Liddle}}]{Wands:2000dp}%
  \BibitemOpen
  \bibfield  {author} {\bibinfo {author} {\bibfnamefont {D.}~\bibnamefont
  {Wands}}, \bibinfo {author} {\bibfnamefont {K.~A.}\ \bibnamefont {Malik}},
  \bibinfo {author} {\bibfnamefont {D.~H.}\ \bibnamefont {Lyth}}, \ and\
  \bibinfo {author} {\bibfnamefont {A.~R.}\ \bibnamefont {Liddle}},\ }\href
  {\doibase 10.1103/PhysRevD.62.043527} {\bibfield  {journal} {\bibinfo
  {journal} {Phys. Rev. D}\ }\textbf {\bibinfo {volume} {62}},\ \bibinfo
  {pages} {043527} (\bibinfo {year} {2000})},\ \Eprint
  {http://arxiv.org/abs/astro-ph/0003278} {arXiv:astro-ph/0003278} \BibitemShut
  {NoStop}%
\bibitem [{\citenamefont {Weinberg}(2003)}]{Weinberg:2003sw}%
  \BibitemOpen
  \bibfield  {author} {\bibinfo {author} {\bibfnamefont {S.}~\bibnamefont
  {Weinberg}},\ }\href {\doibase 10.1103/PhysRevD.67.123504} {\bibfield
  {journal} {\bibinfo  {journal} {Phys. Rev. D}\ }\textbf {\bibinfo {volume}
  {67}},\ \bibinfo {pages} {123504} (\bibinfo {year} {2003})},\ \Eprint
  {http://arxiv.org/abs/astro-ph/0302326} {arXiv:astro-ph/0302326} \BibitemShut
  {NoStop}%
\bibitem [{\citenamefont {Mukhanov}\ \emph {et~al.}(1992)\citenamefont
  {Mukhanov}, \citenamefont {Feldman},\ and\ \citenamefont
  {Brandenberger}}]{Mukhanov:1990me}%
  \BibitemOpen
  \bibfield  {author} {\bibinfo {author} {\bibfnamefont {V.~F.}\ \bibnamefont
  {Mukhanov}}, \bibinfo {author} {\bibfnamefont {H.}~\bibnamefont {Feldman}}, \
  and\ \bibinfo {author} {\bibfnamefont {R.~H.}\ \bibnamefont
  {Brandenberger}},\ }\href {\doibase 10.1016/0370-1573(92)90044-Z} {\bibfield
  {journal} {\bibinfo  {journal} {Phys. Rept.}\ }\textbf {\bibinfo {volume}
  {215}},\ \bibinfo {pages} {203} (\bibinfo {year} {1992})}\BibitemShut
  {NoStop}%
\bibitem [{\citenamefont {Crotty}\ \emph {et~al.}(2003)\citenamefont {Crotty},
  \citenamefont {Garcia-Bellido}, \citenamefont {Lesgourgues},\ and\
  \citenamefont {Riazuelo}}]{Crotty:2003rz}%
  \BibitemOpen
  \bibfield  {author} {\bibinfo {author} {\bibfnamefont {P.}~\bibnamefont
  {Crotty}}, \bibinfo {author} {\bibfnamefont {J.}~\bibnamefont
  {Garcia-Bellido}}, \bibinfo {author} {\bibfnamefont {J.}~\bibnamefont
  {Lesgourgues}}, \ and\ \bibinfo {author} {\bibfnamefont {A.}~\bibnamefont
  {Riazuelo}},\ }\href {\doibase 10.1103/PhysRevLett.91.171301} {\bibfield
  {journal} {\bibinfo  {journal} {Phys. Rev. Lett.}\ }\textbf {\bibinfo
  {volume} {91}},\ \bibinfo {pages} {171301} (\bibinfo {year} {2003})},\
  \Eprint {http://arxiv.org/abs/astro-ph/0306286} {arXiv:astro-ph/0306286}
  \BibitemShut {NoStop}%
\bibitem [{\citenamefont {Li}\ \emph {et~al.}(2011)\citenamefont {Li},
  \citenamefont {Liu}, \citenamefont {Xia},\ and\ \citenamefont
  {Cai}}]{Li:2010yb}%
  \BibitemOpen
  \bibfield  {author} {\bibinfo {author} {\bibfnamefont {H.}~\bibnamefont
  {Li}}, \bibinfo {author} {\bibfnamefont {J.}~\bibnamefont {Liu}}, \bibinfo
  {author} {\bibfnamefont {J.-Q.}\ \bibnamefont {Xia}}, \ and\ \bibinfo
  {author} {\bibfnamefont {Y.-F.}\ \bibnamefont {Cai}},\ }\href {\doibase
  10.1103/PhysRevD.83.123517} {\bibfield  {journal} {\bibinfo  {journal} {Phys.
  Rev. D}\ }\textbf {\bibinfo {volume} {83}},\ \bibinfo {pages} {123517}
  (\bibinfo {year} {2011})},\ \Eprint {http://arxiv.org/abs/1012.2511}
  {arXiv:1012.2511 [astro-ph.CO]} \BibitemShut {NoStop}%
\bibitem [{\citenamefont {V\"aliviita}\ and\ \citenamefont
  {Muhonen}(2003)}]{PhysRevLett.91.131302}%
  \BibitemOpen
  \bibfield  {author} {\bibinfo {author} {\bibfnamefont {J.}~\bibnamefont
  {V\"aliviita}}\ and\ \bibinfo {author} {\bibfnamefont {V.}~\bibnamefont
  {Muhonen}},\ }\href {\doibase 10.1103/PhysRevLett.91.131302} {\bibfield
  {journal} {\bibinfo  {journal} {Phys. Rev. Lett.}\ }\textbf {\bibinfo
  {volume} {91}},\ \bibinfo {pages} {131302} (\bibinfo {year}
  {2003})}\BibitemShut {NoStop}%
\bibitem [{\citenamefont {Salpeter}(1964)}]{Salpeter:1964kb}%
  \BibitemOpen
  \bibfield  {author} {\bibinfo {author} {\bibfnamefont {E.}~\bibnamefont
  {Salpeter}},\ }\href {\doibase 10.1086/147973} {\bibfield  {journal}
  {\bibinfo  {journal} {Astrophys. J.}\ }\textbf {\bibinfo {volume} {140}},\
  \bibinfo {pages} {796} (\bibinfo {year} {1964})}\BibitemShut {NoStop}%
\bibitem [{\citenamefont {Volonteri}(2010)}]{Volonteri:2010}%
  \BibitemOpen
  \bibfield  {author} {\bibinfo {author} {\bibfnamefont {M.}~\bibnamefont
  {Volonteri}},\ }\href {\doibase 10.1007/s00159-010-0029-x} {\bibfield
  {journal} {\bibinfo  {journal} {The Astronomy and Astrophysics Review}\
  }\textbf {\bibinfo {volume} {18}},\ \bibinfo {pages} {279–315} (\bibinfo
  {year} {2010})}\BibitemShut {NoStop}%
\bibitem [{\citenamefont {Arvanitaki}\ \emph {et~al.}(2015)\citenamefont
  {Arvanitaki}, \citenamefont {Baryakhtar},\ and\ \citenamefont
  {Huang}}]{Arvanitaki:2014wva}%
  \BibitemOpen
  \bibfield  {author} {\bibinfo {author} {\bibfnamefont {A.}~\bibnamefont
  {Arvanitaki}}, \bibinfo {author} {\bibfnamefont {M.}~\bibnamefont
  {Baryakhtar}}, \ and\ \bibinfo {author} {\bibfnamefont {X.}~\bibnamefont
  {Huang}},\ }\href {\doibase 10.1103/PhysRevD.91.084011} {\bibfield  {journal}
  {\bibinfo  {journal} {Phys. Rev. D}\ }\textbf {\bibinfo {volume} {91}},\
  \bibinfo {pages} {084011} (\bibinfo {year} {2015})},\ \Eprint
  {http://arxiv.org/abs/1411.2263} {arXiv:1411.2263 [hep-ph]} \BibitemShut
  {NoStop}%
\bibitem [{\citenamefont {Yoshino}\ and\ \citenamefont
  {Kodama}(2015)}]{Yoshino:2015nsa}%
  \BibitemOpen
  \bibfield  {author} {\bibinfo {author} {\bibfnamefont {H.}~\bibnamefont
  {Yoshino}}\ and\ \bibinfo {author} {\bibfnamefont {H.}~\bibnamefont
  {Kodama}},\ }\href {\doibase 10.1088/0264-9381/32/21/214001} {\bibfield
  {journal} {\bibinfo  {journal} {Class. Quant. Grav.}\ }\textbf {\bibinfo
  {volume} {32}},\ \bibinfo {pages} {214001} (\bibinfo {year} {2015})},\
  \Eprint {http://arxiv.org/abs/1505.00714} {arXiv:1505.00714 [gr-qc]}
  \BibitemShut {NoStop}%
\bibitem [{\citenamefont {Kormendy}\ and\ \citenamefont
  {Richstone}(1995)}]{Kormendy:1995er}%
  \BibitemOpen
  \bibfield  {author} {\bibinfo {author} {\bibfnamefont {J.}~\bibnamefont
  {Kormendy}}\ and\ \bibinfo {author} {\bibfnamefont {D.}~\bibnamefont
  {Richstone}},\ }\href {\doibase 10.1146/annurev.aa.33.090195.003053}
  {\bibfield  {journal} {\bibinfo  {journal} {Ann. Rev. Astron. Astrophys.}\
  }\textbf {\bibinfo {volume} {33}},\ \bibinfo {pages} {581} (\bibinfo {year}
  {1995})}\BibitemShut {NoStop}%
\bibitem [{\citenamefont {Magorrian}\ \emph {et~al.}(1998)\citenamefont
  {Magorrian} \emph {et~al.}}]{Magorrian:1997hw}%
  \BibitemOpen
  \bibfield  {author} {\bibinfo {author} {\bibfnamefont {J.}~\bibnamefont
  {Magorrian}} \emph {et~al.},\ }\href {\doibase 10.1086/300353} {\bibfield
  {journal} {\bibinfo  {journal} {Astron. J.}\ }\textbf {\bibinfo {volume}
  {115}},\ \bibinfo {pages} {2285} (\bibinfo {year} {1998})},\ \Eprint
  {http://arxiv.org/abs/astro-ph/9708072} {arXiv:astro-ph/9708072} \BibitemShut
  {NoStop}%
\bibitem [{\citenamefont {Heitmann}\ \emph {et~al.}(2006)\citenamefont
  {Heitmann}, \citenamefont {Lukic}, \citenamefont {Habib},\ and\ \citenamefont
  {Ricker}}]{Heitmann:2006eu}%
  \BibitemOpen
  \bibfield  {author} {\bibinfo {author} {\bibfnamefont {K.}~\bibnamefont
  {Heitmann}}, \bibinfo {author} {\bibfnamefont {Z.}~\bibnamefont {Lukic}},
  \bibinfo {author} {\bibfnamefont {S.}~\bibnamefont {Habib}}, \ and\ \bibinfo
  {author} {\bibfnamefont {P.~M.}\ \bibnamefont {Ricker}},\ }\href {\doibase
  10.1086/504868} {\bibfield  {journal} {\bibinfo  {journal} {Astrophys. J.
  Lett.}\ }\textbf {\bibinfo {volume} {642}},\ \bibinfo {pages} {L85} (\bibinfo
  {year} {2006})},\ \Eprint {http://arxiv.org/abs/astro-ph/0601233}
  {arXiv:astro-ph/0601233} \BibitemShut {NoStop}%
\bibitem [{\citenamefont {Lukic}\ \emph {et~al.}(2007)\citenamefont {Lukic},
  \citenamefont {Heitmann}, \citenamefont {Habib}, \citenamefont {Bashinsky},\
  and\ \citenamefont {Ricker}}]{Lukic:2007fc}%
  \BibitemOpen
  \bibfield  {author} {\bibinfo {author} {\bibfnamefont {Z.}~\bibnamefont
  {Lukic}}, \bibinfo {author} {\bibfnamefont {K.}~\bibnamefont {Heitmann}},
  \bibinfo {author} {\bibfnamefont {S.}~\bibnamefont {Habib}}, \bibinfo
  {author} {\bibfnamefont {S.}~\bibnamefont {Bashinsky}}, \ and\ \bibinfo
  {author} {\bibfnamefont {P.~M.}\ \bibnamefont {Ricker}},\ }\href {\doibase
  10.1086/523083} {\bibfield  {journal} {\bibinfo  {journal} {Astrophys. J.}\
  }\textbf {\bibinfo {volume} {671}},\ \bibinfo {pages} {1160} (\bibinfo {year}
  {2007})},\ \Eprint {http://arxiv.org/abs/astro-ph/0702360}
  {arXiv:astro-ph/0702360} \BibitemShut {NoStop}%
\bibitem [{\citenamefont {Tacchella}\ \emph {et~al.}(2018)\citenamefont
  {Tacchella}, \citenamefont {Bose}, \citenamefont {Conroy}, \citenamefont
  {Eisenstein},\ and\ \citenamefont {Johnson}}]{Tacchella:2018qny}%
  \BibitemOpen
  \bibfield  {author} {\bibinfo {author} {\bibfnamefont {S.}~\bibnamefont
  {Tacchella}}, \bibinfo {author} {\bibfnamefont {S.}~\bibnamefont {Bose}},
  \bibinfo {author} {\bibfnamefont {C.}~\bibnamefont {Conroy}}, \bibinfo
  {author} {\bibfnamefont {D.~J.}\ \bibnamefont {Eisenstein}}, \ and\ \bibinfo
  {author} {\bibfnamefont {B.~D.}\ \bibnamefont {Johnson}},\ }\href {\doibase
  10.3847/1538-4357/aae8e0} {\bibfield  {journal} {\bibinfo  {journal}
  {Astrophys. J.}\ }\textbf {\bibinfo {volume} {868}},\ \bibinfo {pages} {92}
  (\bibinfo {year} {2018})},\ \Eprint {http://arxiv.org/abs/1806.03299}
  {arXiv:1806.03299 [astro-ph.GA]} \BibitemShut {NoStop}%
\bibitem [{\citenamefont {Ishiyama}\ \emph {et~al.}(2020)\citenamefont
  {Ishiyama} \emph {et~al.}}]{Ishiyama:2020vao}%
  \BibitemOpen
  \bibfield  {author} {\bibinfo {author} {\bibfnamefont {T.}~\bibnamefont
  {Ishiyama}} \emph {et~al.},\ }\href@noop {} {\  (\bibinfo {year} {2020})},\
  \Eprint {http://arxiv.org/abs/2007.14720} {arXiv:2007.14720 [astro-ph.CO]}
  \BibitemShut {NoStop}%
\bibitem [{\citenamefont {Balberg}\ \emph {et~al.}(2002)\citenamefont
  {Balberg}, \citenamefont {Shapiro},\ and\ \citenamefont
  {Inagaki}}]{Balberg:2002ue}%
  \BibitemOpen
  \bibfield  {author} {\bibinfo {author} {\bibfnamefont {S.}~\bibnamefont
  {Balberg}}, \bibinfo {author} {\bibfnamefont {S.~L.}\ \bibnamefont
  {Shapiro}}, \ and\ \bibinfo {author} {\bibfnamefont {S.}~\bibnamefont
  {Inagaki}},\ }\href {\doibase 10.1086/339038} {\bibfield  {journal} {\bibinfo
   {journal} {Astrophys. J.}\ }\textbf {\bibinfo {volume} {568}},\ \bibinfo
  {pages} {475} (\bibinfo {year} {2002})},\ \Eprint
  {http://arxiv.org/abs/astro-ph/0110561} {arXiv:astro-ph/0110561} \BibitemShut
  {NoStop}%
\bibitem [{\citenamefont {Balberg}\ and\ \citenamefont
  {Shapiro}(2002)}]{Balberg:2001qg}%
  \BibitemOpen
  \bibfield  {author} {\bibinfo {author} {\bibfnamefont {S.}~\bibnamefont
  {Balberg}}\ and\ \bibinfo {author} {\bibfnamefont {S.~L.}\ \bibnamefont
  {Shapiro}},\ }\href {\doibase 10.1103/PhysRevLett.88.101301} {\bibfield
  {journal} {\bibinfo  {journal} {Phys. Rev. Lett.}\ }\textbf {\bibinfo
  {volume} {88}},\ \bibinfo {pages} {101301} (\bibinfo {year} {2002})},\
  \Eprint {http://arxiv.org/abs/astro-ph/0111176} {arXiv:astro-ph/0111176}
  \BibitemShut {NoStop}%
\end{thebibliography}%

\end{document}